\documentclass{aa}  

\usepackage[]{natbib}
\usepackage{lscape}
\usepackage{longtable}
\usepackage{float}
\usepackage{cuted}
\usepackage{caption}
\usepackage{multirow}
\usepackage{threeparttable}
\usepackage{nicefrac}
\usepackage{graphicx}
\usepackage{txfonts}
\usepackage{upgreek}

\begin{document} 

   \title{CHANG-ES XIX: Galaxy NGC~4013 -- a diffusion-dominated radio halo with plane-parallel disk and vertical halo magnetic fields }

  \author{Y. Stein \inst{1,2}
          \and
          R.-J. Dettmar \inst{2,3}
          \and
          M. We\.zgowiec \inst{4} 
            \and 
          J. Irwin \inst{5}
           \and
          R. Beck \inst{6}
          \and
          T. Wiegert \inst{5}
          \and
          M. Krause \inst{6}
          \and
          J.-T. Li \inst{7}
          \and
          V. Heesen \inst{8}
          \and
          A. Miskolczi \inst{2}
          \and
          S. MacDonald \inst{5}
          \and 
          J. English \inst{9}
          }

   \institute{Observatoire astronomique de Strasbourg, Universit\'e de Strasbourg, CNRS, UMR 7550, 11 rue de l'Universit\'e, \\ 67000 Strasbourg, France; \email{yelena.stein@astro.unistra.fr}
   		\and
   	Ruhr-Universit\"at Bochum, Fakultät für Physik und Astronomie, Astronomisches Institut (AIRUB), 44780 Bochum, Germany
     \and
     Ruhr-Universit\"at Bochum, Research Department: Plasmas with Complex Interactions,  44780 Bochum, Germany
     \and
	Obserwatorium Astronomiczne Uniwersytetu Jagiello\'nskiego, ul. Orla 171, 30-244 Krak\'ow, Poland
      \and
     Department of Physics, Engeneering Physics, \& Astronomy, Queen's University, Kingston, Ontario, Canada, K7L 3N6
        \and
        Max-Planck-Institut f\"ur Radioastronomie, Auf dem H\"ugel 69, 53121 Bonn, Germany
         \and
        Department of Astronomy, University of Michigan, 311 West Hall, 1085 S. University Ave, Ann Arbor, MI, 48109-1107, U.S.A.
        \and
        Universität Hamburg, Hamburger Sternwarte, Gojenbergsweg 112, 21029 Hamburg, Germany
         \and
       Department of Physics and Astronomy, University of Manitoba, Winnipeg, Manitoba, R3T 2N2, Canada 
          }

   \date{Received March 27, 2019}

 
  \abstract
   {The radio continuum halos of edge-on spiral galaxies have diverse morphologies, with different magnetic field properties and cosmic ray (CR) transport processes into the halo.  }
   { Using the  Continuum HAloes in Nearby Galaxies – an EVLA Survey (CHANG-ES) radio continuum data from the Karl G. Jansky Very Large Array (VLA) in two frequency bands, 6\,GHz (C-band) and 1.5 GHz (L-band), we analyzed the radio properties, including polarization and the transport processes of the CR electrons (CREs), in the edge-on spiral galaxy NGC~4013. Supplementary LOw-Frequency ARray (LOFAR) data at 150\,MHz are used to study the low-frequency properties of this galaxy and X-ray data are used to investigate the central region.}
   {We determined the total radio flux densities (central source, disk, halo and total) as well as the radio scale heights of the radio continuum emission at both CHANG-ES frequencies and at the LOFAR frequency. We derived the magnetic field orientation from CHANG-ES polarization data and rotation measure synthesis (RM synthesis). Furthermore, we used the revised equipartition formula to calculate the magnetic field strength. Lastly, we modeled the processes of CR transport into the halo with the 1D \textsc{spinnaker} model.}
 {The central point source dominates the radio continuum emission with a mean of $\sim$~35\% of the total flux density emerging from the central source in both CHANG-ES bands. Complementary X-ray data from \emph{Chandra} show one dominant point source in the central part. The XMM-Newton spectrum shows hard X-rays, but no clear AGN classification is possible at this time. The radio continuum halo of NGC~4013 in C-band is rather small, while the low-frequency LOFAR data reveal a large halo. The scale height analysis shows that Gaussian fits, with halo scale heights of 1.2\,kpc in C-band, 2.0\,kpc in L-band, and 3.1\,kpc at 150\,MHz, better represent the intensity profiles than do exponential fits. The frequency dependence gives clear preference to diffusive CRE transport. The radio halo of NGC~4013 is relatively faint and contributes only 40\% and 56\% of the total flux density in C-band and L-band, respectively. This is less than in galaxies with wind-driven halos. While the \textsc{spinnaker} models of the radio profiles show that advection with a launching velocity of $\sim$ 20\,km~s$^{-1}$ (increasing to  $\sim$ 50\,km~s$^{-1}$ at 4\,kpc height) fits the data equally well or slightly better, diffusion is the dominating transport process up to heights of 1--2\,kpc. The polarization data reveal plane-parallel, regular magnetic fields within the entire disk and vertical halo components indicating the presence of an axisymmetric field having a radial component pointing outwards. The mean magnetic field strength of the disk of NGC~4013 of 6.6\,$\upmu$G is rather small. Large-scale vertical fields are observed in the halo out to heights of about 6\,kpc.}
   {The  interaction  and  the  low  star formation rate (SFR)  across the disk of NGC~4013 probably influence the appearance of its radio continuum halo and are correlated with the low total magnetic field strength. Several observable quantities give consistent evidence that the CR transport in the halo of NGC~4013 is diffusive: the frequency dependence of the synchrotron scale height, the disk/halo flux density ratio, the vertical profile of the synchrotron spectral index, the small propagation speed measured modeled with \textsc{spinnaker}, and the low temperature of the X-ray emitting hot gas.
   }
  
   \keywords{galaxies: magnetic fields - individual galaxies: NGC 4013 - galaxies: halos}
   
   \titlerunning{CHANG-ES XIX: The galaxy NGC 4013}
  \authorrunning{Stein et al.}
   
   \maketitle
%

\section{Introduction}
Studies of star-forming galaxies in the local Universe provide knowledge of evolutionary properties and their driving physical processes. The investigation of
cosmic rays (CRs) transport mechanisms is crucial to understand the disk-halo interaction and the formation of radio halos in galaxies \citep[e.g.,][]{parker1992}. Whereas diffusion-dominated CR transport in galaxies leads to small radio halos \citep[e.g., NGC~7462;][]{heesenetal2016}, advection-dominated galaxies show large radio halos \citep[e.g., NGC~4666;][]{steinetal2019}. 
Polarization observations reveal large-scale magnetic fields in many spiral galaxies \citep[e.g., NGC~4666;][]{steinetal2019}. One proposed mechanism to explain these large-scale fields is the mean-field $\upalpha$-$\upomega$ dynamo \citep[see e.g.,][]{ruzmaikinetal1988, becketal1996, chamandy2016, henriksen2017, henriksenetal2018}. Turbulent gas motions, for example driven by supernova explosions, and the Coriolis force ($\upalpha$-process), combined with shear motions due to differential rotation ($\upomega$-process) amplify and order the magnetic field.

In this paper, the radio halo of the edge-on spiral galaxy, NGC~4013, is analyzed using data from the Continuum HAlos in Nearby Galaxies - an Evla Survey (CHANG-ES) program observed with the Karl G. Jansky Very Large Array (VLA). NGC~4013 is a spiral galaxy with a nearly edge-on orientation at a distance of 16\,Mpc. The basic galaxy parameters are listed in Table~\ref{tab:bparameter}. It exhibits a boxy bulge in optical light \citep{tullyetal1996}. There is strong evidence that the galaxy harbours a bar potential from CO and optical wavelength measurements \citep{garcia-burilloetal1999}. \citet{martinezetal2009} found a tidal stellar stream extending 3\,kpc above the plane of the disk, which they claim to be caused by a minor merger. NGC~4013 has a a strongly warped outer H{\sc i} gaseous disk \citep{bottema1996} and a lagging H{\sc i} halo \citep{zschaechnerrand2015}. A lagging halo can be caused by an active disk--halo interface, such as expected for a "galactic fountain". In spite of clear hints of interaction, there is no obvious partner in the group of 16 galaxies of which NGC~4013 is a member. \citet{wangetal2015} suggest that a one-to-three major merger would explain all observable features, meaning the boxy bulge, the huge H{\sc i} warp and the tidal stellar streams.

\citet{comeronetal2011} found a second thick stellar disk (in \emph{Spitzer} 3.6\,$\upmu$m-band data), which is smoothly distributed. They suggest that this third component was formed during galaxy formation, when an initially thick disk was dynamically heated by a past merger and then subsequently a new thick disk formed within. Follow-up analysis by \citet{comeronet2018} confirmed the existence of the two thick disks in NGC~4013.

\citet{hoetal1997} classified the nucleus of NGC~4013 as a "transition object". This class of objects was described by \citet{hoetal1993} as having an emission line nucleus with [OI] strengths intermediate between those of H{\sc ii} nuclei and low-ionization nuclear emission-line regions (LINERs). \citet{hoetal1993} proposed that transition objects can be explained as LINERs whose integrated spectra are weakened or contaminated by neighboring H{\sc ii} regions.

In the VLA 1.49\,GHz ($\lambda$20\,cm) image from \citet{condon1987}, the galaxy shows a boxy shape, suggesting the existence of a radio halo. But the radio structure is blurred due to the large beam size of 48", making a distinction between disk and halo emission difficult. Conversely, the $\lambda$6\,cm VLA+Effelsberg radio map by \citet[][]{hummeletal1991} shows a strong central source to be visible as well as a thin radio disk. \citet{baldietal2018} investigated eMERLIN array data of NGC~4013 and found radio emission in a circular shape with a diameter of 2.5" – 3" around the optical center. They state that this emission is probably associated with a nuclear star-forming ring and thus internal to the galaxy.

In this paper, we investigate the radio continuum emission of NGC~4013 of the halo and its central source. We present flux density measurements at the two frequencies ($1.5$ and 6\,GHz) of CHANG-ES as well as measurements of supplemental radio continuum data at 150\,MHz from the LOw-Frequency ARray (LOFAR) Two-metre Sky Survey \citep[LoTSS,][]{shimwelletal2017}. With supplementary X-ray data from  \emph{Chandra} and XMM-Newton we have hints about the central source of NGC~4013. Interestingly, the XMM-Newton data show an extended X-ray halo confined to the middle part of the galaxy where most of the polarized radio emission was detected. We further analyze radio scale heights and the linear polarization. The synchrotron map (with thermal emission subtracted) was used to determine the magnetic field strength. The CR transport was analyzed assuming 1D pure diffusion and advection for the CR electrons.

This  paper  is  organized  as  follows.  Section  2 gives an overview of the radio and X-ray data. In Section 3, we present our results. This section contains the Stokes I maps, the scale height analysis, the polarized intensity maps (created both with and without rotation measure (RM) synthesis), the map of the magnetic field strength using equipartition, and the 1D cosmic-ray transport model. In Section 4 main results are discussed and in Section 5 the results are summarized and conclusions are drawn.

\begin{table}
\centering
\begin{threeparttable}
\caption{Basic galaxy parameters of NGC~4013.}  
\label{tab:bparameter}
\begin{tabular}{lc} \hline \hline
RA 						& 11h 58m 31.38s \tnote{1} 	\\
Dec 					& +43d 56m 47.7s \tnote{1} 	\\
Distance (Mpc)			& 16.0 \tnote{1} 			\\
Inclination ($^\circ$)		&	89~$\pm$~2 \tnote{2} \\
PA ($^\circ$)			& 65~$\pm$~3 \tnote{2}		\\
Major Axis (arcmin)			& 5.2 \tnote{3}		 \\
Minor Axis (arcmin)			& 1.0 \tnote{3}						\\
v$_{\text{sys}}$ (km s$^{-1}$) 	&	831 \tnote{3}	\\
v$_{\text{rot}}$ (km s$^{-1}$) 	&	195 \tnote{4}	  \\
SFR (M$_{\odot}$~yr$^{-1}$)	& 0.5 \tnote{5}					\\
SFRD (10$^{-3}$\,M$_{\odot}$~yr$^{-1}$~kpc$^{-2}$)& 2.4 \tnote{5}		\\			
Classification 			  	&	H{\sc ii}, LINER \tnote{1}					 \\
\hline
\end{tabular}
\begin{tablenotes}
\footnotesize
\item References:\\
\item[1] \citet{irwinetal2012}
\item[2] This work
\item[3] NASA/IPAC Extragalactic Database (NED, https://ned.ipac.caltech.edu)
\item[4] \citet{bottema1996}
\item[5] Star formation rate (SFR) and star formation rate surface density (SFRD) from \citet{wiegertetal2015}
\end{tablenotes}
\end{threeparttable}
\end{table}
\normalsize

\section{Data}

\subsection{VLA}
The radio continuum data are part of the CHANG-ES survey \citep[][]{irwinetal2012} observed with the Karl G. Jansky Very Large Array (VLA) . 
Observations were obtained in B-, C-, and D-configurations at L-band (1.5\,GHz, 500\,MHz bandwidth, with a gap of 144\,MHz width where strong radio frequency interference is located), and in the C- and D-configurations at C-band (6\,GHz, 2\,GHz bandwidth). We used 2048 spectral channels in 32 spectral windows at 1.5\,GHz and 1024 channels in 16 spws at 6\,GHz. All polarization products (Stokes I, Q, U, and V) were obtained. The D-configuration data \citep{wiegertetal2015} are public\footnote{CHANG-ES data release I available at www.queensu.ca/changes} and the B- and C-configuration data will become public soon (Irwin et al. in prep., Walterbos et al. in prep).

\begin{table}
\centering
\caption{Observation parameters of NGC 4013.} 
\label{tab:oparameter}
\begin{tabular}{l c c c} 
\hline\hline   
Dataset	&Observing Date		& Time on Source\\
			 & \           	& (before flagging)\\ 
\hline
L-band B-configuration &  11. Aug 2012  & 2 hr\\
L-band C-configuration &  31. Mar 2012    & 30 min\\
L-band D-configuration &  18. Dec 2012   & 20 min\\
C-band C-configuration&  19. Feb 2012   & 3 hr\\
C-band D-configuration & 27. Dec 2012   & 40 min\\
\hline
\end{tabular} 
\end{table}

The data reduction for Stokes I (total power) and Stokes Q and U (linear polarization) was carried out for all five data sets of NGC~4013 (see Table~\ref{tab:oparameter}) separately, using the Common Astronomy Software Applications (CASA) package \citep[][]{mcmullinetal2007} and following the calibration procedures as described in the CHANG-ES paper by \citet{irwinetal2013}. We used J1331+3030 (3C286) as the primary calibrator (also as the bandpass and polarization angle calibrator), J1246-0730 as the secondary calibrator, and J1407+2827 as the zero polarization calibrator. The calibrated data from the different configurations were then combined for C-band and L-band and were used for imaging Stokes I, Q, and U.

Large-scale  flux  density for extended  structures is missed by the VLA at C-band in C- and D-configuration for structures larger than 240". In L-band B-configuration missing flux density is expected for structures larger than 120" and C- and D-configuration for structures larger than 970". As the radio extent of NGC~4013 in the CHANG-ES bands is 205", we expect no significant influence by missing spacings in the configuration-combined images of both bands (potential missing spacings due to the B-configuration in the L-band setup are compensated by the inclusion of the other two configurations).  

The Stokes I maps were produced by cleaning with a robust weighting \citep{briggs1995} parameter of zero, as implemented in CASA and uv-tapering of 18~k$\lambda$. The polarization and magnetic field orientation maps were created from the Stokes Q and U maps, which were cleaned with a robust parameter of two in order to be more sensitive to faint structures.

Table~\ref{tab:rms4013} shows all rms values for the different maps produced for this study. The achieved rms from the combined (C- and D-configuration) C-band data for Stokes I (rob 0, 18~k$\lambda$ taper) is 4.3\,$\mu$Jy/beam with a beam size (i.e. resolution/FWHM) of 5.2"~$\times$~5.3". The resulting rms of the combined (B-, C- and D-configuration) L-band data for Stokes I (rob 0, 18~k$\lambda$ taper) is 16.7\,$\upmu$Jy/beam with a beam size of 5.9"~$\times$~6.0". Smoothed images of Stokes I of both bands were produced for the scale height analysis and the thermal and nonthermal separation and the following determination of the magnetic field strength and the CR transport analysis. For RM-synthesis, Stokes Q and Stokes U were imaged for each spw with a robust weighting parameter of two. The technique was performed following \citet{steinetal2019}.

\subsection{LOFAR}
 The LoTSS is a sensitive, high-resolution, low-frequency (120-168\,MHz) survey of the northern sky using LOFAR \citep{shimwelletal2017}. LoTSS data of NGC~4013 were obtained from data release 1 \citep[DR1,][]{shimwelletal2019}\footnote{LoTSS data are publicly available at: https://lofar-surveys.org}. In DR1, images with two resolutions of 6" and 20" are provided. We use the 6" image of NGC 4013 and applied Gaussian smoothing of 10" for the scale height analysis and of 18" to show the extent in comparison to C-band.

\subsection{XMM-Newton}
The X-ray analysis of the emission from NGC~4013 was performed with the use of the XMM-Newton archive data (see Table~\ref{xrays}) processed with the SAS 15.0.0 package \citep{sas}. Using the standard reduction procedures, event lists for two EPIC-MOS cameras \citep{turner} and the EPIC-pn camera \citep{strueder} for each of the two observations were produced and subsequently filtered for periods of high-energy emission.

To increase the sensitivity of the resulting spectrum, information from all three EPIC cameras was used in the spectral analysis. Spectra were extracted from each of the event lists from a circular central region having a radius of 12\,", that allowed a full sampling of the PSF of the instrument. The position of the region was chosen to match the radio center of the galaxy. The background spectra were obtained using blank sky event lists \citep[see][]{carter}, using the same procedures as in the case of the source event lists. For each spectrum ancillary files (response matrices and effective area files) were used to correct for the instrumental effects.
The spectra from all three EPIC cameras of each observation were then merged using the SAS task $epicspeccombine$. Merging of the spectra was accompanied by subtraction of the background spectra. It is important to note, that the subtraction takes place only after both source spectra and background spectra are combined. This leads to higher signal-to-noise ratios in both resulting spectra, assuring best subtraction of the background. XSPEC~12 \citep{xspec} was used to fit model to the final spectrum of the core region of NGC~4013.

\begin{table}
\caption{\label{xrays}Parameters of the XMM-Newton observations of NGC~4013.}
\centering
\begin{tabular}{ll}
\hline\hline
Galaxy                  & NGC~4013     \\    	
ObsID/obs.date          & 0306060201/13.11.2005    \\
                        & 0306060301/15.11.2005    \\
n$_{\rm H}$\tablefootmark{a}  & 1.26\,$\times$\,10$^{20}$\,cm$^{-2}$  \\
MOS filter/obs.mode     & Medium/Full Frame \\
pn filter/obs.mode      & Medium/Full Frame \\
Total/clean MOS time    & 94/79 ks  \\
Total/clean pn time     & 71/67 ks  \\
\hline
\end{tabular}
\tablefoot{
\tablefoottext{a}{Weighted average value after LAB Survey of Galactic \ion{H}{i} \citet{lab}.}
}
\end{table}

\subsection{\emph{Chandra}}
Further supplemental X-ray data were gained from \emph{Chandra}. The \emph{Chandra} data were used to detect point sources reliably due to their high angular resolution. The parameters of the \emph{Chandra} observations are presented in Table~\ref{xrays-chandra}. The data were taken from \citep{liwang2013a}, where the details about the observations and data reduction techniques, including calibration, can be found.

\begin{table*}
\caption{\label{xrays-chandra}Parameters of the  \emph{Chandra} observations of NGC~4013.}
\centering
\begin{tabular}{lcccccc}
\hline\hline
Obs ID &  Instr.  &  Exp. (sec) &  PI  &   Data Mode &   Exp Mode   &  Start Date\\
4739  &  ACIS-S    & 79.10    &   Strickland & FAINT  &  TE  &  2004-02-03 09:31:16   \\
4013  &  ACIS-S    & 4.90    &   Satyapal    &  FAINT &   TE &   2003-03-16 17:09:05  \\
\hline
\end{tabular}
\end{table*}

\section{Data analysis and results}

\subsection{Radio continuum total intensity maps}
\label{sec:radiocont}

\begin{figure*}
	 \begin{minipage}{0.48\textwidth}
		\includegraphics[width=\textwidth]{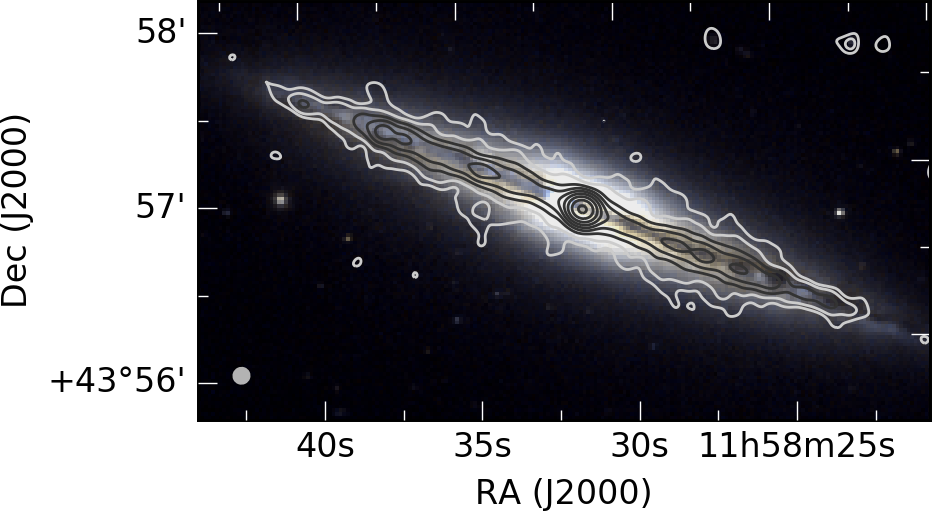}
	\caption{NGC~4013 total intensity image of combined (C- and D-configuration) VLA C-band data at 6\,GHz. Contours start at the 3$\upsigma$ level with an rms noise $\upsigma$ of 3.6\,$\upmu$Jy/beam and increase in powers of 2 (up to 256). The beam size is 5.2" $\times$ 5.3" and is shown in the bottom left corner of the image. A tapering of 18\,k$\lambda$ was used, while the robust parameter was set to zero.}
	\label{fig:N4013-Ccomb-stokesI}
	\end{minipage}
\hspace{0.02\textwidth}
 \begin{minipage}{0.48\textwidth}
		\includegraphics[width=\textwidth]{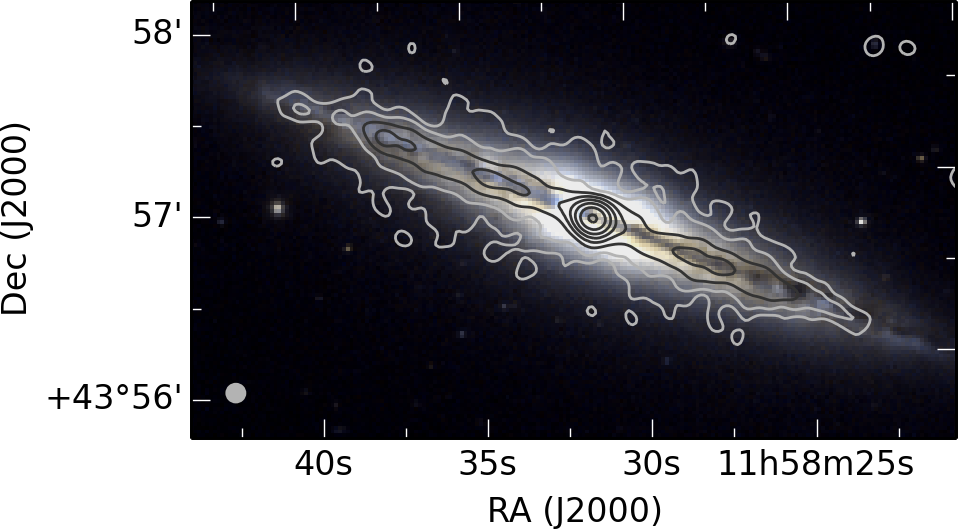}
	\caption{NGC~4013 total intensity image of combined (B-, C- and D-configuration) VLA L-band data at $1.6$\,GHz. Contours start at the 3$\upsigma$ level with a $\upsigma$ of 16.7\,$\upmu$Jy/beam and increase in powers of 2 (up to 128). The beam size is 5.9~$\times$ 6.0" and is shown in the bottom left corner of the image. A tapering of 18\,k$\lambda$ was used, while the robust parameter was set to zero.}
	\label{fig:N4013-Lcomb-stokesI}
	\end{minipage}
	\end{figure*}
	
\begin{figure}
\centering
	\includegraphics[width=0.48\textwidth]{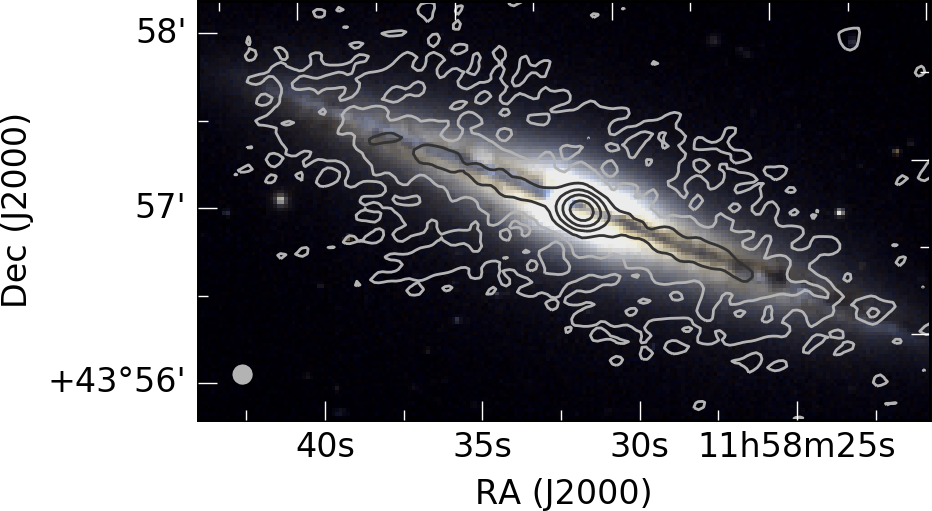}
	\caption{NGC~4013 total intensity image of LOFAR data at 150\,MHz. Contours start at the 3$\upsigma$ level with a $\upsigma$ of 204\,$\upmu$Jy/beam and increase in powers of 2 (up to 32). The beam size is 6.0" $\times$ 6.0" and is shown in the bottom left corner of the image.}
	\label{fig:N4013-Lofar-stokesI}
\end{figure}

The total power contours from Stokes I are overlaid on the optical SDSS images made from the ugr-filters using the formulas of Lupton (2005)\footnote{www.sdss.org/dr12/algorithms/sdssUBVRITransform/\#Lupton2005}. The Stokes I contour map of C-band is displayed in Fig.~\ref{fig:N4013-Ccomb-stokesI}, the one of L-band in Fig.~\ref{fig:N4013-Lcomb-stokesI} and the total intensity LOFAR image in Fig.~\ref{fig:N4013-Lofar-stokesI}. The radio emission is dominated by the central point source, which is seen in all frequencies. Especially in C-band, the emission from the disk is weak and only a thin radio disk and a small radio halo are observed. The presented CHANG-ES images are produced with a $uv$-tapering of 18\,k$\lambda$ to enhance the extended flux of the galaxy and to achieve similar resolutions for a quantitative comparison.

The L-band radio halo shows an extent of up to 2\,kpc above and below the mid plane and the LOFAR radio halo is extended to 3\,kpc, where both values were measured within the 3$\upsigma$-contours. In L-band the radio halo is observable further out in comparison to the C-band map and at 150\,MHz even further than in L-band. This is the expected behaviour for spectral ageing, where older cosmic ray electrons (CREs) corresponding to lower energies (emitting at lower frequencies) are transported away from their sources within the disk of the galaxy, while losing energy due to various loss processes. No boxy structure is found, in contrast to what the observations of \citet{condon1987} suggest, but a slightly dumbbell-shaped thick disk can be seen in the L-band and 150\,MHz maps.  

\begin{table}
\begin{threeparttable}
\centering
\caption{Integrated flux densities of NGC~4013.}  
\label{tab:fluxmeasuren4013}
\begin{tabular}{ccc} 
\hline\hline
Observation &Freq&   Flux density                                    	\\
\		    & (GHz)	&(Jy)                                       \\
\hline 
This work\tnote{1}		& 0.15    & 0.218~$\pm$~0.021	             	\\
\citet{condonetal1998}\tnote{2}	& 1.40    &	0.0405~$\pm$~0.0023     	 \\
This work\tnote{3}		& 1.58    & 0.0368~$\pm$~0.0018	             	\\
This work\tnote{3} 		& 6.00    &	0.0118~$\pm$~0.0006 	            \\
\hline
\end{tabular}
\begin{tablenotes}
\footnotesize
\item Abreviations:
\item[1] The LOFAR Two-metre Sky Survey (LoTSS), measured on the 10" resolution image
\item[2] The NRAO VLA Sky Survey (NVSS)
\item[3] CHANG-ES with effective frequency after flagging measured on the 10" resolution image
\end{tablenotes}
\end{threeparttable}
\end{table}
\normalsize

With the total flux density of 11.8~$\pm$~0.6\,mJy in C-band and 36.8~$\pm$~1.8\,mJy in L-band the radio spectral index is $\upalpha~=~-0.85~\pm~0.05$. The spectral index between 150\,MHz and 1.58\,GHz is $\upalpha~=~-0.76~\pm~0.05$. There is only one additional flux density measurement in the literature, which is from NVSS (see Table~\ref{tab:fluxmeasuren4013}). This value of 40.5~$\pm$~2.3\,mJy at 1.4\,GHz is in agreement with the flux densities obtained in this work. The observed radio spectral indices are in good agreement with expected values for synchrotron emission. The low-frequency radio spectral index below $1.5$~GHz is slightly steeper (with a difference of $2\upsigma$) than the value of $\upalpha = -0.57\pm 0.10$ \footnote{Quoted uncertainties for results by \citet{chyzyetal2018} refer to the standard deviation rather than the smaller standard deviation of the mean.} found by \citet{chyzyetal2018}. Also, we see only a small break in the radio continuum spectrum with $\Delta\upalpha = \upalpha_{\rm low}-\upalpha_{\rm high}$ = 0.09~$\pm$~0.05, whereas \citet{chyzyetal2018} found $\Delta\upalpha = 0.18\pm 0.20$. This suggests different CR electron propagation and energy losses in NGC~4013 compared to other spiral galaxies.

\subsection{Central point source}
The galaxy NGC~4013 was classified as a LINER by \citet{hoetal1997} and \citet{irwinetal2012}. There is an additional X-ray observation published by \citet{dudiketal2005}. In their study of high-resolution X-ray imaging of nearby LINERs observed by  \emph{Chandra}, a nuclear point source of NGC~4013 was not detected. They concluded either a lack of an energetically significant AGN or a highly obscured AGN with column densities of 11~$\times$~$10^{23}$~cm$^{-2}$ to 10.2~$\times$~$10^{24}$~cm$^{-2}$, implying that the AGN has a luminosity between 2~$\times$~10$^{38}$~erg\,s$^{-1}$ and 9.5~$\times$~10$^{42}$~erg\,s$^{-1}$. Here, we investigate the "central point source", meaning a centrally peaked source, although it is not an unresolved source in both dimensions as defined in \citet{irwinetal2019}. Using the radio continuum data of both CHANG-ES bands and LOFAR data as well as the X-ray data from  \emph{Chandra} and XMM-Newton we further constrain the classification of this galaxy and its central source.

\subsubsection{Radio continuum}
A clear detection of a central point source is visible in both bands of the CHANG-ES data as well as in the LOFAR data. To further investigate the central source, variability in the total intensity of the central source was checked. Therefore, flux densities of the central source were measured in images from the two different VLA array configurations of the C-band data, which are observed at different times (see Table \ref{tab:centralsourcemeasure}). As one C-band observation was done in the beginning of 2012 and the other in the end of 2012, time variability could be tested. Additionally, the flux density of the central source in B-configuration L-band and LOFAR 150\,MHz was measured. To do so, all images were smoothed to the same beam size of 10\,arcsec in order to overcome beam-dependent errors in the flux density measurement. The flux density was then measured via a Gaussian profile fit to the intensity, using CASA.

\begin{table}
\centering
\begin{threeparttable}
\caption{NGC~4013 flux density measurement of the central source.}  
\label{tab:centralsourcemeasure}
\begin{tabular}{lcccc} \hline \hline
 \ & Flux	             & Beam & Region & Obs. date\\[-0.5em]
  \ & 	             & size & size & \\[-0.1em]

\           & (mJy)	           & (")	& (")				&							\\
\hline
BL				&  10.09 $\pm$ 0.44		 &10		&  	20	  & Aug. 11/2012 \\
DC				&  4.52 $\pm$ 0.20 		 &10		&  	20		& Dec. 27/2012 \\
CC				&  4.52 $\pm$ 0.20 		 &10		&  	20		& Feb. 19/2012 \\
LOFAR           &  28.3 $\pm$ 1.7 	 &10		&  	20		& Jan. 21/2015 \\
\hline
\end{tabular}
\begin{tablenotes}
\item \textbf{Notes.} Measurements of the different array configurations B-configuration L-band (BL), D-configuration C-band (DC) and C-configuration C-band (CC)
\end{tablenotes}
\end{threeparttable}
\end{table}

We chose a circular region two times larger than the beam size. To measure the error, the chosen region was varied from the beam size to four times the beam size. The difference in the flux density measurements were then used to calculate a mean error.
The integrated flux density of the central source of both C-band configurations is 4.52~$\pm$~0.20\,mJy. Therefore, no measurable variability is found between our C-band observations. The flux density level of the galaxy was 0.14 and 0.17\,mJy/beam, respectively. The integrated flux density of the central source in B-configuration L-band is 10.09~$\pm$~0.44\,mJy with a galaxy flux density level of 0.35\,mJy/beam. The flux density of the LOFAR observations is 28.33~$\pm$~1.73\,mJy with a base level of 2.28\,mJy/beam.

A resulting radio spectral index of $\upalpha~=~-0.66~\pm~0.05$ is calculated between the C-band flux density and the B-configuration L-band flux density as well as $\upalpha~=~-0.44~\pm~0.05$ between the L-band and the LOFAR flux densities. The values suggest a flattening of the spectral index towards lower frequencies due to synchrotron self-absorption, which requires a very compact object and hints at an AGN. As some AGN show circular polarization \citep{irwinetal2018}, this was investigated as well. In the Stokes $V$ image, no circular polarized emission could be found. In summary, apart from the fact that the central source dominates the emission of the galaxy and that the spectral index is relatively flat towards lower frequencies, no strong indication that the central object might be an AGN can be found from its radio properties. 

\subsubsection{X-ray}
\begin{figure}
	\centering
		\includegraphics[width=0.49\textwidth]{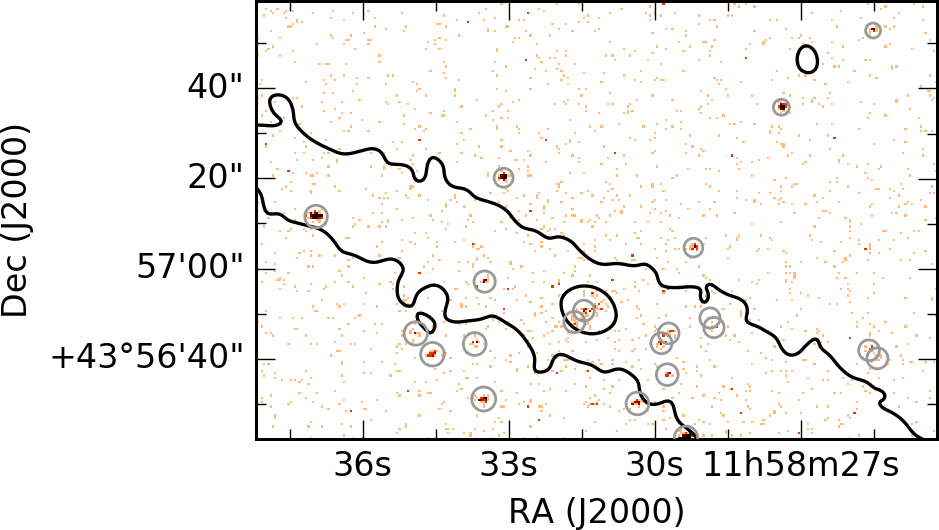}
		\caption{NGC~4013 \emph{Chandra} image with combined (C- and D-configuration) VLA C-band contours (same data as Fig~\ref{fig:N4013-Ccomb-stokesI}) at 3- and 30-$\upsigma$-levels overlaid. The false-color represents the \emph{Chandra} color from 0.3 to 7\,keV. Gray cycles denote \emph{Chandra} point-like sources. The image is off-centered to show the location of the two point-like sources in the northern halo of NGC~4013.}
	\label{fig:N4013-Chandra-Ccomb}
\end{figure}

\textit{Chandra:}\\ 
To further constrain the nature of the central point source, \emph{Chandra} X-ray data \citep{liwang2013a} on Stokes I contours are shown in Figure~\ref{fig:N4013-Chandra-Ccomb}. Clearly, there is a \emph{Chandra} point-like source located at the same position as the central radio source, which is the dominant source within the entire surroundings apart from another less strong point-like source next to it. This finding corroborates the LINER classification since we expect X-ray emission emanating from an AGN but not from a star-burst nucleus.\\

\noindent\textit{XMM-Newton:}\\ 
The core region of NGC~4013 was fit with a complex model consisting of a gaseous component that accounted for the emission from the hot gas and a power-law model, that accounted for the central source. The mekal model is a model of an emission spectrum from hot diffuse gas based on the model calculations of Mewe and Kaastra \citep{mewe,kaastra}. 

Since we assumed that most of the power-law flux would be attributed to the central source (see also Fig.~\ref{fig:N4013-Chandra-Ccomb}) and given the relatively large uncertainties of its value, we used this flux to calculate the luminosity of the central source. We obtained a value of 7.1$^{+5.2}_{-3.0}$\,$\times$\,10$^{38}$\,{\rm erg}\,s$^{-1}$ (Table~\ref{tab:XMMfluxes_n4013n4666}).

\begin{figure}
\centering
\includegraphics[angle=-90,width=0.49\textwidth]{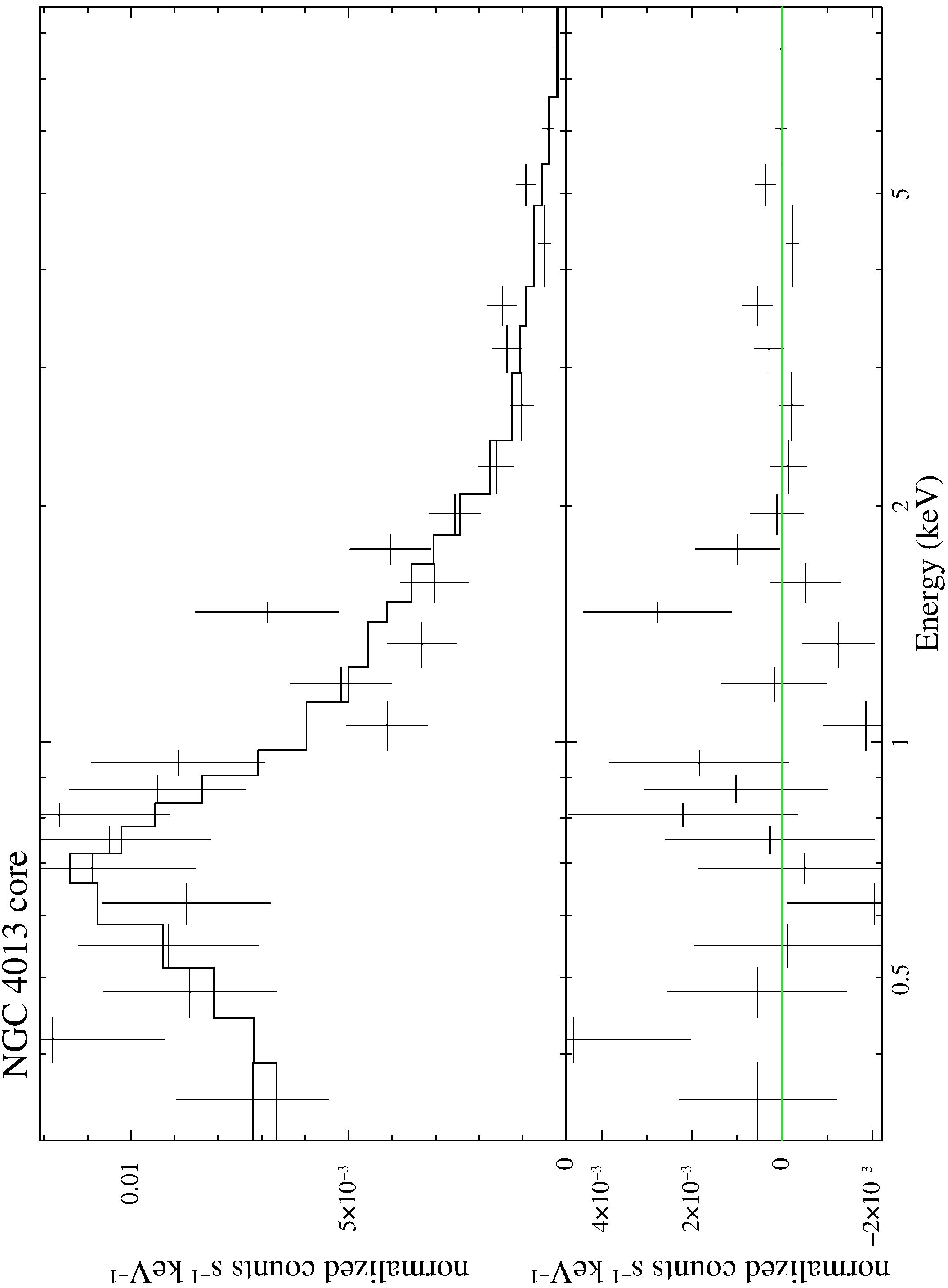}
\caption{XMM-Newton EPIC spectrum of the core region of NGC~4013.}
\label{fig:N4013_XMM_spectrum}
\end{figure}

The spectrum of the core region of NGC~4013 is presented in Figure~\ref{fig:N4013_XMM_spectrum}. No clear AGN characteristics are visible, but significant hard emission (above 2\,keV) and a relatively flat photon index suggest a prominent central source.

The average temperature of the hot gas in the core region of NGC~4013 is 0.27$^{0.06}_{0.04}$\,keV. Assuming a spherical emitting volume of the region, the model fit yields a number density of 5.37$^{+0.85}_{-1.07}$\,$\times$\,10$^{-3}$\,${\rm cm}^{-3}$ and the corresponding thermal energy density of 3.50$^{+1.44}_{-1.11}$\,$\times$\,10$^{-12}$\,{\rm erg}\,s$^{-1}$.

\begin{table}
\caption{Physical parameters derived from the model fit to the spectrum of the NGC~4013 core region.}  
\label{tab:XMMfluxes_n4013n4666}
\begin{tabular}[h]{ll}
\hline\hline
kT          &  0.27$^{+0.06}_{-0.04}$\,keV \,\, (3.13$^{+0.70}_{-0.46}\times$\,10$^6$\,K)    \\
Mekal flux  &  2.00$^{+0.87}_{-0.79}$\,$\times$\,10$^{-15}$\,{\rm erg}\,{\rm cm}$^{-2}$\,s$^{-1}$  \\
Photon Index&  1.10$^{+0.22}_{-0.21}$   \\
Core flux   &  2.31$^{+1.69}_{-0.99}$\,$\times$\,10$^{-14}$\,{\rm erg}\,{\rm cm}$^{-2}$\,s$^{-1}$  \\
Core luminosity & 7.1$^{+5.2}_{-3.0}$\,$\times$\,10$^{38}$\,{\rm erg}\,s$^{-1}$ \\
\hline
\end{tabular}
\end{table}

\begin{figure}
\centering
 \includegraphics[width=0.48\textwidth]{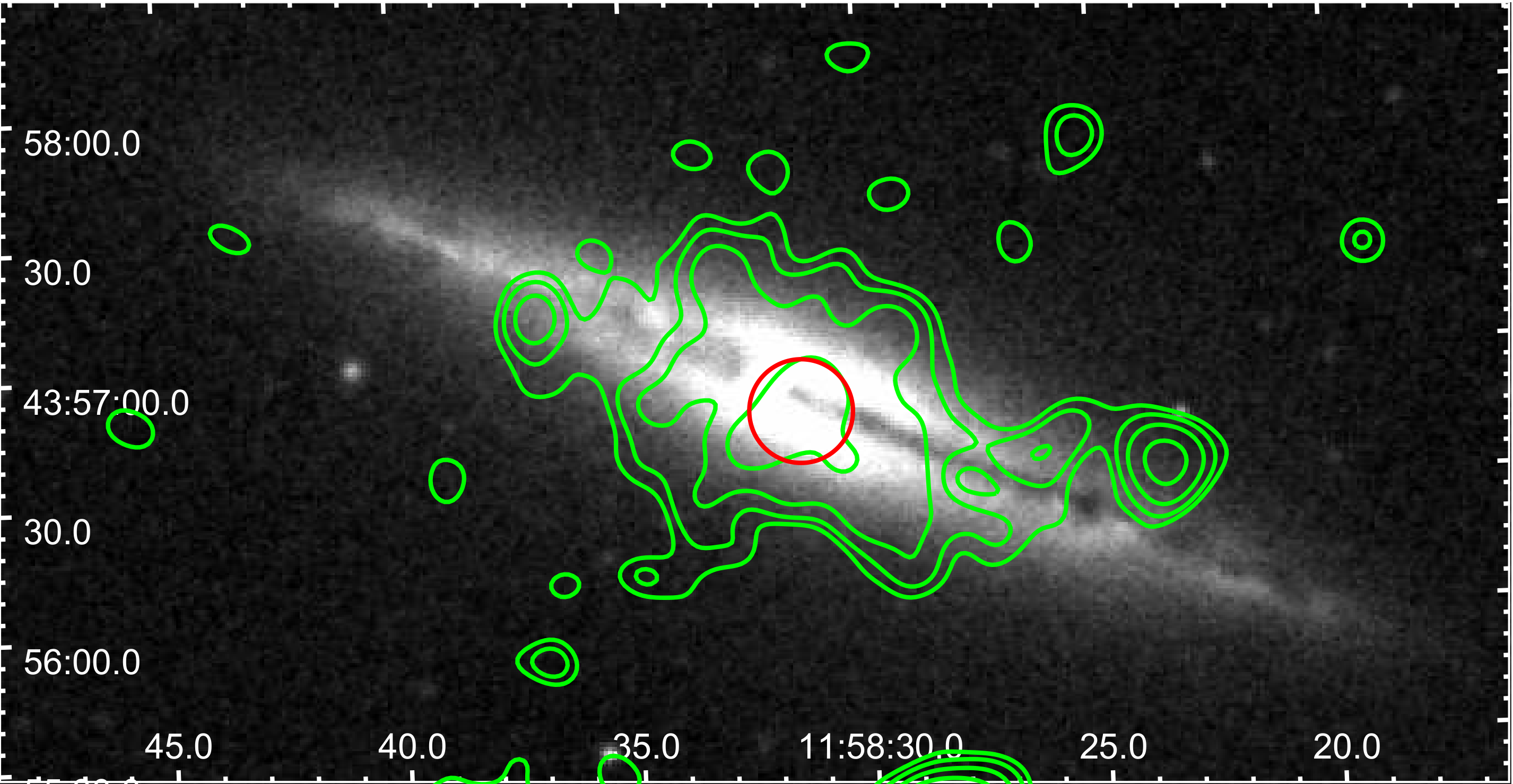}
\caption{XMM-Newton EPIC emission of NGC~4013 in the energy band 0.2-1\,keV (contours) overlaid on the DSS blue image. The contours start at 3, 3.5, 4.5, 6 $\upsigma$ levels. The image is adaptively smoothed with the largest smoothing scale of 10\arcsec and the signal-to-noise ratio of 10. The spectral core region is marked with a red circle. }
\label{fig:n4013_xmm}
\end{figure}

Fig.~\ref{fig:n4013_xmm} shows the extended emission from NGC~4013 in the 0.2-1\,keV energy band. The spectral region used for the analysis is marked with a red circle. Most of the emission of the hot gas comes from the central parts of the galactic disk with some outflows into the halo visible.

\subsection{Intensity profiles of the radio disk and halo}
\label{subsec:strip}
\subsubsection{Scale heights - Strip fitting}
The  total intensity  profiles  perpendicular  to  the  major  axis  of  edge-on spiral galaxies are fit by exponential or Gaussian functions to determine scale heights of the thin and thick disk. The thin disk refers to the disk and the thick disk refers to the radio continuum halo of the galaxy, which are the used terms in this work. Following \citet{dumke1995}, the observed intensity profile can be described by the convolution of a Gaussian beam profile with the true intensity profile of the galaxy, which could be either exponential or Gaussian.

\begin{figure}
\centering
\includegraphics[width=0.5\textwidth]{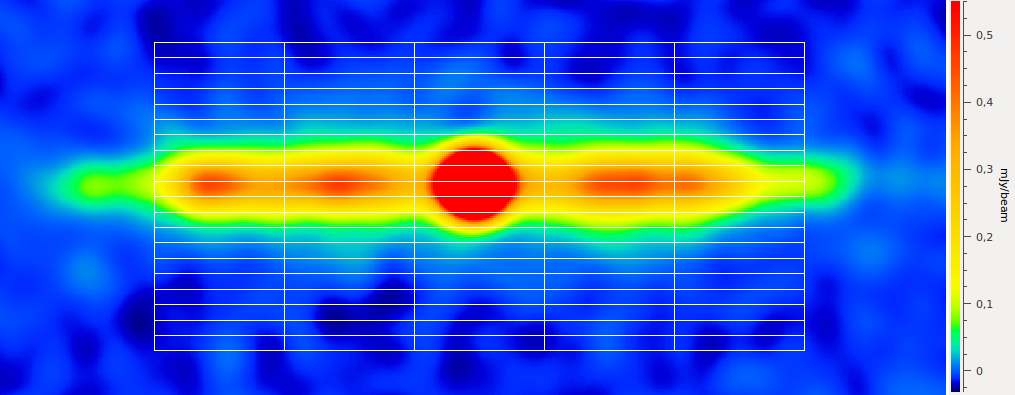}
 \caption{NGC~4013 screenshot of the strips used with NOD3 (combined C-band).}
 \label{fig:N4013_nod3_screenshot}
\end{figure}

We used the "BoxModels" tool within the NOD3 software package \citep{mulleretal2017} with the combined configuration maps of C-band and L-band as well as the LOFAR map, which were smoothed to the same beam size of 10"~$\times$~10" in order to compare the results. The scale height analysis was carried out in an analogous way to \citet{krauseetal2018} using five strips perpendicular to the major axis (see~Fi\-gure~\ref{fig:N4013_nod3_screenshot}). The box height was chosen to be approximately half of the beam size, which was 4" in both bands. The box width was 35" to cover the major axis with the five strips. We note that with the LOFAR data it would have been possible to measure seven strips due to the larger extent of the major axis. A two-component exponential and Gaussian fit was executed for all three frequencies to the five strips of radio intensity profiles of NGC~4013. We achieved a fit with similar $\upchi^2$ using Gaussian distributions in comparison with exponentials for the C-band intensity profile, but better results for a Gaussian fit for the L-band and LOFAR intensity profiles.

\begin{figure*}
 \centering
 \includegraphics[width=0.7\textwidth]{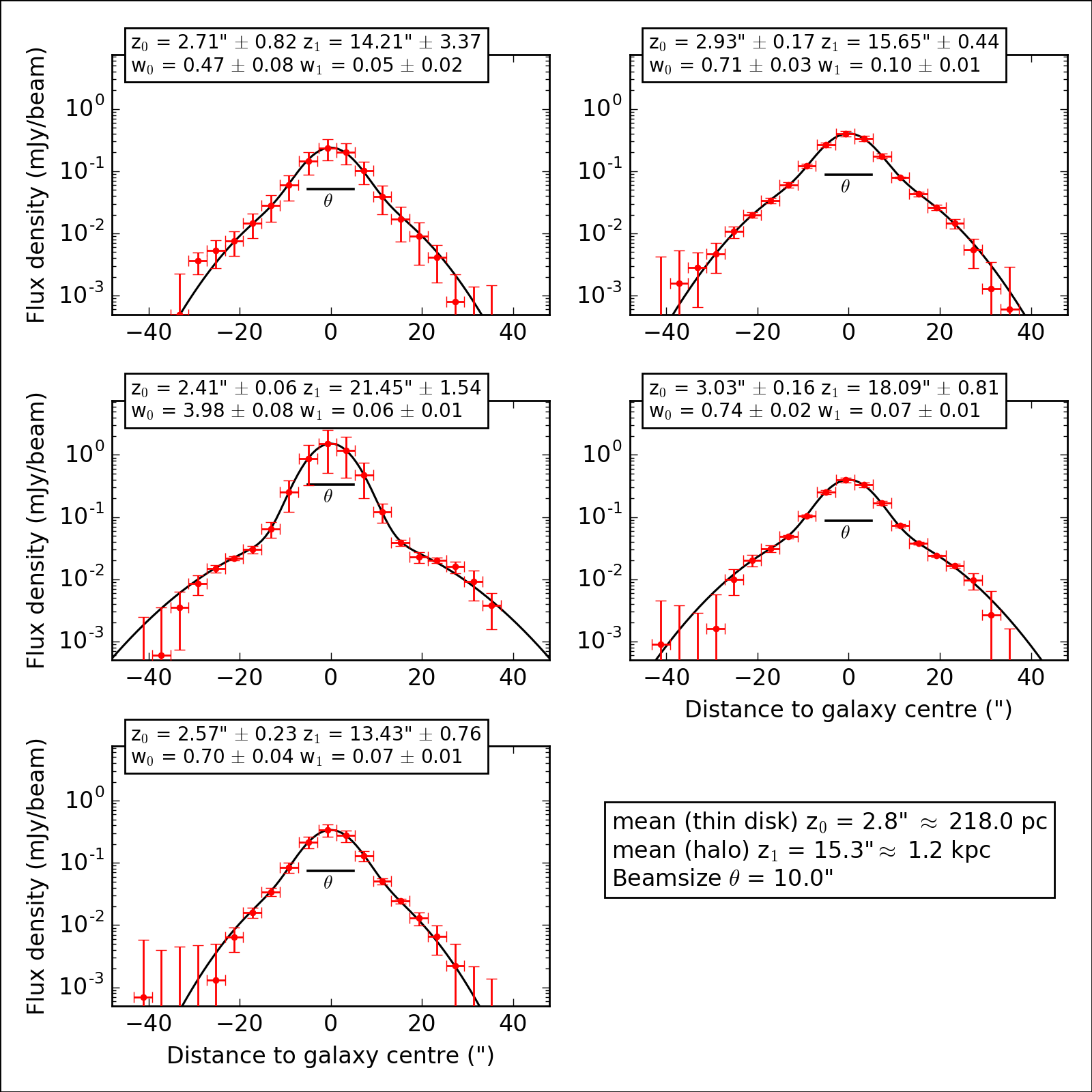}
 \caption{Strip fitting with NOD3 for five strips of NGC~4013 on the combined (C- and D-configuration) C-band data with a two-component Gaussian fit ($\upchi_{\text{total}}^2$~=~0.05). The first plot is at 75" on the major axis and the last plot is at -75" on the major axis from the center. The red dots represent the mean intensity of each box and the black line is the fit to the data done by NOD3. The black horizontal line shows the size of the beam~$\theta$. The parameters of the exponential fit function are given in the boxes on top of each strip. The mean of the four scale heights omitting the central strip as well as the beam size are given in the box to the lower right.}
 \label{fig:N4013_nod3_C}
\end{figure*}

\begin{figure*}
 \centering
 \includegraphics[width=0.69\textwidth]{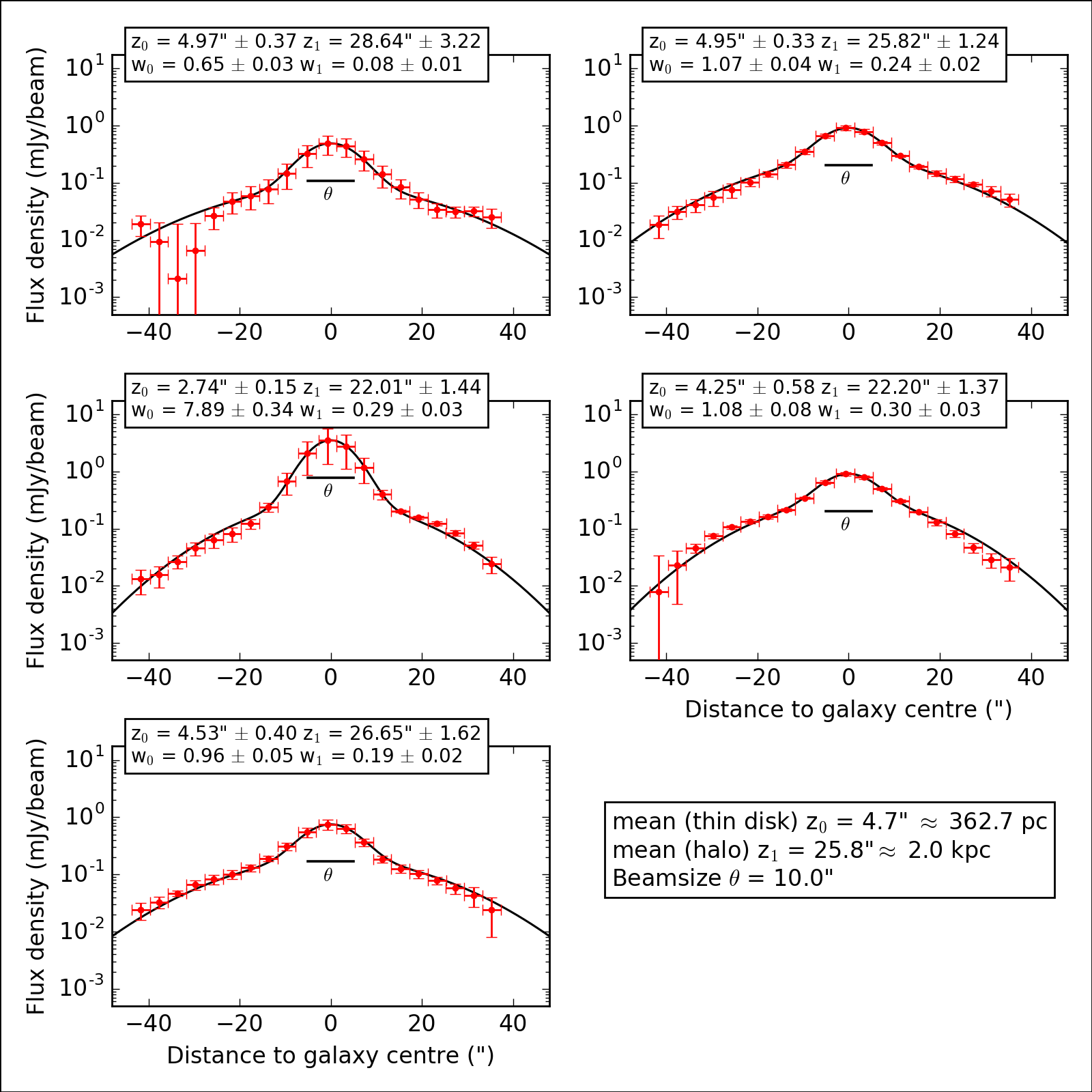}
 \caption{Strip fitting with NOD3 for five strips of NGC~4013 on the combined (B-, C- and D-configuration) L-band data with a two-component Gaussian fit ($\upchi_{\text{total}}^2$~=~0.92). Designations are the same as for Figure~\ref{fig:N4013_nod3_C}.}
 \label{fig:N4013_nod3_L}
\end{figure*}

\begin{figure*}
 \centering
 \includegraphics[width=0.69\textwidth]{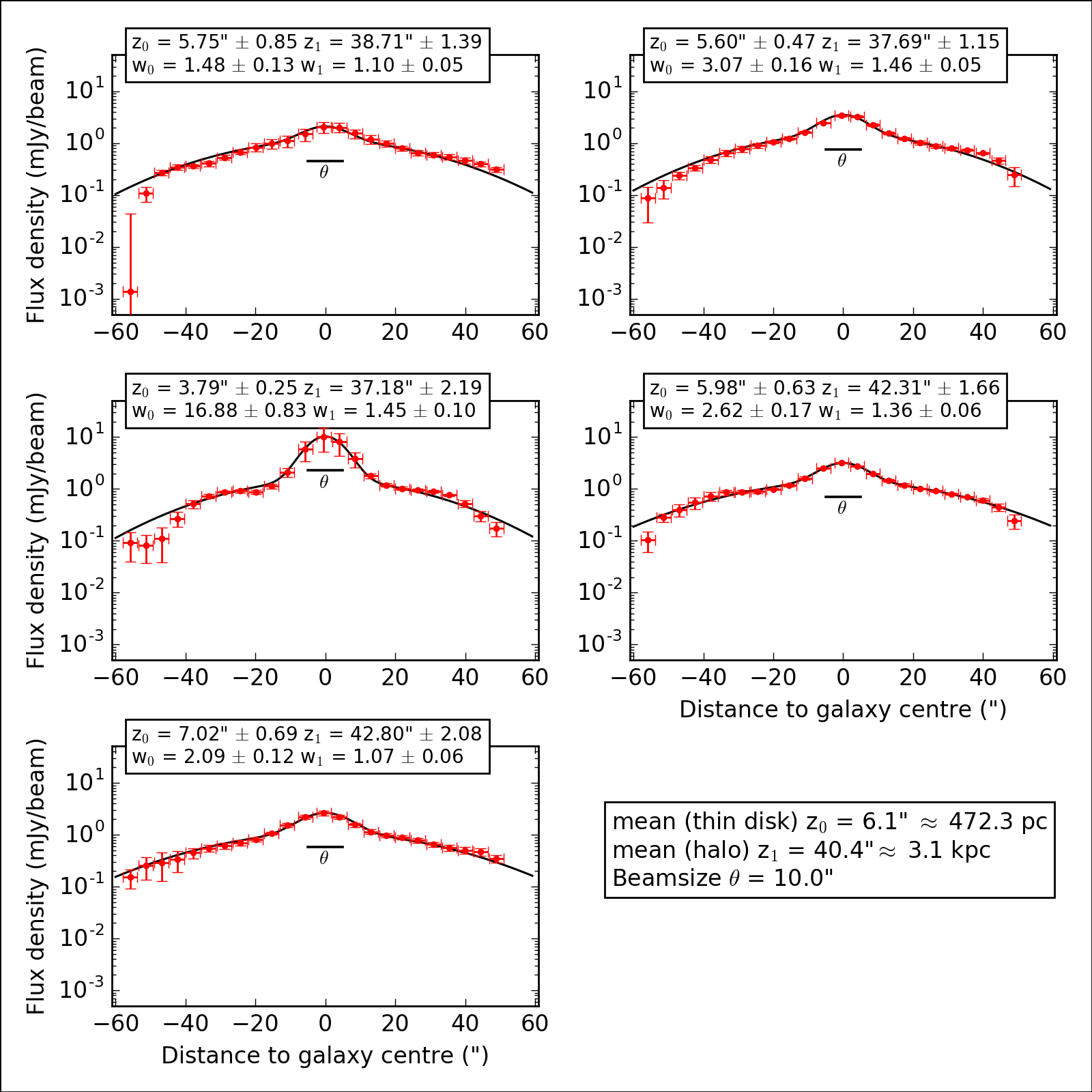}
 \caption{Strip fitting with NOD3 for five strips of NGC~4013 on the LOFAR data with a two-component Gaussian fit ($\upchi_{\text{total}}^2$~=~1.07). Designations are the same as for Figure~\ref{fig:N4013_nod3_C}.}
 \label{fig:N4013_nod3_Lofar}
\end{figure*}

The parameters of NOD3 are summarized in Table~\ref{tab:N4013nod3para}. The intensity profiles of the five strips and the two-component Gaussian fits are displayed in Figure~\ref{fig:N4013_nod3_C} for C-band in Figure~\ref{fig:N4013_nod3_L} for L-band and in Figure~\ref{fig:N4013_nod3_Lofar} for the 150\,MHz LOFAR data.

The resulting Gaussian scale heights of the strip fitting are presented in Table~\ref{tab:N4013scaleheight}. For the final scale height determination, the central strip was omitted in order to exclude contamination by the central source. The scale height analysis of NOD3 shows an increasing thin disk component from higher to lower frequencies. The ratio between the thin disk scale height and the halo scale height is similar for C-band and L-band and decreases slightly towards the lower frequency range of LOFAR. For the thin disk ("disk") scale heights, we find 2.8"~$\pm$~0.7" (220\,pc) in C-band, 4.7"~$\pm$~0.3" (360\,pc) in L-band, and 6.1"~$\pm$~0.6" (470\,pc) at 150\,MHz. For the thick disk ("halo") component, the scale heights are 15.3"~$\pm$~3.0" (1.2\,kpc) in C-band, 25.8"~$\pm$~1.5" (2.0\,kpc) in L-band, and 40.4"~$\pm$~4.0" (3.1\,kpc) at 150\,MHz. Interestingly, the halo scale heights are similar to the measured 3$\upsigma$ extent measured in Section~\ref{sec:radiocont}. This underlines the faintness of the radio halo, which we can observe only to an extent of the scale height.

The radio scale heights are influenced by the contribution from  thermal emission. Whereas exponential scale heights seem to be the same within the errors in C-band for NGC~4565, the Gaussian scale heights of the total radio continuum emission in the halo are about 1\,kpc smaller compared to nonthermal scale heights (Schmidt et al. 2019, submitted). C-band data are most affected by the thermal contribution, so that the halo scale height of 15.3" (1.2\,kpc) is a lower limit. (The profiles of the nonthermal emission for C-band (see Section~\ref{sec:nonthermal}) could only be fit with large uncertainties with a resulting Gaussian halo scale height of 19.8"~$\pm$~4.0".)
The thermal contribution diminishes at lower frequencies, making the ratio of scale heights at L-band and 150\,MHz suited for studying CRE propagation (see below).

Our radio scale heights differ from the ones determined by \citet{krauseetal2018}, the main reason being that we fit double Gaussian functions while they fit exponential ones. In order to make a comparison possible, we show the fitting of two-component exponential functions in Figs~\ref{fig:N4013_nod3_C_exp}, \ref{fig:N4013_nod3_L_exp}, and \ref{fig:N4013_nod3_Lofar_exp}, for C-band, L-band, and LOFAR, respectively. With the combined configuration C-band data, the two-component exponential fits are comparable to the D-configuration C-band fit from \citet{krauseetal2018}. With the exponential fit in L-band (Fig.~\ref{fig:N4013_nod3_L_exp}) we achieve slightly larger scale heights but they are within the uncertainties of \citet{krauseetal2018}. We conclude that Gaussian scale heights cannot easily be compared to exponential scale heights. 

We now compare the exponential scale heights with other values from the literature. \citet{hummeletal1991} found a radio scale height of 4" from VLA C-band observations. This is half of our exponential scale height measured for the halo component of $8.1$", which, however, equates only to $0.6$\,kpc and thus shows the halo to be difficult to detect at C-band. With H{\sc i} data of NGC~4013, a scale height of 3" (210\,pc) was found by \citet{zschaechnerrand2015}, which is in good agreement with the Gaussian C-band scale height of the thin disk and is also similar to the mean scale height of the thin disk at 5~GHz, for which \citet{heesenetal2018} found $0.25\pm 0.13$\,kpc. \emph{Spitzer} data at 3.6\,$\upmu$m, analyzed by \citet{comeronetal2011}, suggest a thin disk of 122.5\,pc and a thick disk of 557.5\,pc. These values are roughly half the values found here for the Gaussian C-band radio scale heights. 

In summary, the scale height analysis shows Gaussian functions to better represent the vertical radio intensity profiles than exponential functions at L-band and the LOFAR frequency. For C-band, the fit quality is similar. Nevertheless, the Gaussian profile represents the data slightly better visually.

\begin{table}
\caption{NGC~4013 parameters for the scale height fitting.}  
\label{tab:N4013nod3para}
\begin{tabular}{lccc}
\hline \hline
Parameter 						&C-band			& L-band & LOFAR\\
\hline
Beam size (")					& 10.0			& 10.0  & 10.0 \\
Effective beam size (")			& 10.2			& 10.2 & 10.2 \\
Inclination	($^\circ$)			&  89   		& 89	& 89  \\
Position Angle ($^\circ$)		&  65   		& 65  & 65		  \\	
rms	(mJy/beam)					&  0.010		& 0.025 & 0.10		\\
Galaxy diameter (") 			& 205			& 205	& 230		\\
Box width (")					& 35    		& 35   & 35		  \\
Box height (")					&  4   			& 4 	& 4   \\
Number of boxes in X			&  5   		    & 5		&5  \\
Number of boxes in Y			&  20   	    & 20	& 24	  \\
\hline
\end{tabular}
\end{table}

\subsubsection{Amplitudes and integrated flux densities of Gaussian profiles}

With the Gaussian fit to the intensity profile, we gather the scale heights as well as the amplitudes of the Gaussian functions in each of the five strips. In Table~\ref{tab:N4013albflux_guassian} we present the mean disk amplitude and halo amplitude as well as the the ratio of them together with the mean scale height values. As expected, the disk amplitude dominates in all cases. But even more interesting is the question of which component of the total flux density is the dominating one, i.e. the disk or the halo. With the parameters of the Gaussian fit, we determine the integrated flux density, i.e. calculate the integral of the Gaussian function. To do so, we derive the mean amplitudes and mean scale heights. This is done for four strips omitting the central strip, because we want to find the general Gaussian function describing the galaxy without the central source. 

With the Gaussian distribution of the intensity profile for the mean of the four strips:
\begin{align}
w(z) = \bar{w_0} \ \text{exp}\,(-\,z^2/\bar{z_0}^2)    
\end{align}
the integrated flux density becomes: 
\begin{align}
W \, (\text{mJy $\cdot$ arcsec /\,beam}) = \sqrt{\pi\, } \, \bar{w_0} \, \bar{z_0} \, ,   
\end{align}
where the mean amplitude $\bar{w_0}$ is measured in mJy per beam area and the mean scale height $\bar{z_0}$ in arcsec.
We divide $W$ by the beam area $b$ in arcsec ($b$=113.3\,arcsec$^2$) to obtain the integrated flux density $W$ in mJy per arcsec, i.e. the flux density for a layer of one arcsec thickness perpendicular to the strip, and averaged the values for all strips. To determine the total flux density of the thin disk and the halo, which is to be compared with the measured total integrated flux densities, we multiply with the extent $e$ along the major axis:
\begin{align}
F (\text{mJy}) =  \sqrt{\pi\, } \, \bar{w_0}  \, \bar{z_0} \cdot e\, /\, b    
\end{align}

The extent $e$ is chosen to best represent the mean Gaussian profile of the galaxy. First, we estimate the extent of the galaxy at each frequency (Table~\ref{tab:N4013nod3para}), then we lower this extent to account for the contributing size of the central source. In the last step we consider the total flux densities of Table~\ref{tab:fluxmeasuren4013} as a reference point and increase the value until the calculated and measured total flux densities are similar. We obtain $e$~=~ 160"~$\pm$~15" in C-band, $e$~=~190"~$\pm$~15" in L-band, and $e$~=~195"~$\pm$~15" at 150\,MHz. The total emission of the "disk + halo" is the integrated flux density of Table~\ref{tab:fluxmeasuren4013} minus the central source flux density of Table~\ref{tab:centralsourcemeasure}, which is (11.8~$\pm$~0.6\,mJy) - (4.5~$\pm$~0.2\,mJy) = 7.3~$\pm$~0.6\,mJy in C-band, (36.8~$\pm$~1.8\,mJy) - (10.1~$\pm$~0.4\,mJy) = 26.7~$\pm$~2.2\,mJy in L-band, and (218~$\pm$~21\,mJy) - (28.3~$\pm$~1.7\,mJy) = 190~$\pm$~21\,mJy for the LOFAR data. These values are presented on the right-hand side of Table~\ref{tab:N4013albflux_guassian} along with the flux densities of the central source (from Table~\ref{tab:centralsourcemeasure}) for comparison. The results of the integrated flux densities of the disk and halo as well as their ratio are presented. As a consistency check, we added up the flux densities of the disk and halo Gaussian intensity models and confirmed that they agree with the measured total flux densities.
  
The C-band total flux densities are approximately the same for the disk and the central source, whereas the halo flux density is lower. In L-band, the halo flux density dominates mildly, but still agrees with the contribution from the disk within the uncertainties, while the flux density of the central source is slightly smaller. At 150\,MHz, the contribution from the halo to the flux density clearly dominates. The ratio of the integrated flux density of the disk to the integrated flux density of the halo shows a trend with frequency: from 6\,GHz to 150\,MHz, the ratio decreases from $1.7$ to $0.3$. The amplitude ratios follow the same trend, decreasing from $9.4$ to $1.9$. The CREs emitting at lower frequencies have longer lifetimes and thus can be transported into the halo more easily.

\begin{table*}
\centering
\begin{threeparttable}
 \captionof{table}{Integrated flux densities and scale heights of the thin disk and halo of NGC~4013.}  
\label{tab:N4013albflux_guassian}
\label{tab:N4013scaleheight}
\begin{tabular}{lccccc|cc}
\hline \hline
\multicolumn{6}{c}{Gaussian fits} & \multicolumn{2}{|c}{From total flux densities}\\  
     & $\bar{w}_0$ $^1$ & $\bar{z}_0$ $^1$ & $\bar{z}_0$& flux density& disk + halo & disk + halo & central source \\
     &(mJy/beam) & (") & (kpc) &  (mJy) &  (mJy) &  (mJy) &  (mJy) \\
\hline
C-band &        &     &   & && & \\
Disk  & 0.66~$\pm$~0.08 & 2.8 $\pm$ 0.7 & 0.2 $\pm$ 0.1 & 4.6 $\pm$ 1.7 & \multirow{2}{*}{ 7.3 $\pm$ 2.6}& \multirow{2}{*}{7.3 $\pm$ 0.6}& \multirow{2}{*}{4.5 $\pm$ 0.2} \\
Halo & 0.07~$\pm$~0.03 & >15.3 $\pm$ 3.0 $^2$  & >1.2 $\pm$ 0.2 $^2$ & 2.7 $\pm$ 2.0 & & &\\
Ratio &   9.4  &  < 0.18 $^2$  & & 1.7     & & & \\
\hline
L-band  &        &     & &  & & & \\
Disk  & 0.9~$\pm$~0.2 & 4.7 $\pm$ 0.3  &0.36 $\pm$ 0.05  & 11.5 $\pm$ 1.7 & \multirow{2}{*}{26.4 $\pm$ 4.9} & \multirow{2}{*}{26.7 $\pm$ 2.2}& \multirow{2}{*}{10.1 $\pm$ 0.4} \\
Halo & 0.2~$\pm$~0.1 & 25.8 $\pm$ 1.5  &2.0 $\pm$ 0.1    & 14.9 $\pm$ 4.6 & & & \\
Ratio &  4.5 & 0.18  &      & 0.8   &  &  &   \\
\hline
LOFAR  &        &     & &  & & & \\
Disk  & 2.3~$\pm$~1.1 & 6.1 $\pm$ 0.6 & 0.47 $\pm$ 0.10  & 43 $\pm$ 17 & \multirow{2}{*}{191 $\pm$ 20} & \multirow{2}{*}{190 $\pm$ 21}& \multirow{2}{*}{28.3 $\pm$ 1.7} \\
Halo & 1.2~$\pm$~0.1 & 40.4 $\pm$ 4.0  & 3.1 $\pm$ 0.3  & 148 $\pm$ 11 & & & \\
Ratio &  1.9 & 0.15  &    & 0.3   &  &  &   \\
\hline
\end{tabular}
\begin{tablenotes}
\item \textbf{Notes.} $^1$ $\bar{w}_0$ is the mean amplitude omitting the central strip, $\bar{z}_0$ is the mean scale height omitting the central strip.\\
$^2$ Lower limit if only the synchrotron emission is considered.
\end{tablenotes}
\end{threeparttable}
\end{table*}


\subsection{Polarization and the large-scale magnetic field structure}
\label{subsec:pol}

\begin{figure*}
\begin{minipage}[t]{0.35\textwidth}\vspace{0pt}
\includegraphics[height=5.5cm]{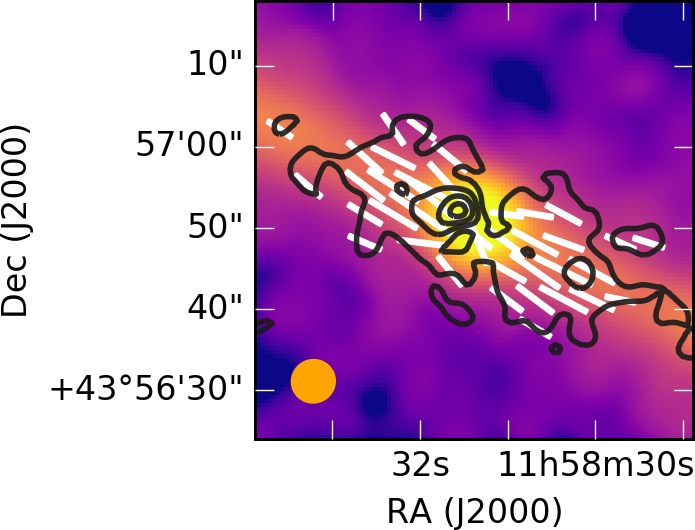}
\end{minipage}\hfill
\begin{minipage}[t]{0.60\textwidth}\vspace{0pt}
\includegraphics[height=5.5cm]{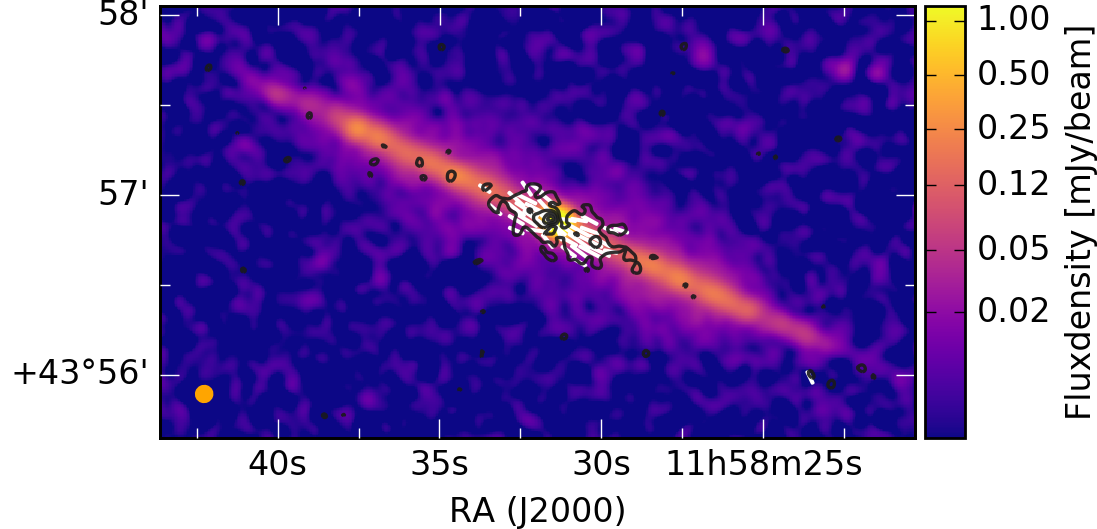}
\end{minipage}
\caption{NGC~4013 combined (C- and D-configuration) C-band Stokes I image (same data shown in Figure~\ref{fig:N4013-Ccomb-stokesI}) with apparent magnetic field orientations and polarized intensity contours at 3, 6, 9, and 12$\upsigma$ levels with a $\upsigma$ of 2.5\,$\upmu$Jy/beam. The beam size is 5.2" $\times$ 5.3" and is shown in the bottom left corner of the image. No tapering was used and the robust parameter was set to two. The left image shows an enlargement of the central region.} 
	\label{fig:N4013-C_I_PIcont_magvec_rob2}
\end{figure*}

\begin{figure*}
\centering
\includegraphics[width=0.69\textwidth]{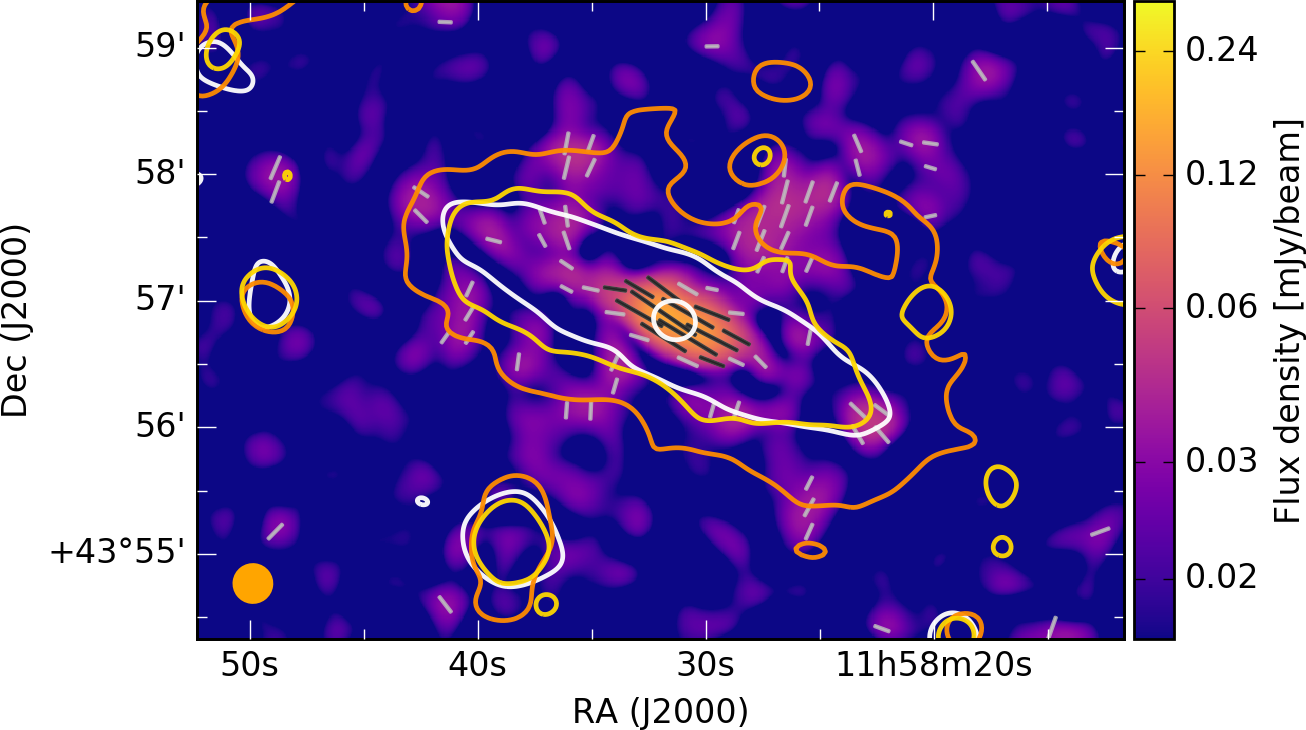}
\caption{NGC~4013 combined (C- and D-configuration) C-band polarized intensity from Figure~\ref{fig:N4013-C_I_PIcont_magvec_rob2}, but smoothed to a beam size of 18"~ $\times$~18" with a $\upsigma$ of 8.6\,$\upmu$Jy/beam. Apparent magnetic field orientations are shown in light grey, where the polarized intensity is at the 2.5$\upsigma$ level or higher, black apparent magnetic field orientations are shown for polarized intensity of 5$\upsigma$ level or higher. Stokes I intensity contours of C-band are displayed in white at 3 and 64\,$\upsigma$ levels with a $\upsigma$ of 13\,$\upmu$Jy/beam obtained with an $uv$-taper of 12\,k$\lambda$ and Gaussian smoothing. The yellow contours of the L-band Stokes I intensity obtained with an $uv$-taper of 18\,k$\lambda$ and Gaussian smoothing are displayed at 3\,$\upsigma$ level with a $\upsigma$ of 50\,$\upmu$Jy/beam. The orange contours represent the Stokes I intensity of LOFAR at 3$\upsigma$ with a $\upsigma$ of 204\,$\upmu$Jy/beam.}
	\label{fig:N4013-C_I_PI_magvec_rob2_smooth18}
\end{figure*}

The results from the C-band polarization data are shown at high and low angular resolution in Figs~\ref{fig:N4013-C_I_PIcont_magvec_rob2} and \ref{fig:N4013-C_I_PI_magvec_rob2_smooth18}, respectively. In Figure~\ref{fig:N4013-C_I_PIcont_magvec_rob2}, a symmetric structure in the disk is visible east and west of the central point-like source with the peak of the polarized emission in the east. The orientation of the magnetic field is disk-parallel everywhere and no large-scale polarized structures, besides that of the disk field, are detected. The low-resolution map (beam size of 18"~$\times$~18") presented in Figure~\ref{fig:N4013-C_I_PI_magvec_rob2_smooth18} reveals new structures.  As before, a plane-parallel magnetic field in the disk is observed, but now many polarized features reaching far into the halo with vertical field orientations are seen as well. Because the total radio intensity is not very high (11.8~$\pm$~0.06\,mJy in C-band), the polarized intensity is even lower (1.63~$\pm$~0.08\,mJy in C-band from Figure~\ref{fig:N4013-C_I_PI_magvec_rob2_smooth18}). 
However, strong smoothing (18") in combination with imaging with a robust weighting parameter of 2 made it possible to reveal the large-scale magnetic field in NGC~4013. For comparison, C-band, L-band and LOFAR Stokes I contours are shown at a 3-$\upsigma$ level in Figure~\ref{fig:N4013-C_I_PI_magvec_rob2_smooth18}. The polarized emission extends as far as the LOFAR emission and the L-band outer contour shows an elongation towards the northern polarization 'blob', lending credibility to the detection. The rotation curve of the galaxy indicates that the eastern part is approaching the observer \citep[e.g.,][]{bottema1996}.
Thus, within the disk, the approaching side shows more polarized emission than the receding side, similar to NGC~4666 \citep{steinetal2019}. 

The vertical magnetic field structure in the halo cannot be explained by diffusion alone and indicates an additional outflow. To investigate this further, we follow the strategy described in \citet{damasetal2016}, who used the projected distance of polarized structures together with an estimate of the CRE lifetime to measure the outflow speed. The maximum extent of the north-western polarized emission blob 
above the disk is about 82" (6.4\,kpc). The synchrotron lifetime follows from: 
\begin{align}
t_\mathrm{syn} = 1.06 \times 10^9 \ B^{-3/2} \ \upnu^{-1/2} ~\rm yr  
\end{align}
with the total field strength $B$ in $\upmu$G and $\upnu$ in GHz. We assume a decline of the magnetic field strength of the disk towards the halo, going from the disk value of B~$\simeq$~6.6\,$\upmu$G (see Section 3.7) to  B~$\simeq$~2\,$\upmu$G at the maximum z-extent of 6.4\,kpc. In order to account for inverse Compton losses, we added the average magnetic field strength $B_\text{halo}$~$\simeq$~4.0\,$\upmu$G encountered by CREs travelling into the halo and the equivalent magnetic field strength $B_\text{CMB}$ in quadrature,  with the latter accounting for the energy density of the CMB: 
\begin{align}
\sqrt{(B_{\text{halo}}^2 + B_{\text{CMB}}^2)} = \sqrt{(4.0^2 + 3.25^2)} \simeq 5\, \upmu\text{G}
\end{align}
The synchrotron lifetime then is $t_\mathrm{syn}~\simeq~3.8~\times~10^7$\,yr (B~$\simeq$~5\,$\upmu$G, $\upnu$=6\,GHz). Thus, the velocity which is needed to transport CREs to 6.4\,kpc height in the halo of NGC~4013 is about 160\,km s$^{-1}$. This is a low advection velocity in comparison to other galaxies \citep{heesenetal2018}.

\begin{figure}
	\centering
\includegraphics[width=0.48\textwidth]{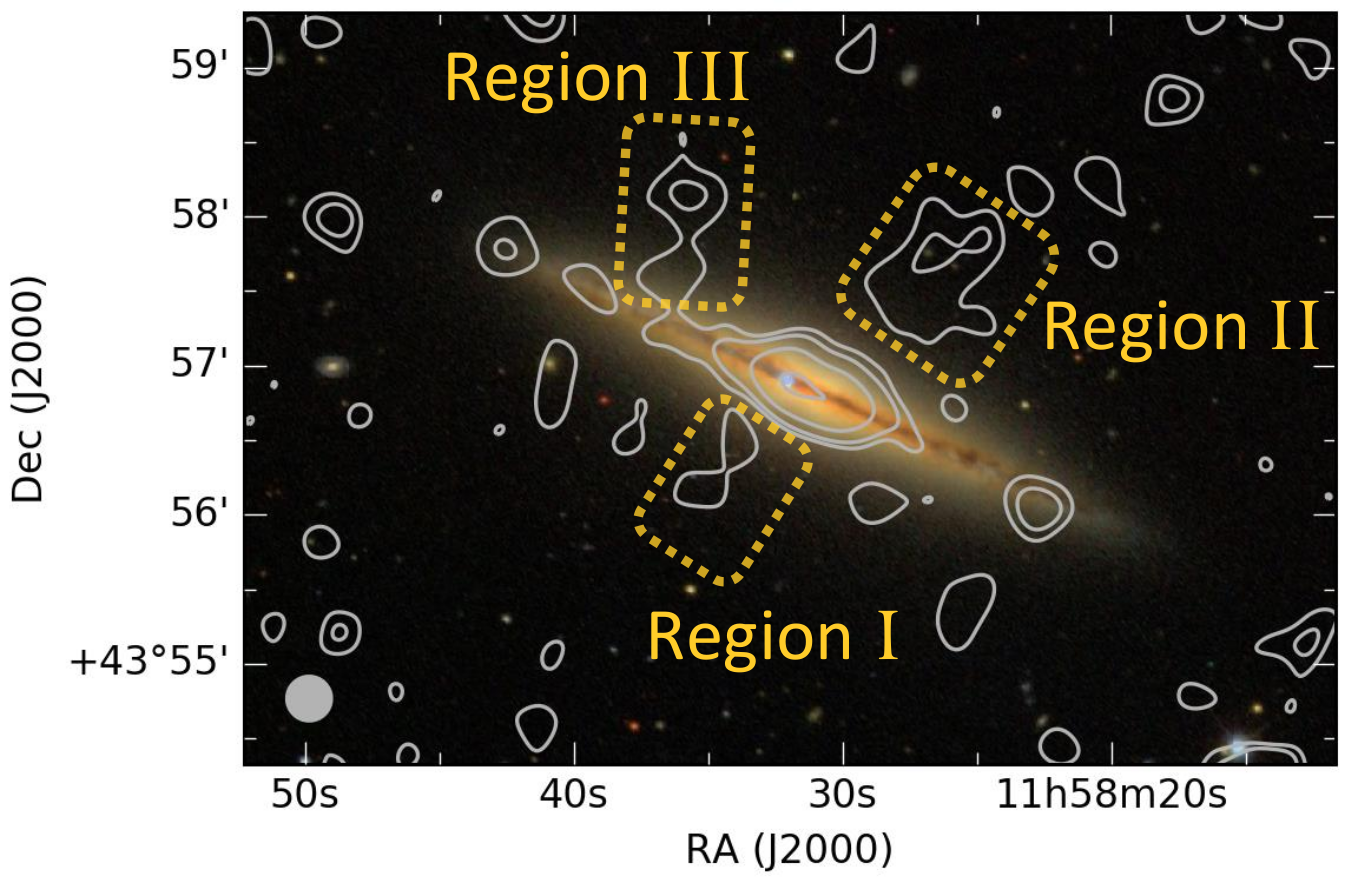}
\caption{NGC~4013 combined (C- and D-configuration) C-band polarized intensity contours from Figure~\ref{fig:N4013-C_I_PI_magvec_rob2_smooth18} with contours at 2.5, 4, 8 and 16$\upsigma$ levels on SDSS color image (retrieved via Aladin sky atlas$^5$). Regions I, II, and III discussed in the text are marked in yellow.}
\label{fig:N4013-pol_SDSS}
\end{figure} 

The X-ray emission, presented in Figure~\ref{fig:n4013_xmm}, is mainly located where we observe the strongest polarized emission of NGC~4013 (yellow/orange color in Fig.~\ref{fig:N4013-C_I_PI_magvec_rob2_smooth18}). The southern polarization structure in the halo pointing roughly towards the center (Region I in Fig.~\ref{fig:N4013-pol_SDSS}) seems to be present in the XMM-Newton image as well. The polarized 'blob' in the halo to the north (Region II) shows the same orientation of the magnetic field vectors in an area of more than two beams. We find two distinct X-ray emission features within that blob, but these could be associated with two point sources confirmed by Chandra (see Fig.~\ref{fig:N4013-Chandra-Ccomb}), probably background sources. Furthermore, a source is located in the upper part of the blob (RA 11:58:26.87, Dec +43:57:46.14), visible in C-band Stokes I (e.g., Fig.~\ref{fig:N4013-C_I_PIcont_magvec_rob2}) and in the optical (e.g., in SDSS) as a diffuse source, possibly a background galaxy. The possibility that these sources could generate an extended polarized blob seems unlikely. Therefore, we assume this polarization feature to originate from NGC~4013. The polarization structure to the northeast of the galaxy (Region III) reaches far into the halo. This location coincides with an extended HI filament \citep{bottema1996} showing that this galaxy seems to have local outflows, which have lower velocities than the escape velocity.

In the L-band data, no polarized emission is detected. Polarized emission in L-band is expected to be affected by depolarization due to Faraday rotation. As this effect is wavelength dependent, the depolarization is stronger in L-band than in C-band. Therefore, images were produced with almost natural weighting ($\tt robust=2$), strong $uv$-tapering, and additional use of RM-Synthesis, but still no polarized flux density within the central disk could be recovered, with the exception of a few patches of low intensity emission. These are at a level of 2$\upsigma$ (with a $\upsigma$ of 29\,$\upmu$Jy/beam) and consistent with the peaks of the C-band emission.

\subsection{Rotation measures and the magnetic field configuration in the disk}
The polarization angle of an electromagnetic wave experiences a frequency dependent rotation while propagating through a magnetized plasma. This effect is called Faraday rotation  with the rotation of the polarization angle $\chi$ being proportional to the wavelength squared and the RM.

We perform RM synthesis following \citet{steinetal2019}, with the important parameters presented in Table~\ref{tab:RMpar}. From the data in \citet{rm2014}, the Galactic foreground RM of NGC~4013 is determined. This value (RM$_{\text{foreground}}~=~-8.1~\pm~1.2$) is subtracted from the final RM-cube to correct for the foreground contribution.
\addtocounter{footnote}{+1}
\footnotetext{https://aladin.u-strasbg.fr}
\addtocounter{footnote}{+1}

\begin{table}
\begin{threeparttable}
\centering
\caption{RM-synthesis parameters.}  
\label{tab:RMpar}
\begin{tabular}{l c c} \hline\hline
 \	                            																						& C-band     & L-band   \\ \hline
Bandwidth [GHz]																																	& 2         & 0.5     \\
$\upnu_{\text{min}}$ to $\upnu_{\text{max}}$ [GHz]						  								       & 5 to 7    & 1.2 to 1.75 [gap]\\
$\Delta \lambda^2$ = $\lambda_1^2 - \lambda_2^2$ [m$^2$] 													& 0.00184 	& 0.0336 	\\
$\delta \phi$ = $\frac{2 \sqrt{3}}{\Delta \lambda^2}$ [rad m$^{-2}$]           & 1882   	  & 103    \\ \hline
$\lambda_{min}$ [m]   	 																														& 0.042     & 0.17    \\
max$_{\text{scale}}$  = $\frac{\pi}{\lambda_{min}^2}$ [rad m$^{-2}$]  						& 1781      & 109    \\ \hline
spw-width $\delta$f [MHz]  																											& 125       & 16        \\
$\delta \lambda^2$ [m$^2$]												                   							 & 0.0001735 & 0.001447    \\
RM$_{\text{max}}$ = $\frac{\sqrt{3}}{\delta \lambda^2} $ [rad m$^{-2}$]											& 9983      & 1197       \\
\hline
\end{tabular}
\begin{tablenotes}
\footnotesize
\item \textbf{Notes:}
\begin{itemize}
\item Channel width: $\delta \lambda^2$, width of the $\lambda^2$ distribution: $\Delta \lambda^2$, shortest wavelength squared: $\lambda_{min}^2$  
\item FWHM of the resolution (RMSF) in $\Phi$ space: $\delta \phi$ (rad m$^{-2}$) 
\item Largest scale in $\Phi$ space to which the observation is sensitive: max-scale (rad m$^{-2}$)
\item Maximum observable RM (resp. $\Phi$): RM$_{\text{max}}$ (rad m$^{-2}$).
(from \cite{brentjensdebruyn2005}).
\end{itemize}
\end{tablenotes}
\end{threeparttable}
\end{table}
\normalsize

\begin{figure}
	\centering
		\includegraphics[width=0.5\textwidth]{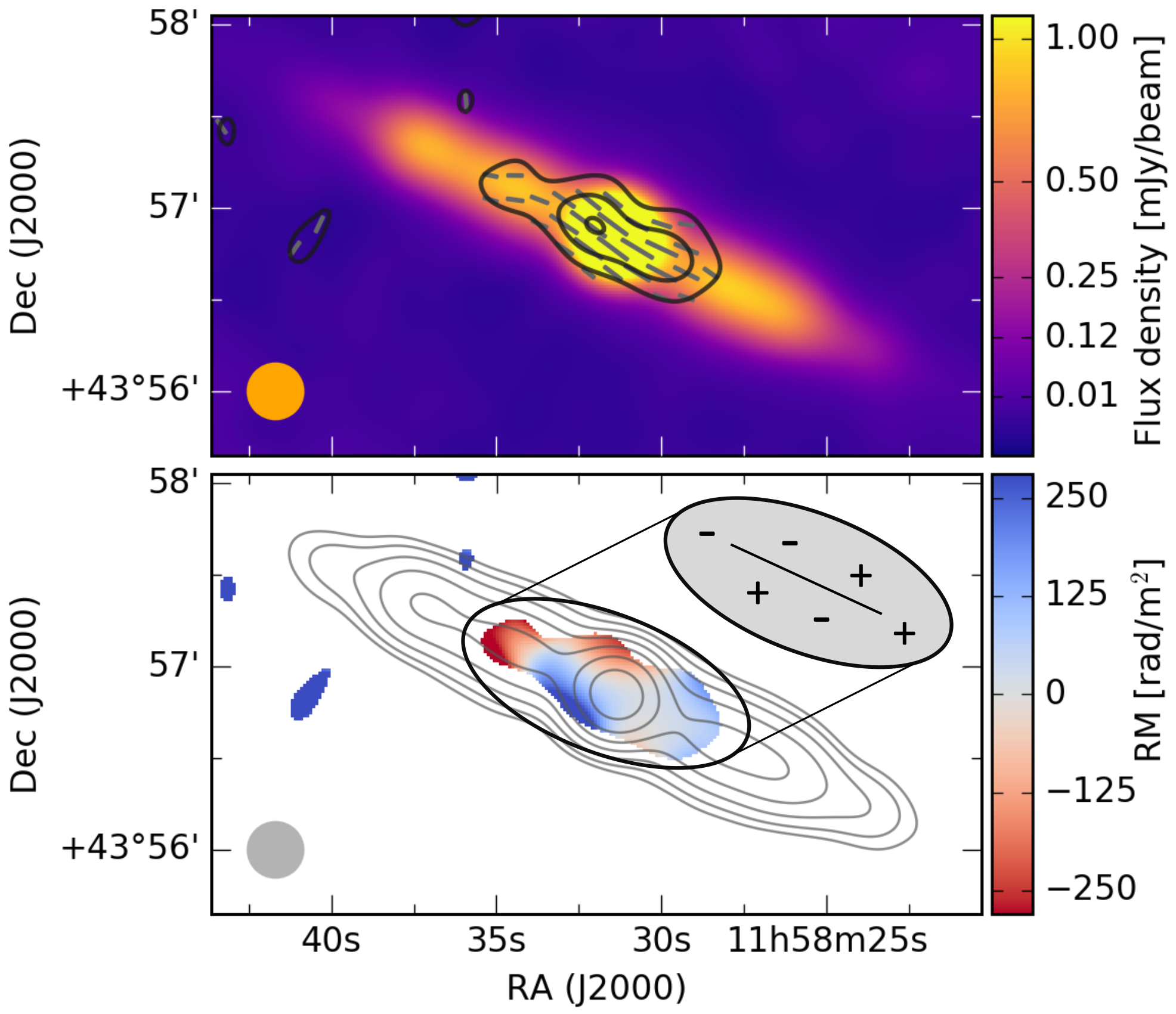}
	\caption{NGC~4013 images from RM-synthesis. Top: total intensity color image with polarization contours at 3, 9, 12 $\upsigma$ levels with $\upsigma$ of 11.6\,$\upmu$Jy. Magnetic field orientations are shown in black. 
	Bottom: RM map of NGC~4013 from C-band on Stokes I contours with a $uv$-taper of 12\,k$\lambda$ and Gaussian smoothing resulting in a beam size of 18" $\times$ 18". Contours start at the 3$\upsigma$ level with a $\upsigma$ of 15.0\,$\upmu$Jy/beam and increase in powers of 2 (up to 256). The RM map is cut at the 3$\upsigma$ level of 35\,$\upmu$Jy/beam of the polarized intensity map. The mean RM error is 25\,rad/m$^2$.}
	\label{fig:N4013-C_RM}
\end{figure}

The maps of polarized intensity and Faraday RM, both produced with RM-synthesis, are shown in Figure~\ref{fig:N4013-C_RM}. We overlay the total intensity map with polarization contours, and the intrinsic magnetic field orientation, corrected for Faraday rotation, and the RM map with total intensity contours. As in Figure~\ref{fig:N4013-C_I_PIcont_magvec_rob2}, the magnetic field orientation is disk-parallel everywhere and the polarized intensity is confined to the central third of the galaxy, which coincides with the inner part of the hot X-ray halo (Fig.~\ref{fig:n4013_xmm}).

The RM map shows values between $-300 \pm 25$\,rad/m$^2$ and $+300 \pm 25$\,rad/m$^2$. The sign of RM changes along the disk from east to west, where three overlapping patches are visible. The magnetic field of the patch furthest to the east of the central source is pointing away from the observer (negative RM), whereas the magnetic field of the westernmost patch is pointing towards the observer (positive RM). This is a signature of a spiral field that is axisymmetric with respect to the center and symmetric (even parity) with respect to the plane. Assuming trailing spiral arms and considering the rotation of the galaxy with the north-eastern side of the disk being the approaching side \citep{bottema1996}, the RM values suggest that the radial component of the magnetic field vectors is pointing outwards. Dynamo theory predicts that the radial components of axisymmetric spiral fields should point either inwards or outwards with equal probability, which is indeed the case \citep[with similar numbers of cases found each so far, see][]{beck2016}.

The field of the central region east of the center shows a field pointing towards the observer in the bottom part (positive RMs) and away from the observer in the upper part (negative RMs). The western side shows hints of the opposite trend, where the upper part points towards the observer and the lower left part points away from the observer. This RM pattern could be caused by an antisymmetric (odd parity) magnetic field structure. A similar pattern, i.e. symmetric in the outer disk and antisymmetric in the inner disk, was proposed for the magnetic field in the Milky Way \citep{han97,han06}.

The signal-to-noise ratios of our polarization data are small, so uncertainties remain about the large-scale structure of the magnetic field. Deeper observations with a wider $\lambda^2$-space distribution are necessary to further investigate the magnetic field structure of this galaxy.  

\subsection{Thermal and nonthermal emission}
\label{sec:nonthermal}
\subsubsection{Thermal  and nonthermal separation}

The separation of thermal and nonthermal (synchrotron) emission in spiral edge-on galaxies was explored by \citet{vargasetal2018}. We followed their recommended approach of calculating the thermal flux by correcting the H$\upalpha$ flux for internal absorption by dust with mid-infrared data, in our case \emph{WISE} 22\,$\upmu$m data, with the revised correction factor for edge-on galaxies of $0.042$ \citep[replacing the $0.031$ in equation~(5) of][]{calzettietal2007}. Nonthermal maps were produced for the CHANG-ES bands with a final resolution of 15"~$\times$~15" \citep[see ][for details]{steinetal2019}. For the LOFAR data, the expected thermal contribution is small and we disregard it.


\subsubsection{Degree of polarization of the nonthermal emission}

\begin{figure}
	\centering
		\includegraphics[width=\hsize]{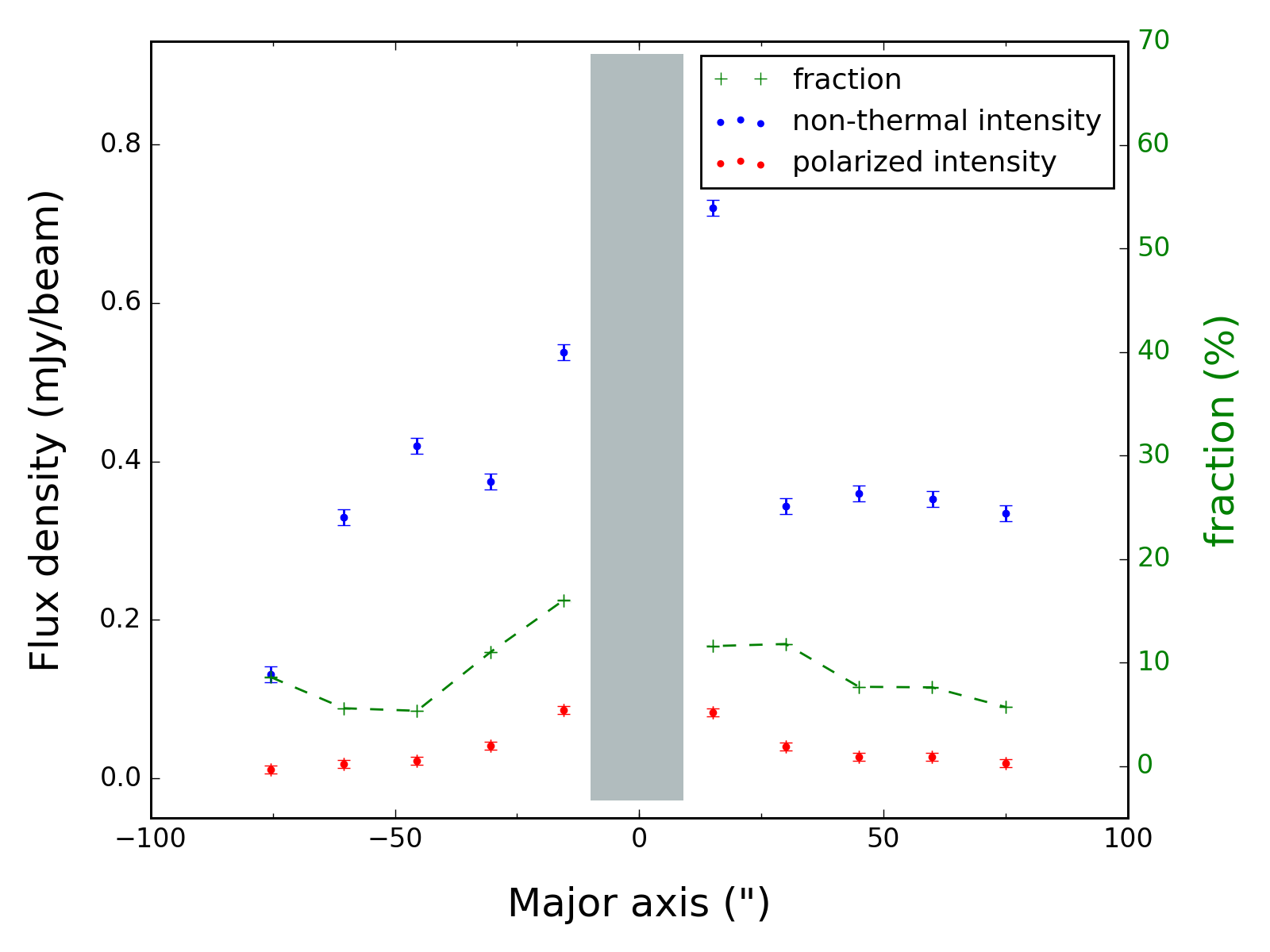}
	\caption{NGC~4013 box integration along the major axis with a box size of 15"$\times$~15", showing the nonthermal intensity (blue data points), the polarized intensity (red data points) as well as the degree of polarization (green) of C-band. The grey box shows the region of the central source, where the data are not considered.}
	\label{fig:N4013degreepol}
\end{figure}

To further investigate the appearance of the polarized emission, we plot C-band flux densities of the nonthermal intensity, polarized intensity as well as the fraction of both values within boxes along the major axis. The box width and height were chosen to be equal to the beam size of 15". The result is shown in Figure~\ref{fig:N4013degreepol}.  The values of the polarized fraction in the disk vary between 4\% and 12\%, which is a normal range for C-band. Generally, the polarized fraction is highest in the center and decreases with increasing distance along the major axis. The steep increase towards the edges can be explained by the lower rms noise of the polarized intensity map in comparison to the total intensity map, so that we detect polarized emission out to larger distances.

\subsubsection{Nonthermal fractions}

\begin{figure}
	\centering
		\includegraphics[width=0.48\textwidth]{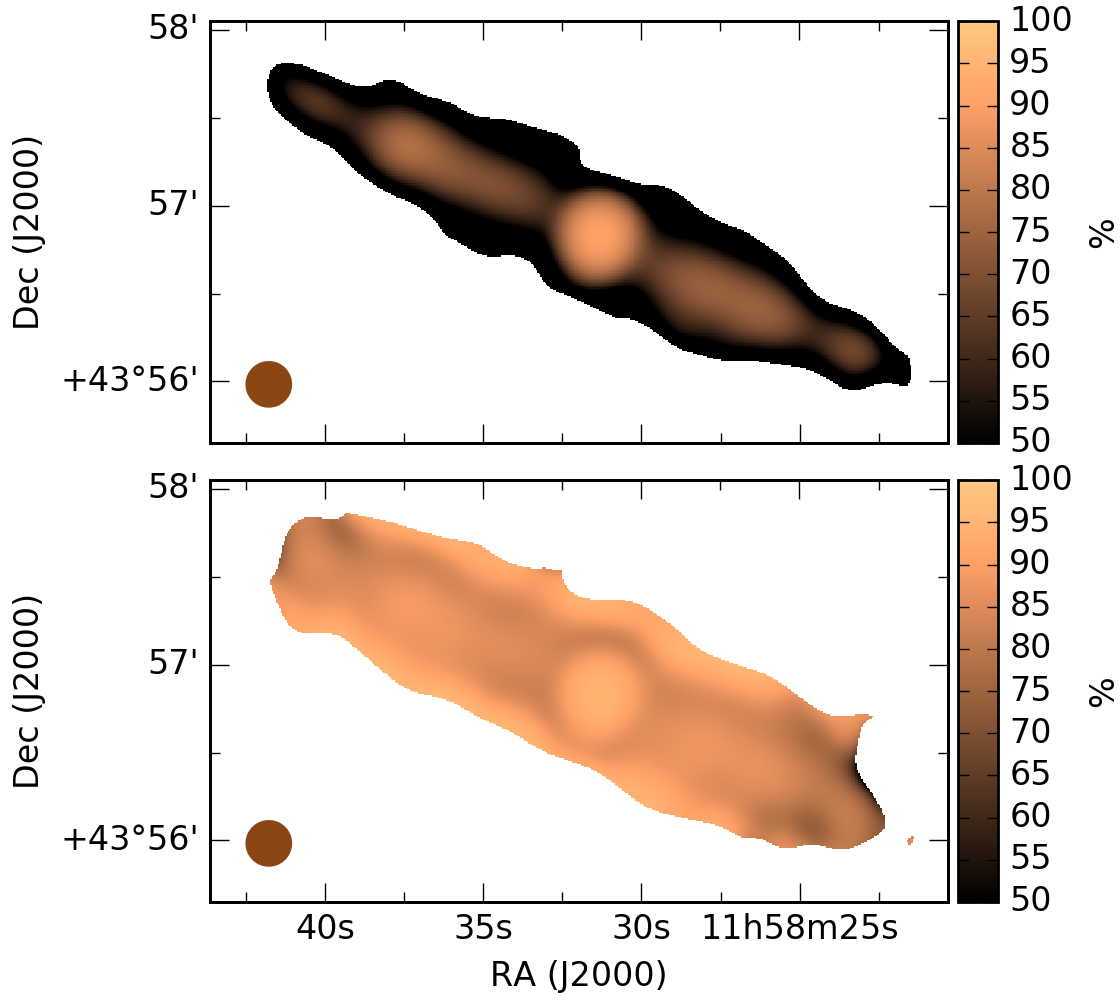}
	\caption{NGC~4013 nonthermal fractions of C-band (top) and L-band (bottom). The beam size of 15"~$\times$~15" is shown in brown.}
	\label{fig:N4013_nonthermalfrac_Lband}
	\label{fig:N4013_nonthermalfrac_Cband}
\end{figure}

Maps of the nonthermal fraction for both C-band and L-band are shown in  Figure~\ref{fig:N4013_nonthermalfrac_Cband}. The central point-like source dominates the galaxy and shows synchrotron emission of high intensity as well as a high nonthermal fraction of 93\% in C-band. Such a high nonthermal fraction suggests a strong synchrotron source including a high magnetic field strength. The nonthermal fraction of the disk has a mean value of 76\%. This implies a thermal fraction of 24\% in C-band, which agrees well with the mean thermal fraction of ($23\pm 13$)\% in late-type galaxies at 5~GHz \citep{tabatabaeietal2017}.

The nonthermal fraction map in L-band looks similar to the C-band map, but with a more extended distribution towards the halo. The central point-like source is again dominating and exhibits a nonthermal fraction of 97\%, which is very high and confirms the C-band results. The entire galaxy disk shows a mean nonthermal fraction of 90\% with a corresponding thermal fraction of 10\% in L-band. This is in very good agreement with the mean thermal fraction of ($10\pm 9$)\% in late-type galaxies at $1.4$~GHz \citep{tabatabaeietal2017}. 
 
\subsubsection{Nonthermal spectral index}

\begin{figure}
	\centering
		\includegraphics[width=1.0\hsize]{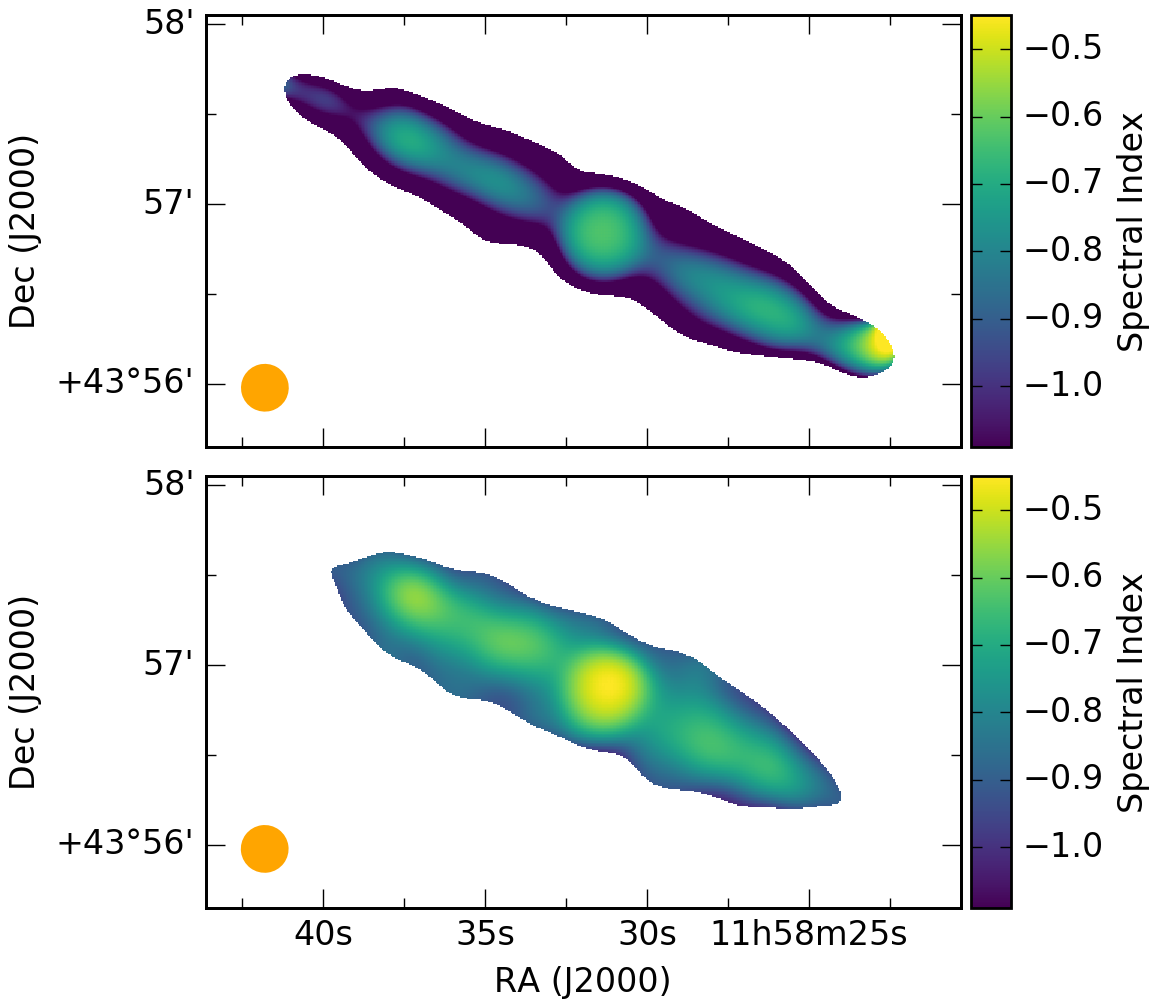}
	\caption{Maps of the nonthermal spectral index. Top: Calculated between C- and L-band. Bottom: Calculated between L-band and LOFAR. The mean error is $0.05$; the error is largest towards the edges with values of up to $0.3$. The beam size of 15"~$\times$~15" is shown in orange.}
	\label{fig:N4013_SpI}
\end{figure}

As in \citet[][]{steinetal2019}, we derive the nonthermal spectral index map, here from the nonthermal maps of C-band and L-band as well as from the nonthermal map of L-band and LOFAR, which are presented in Figure~\ref{fig:N4013_SpI}. We also created a map of the uncertainties, using error propagation, where we found a mean uncertainty of $0.05$. The spectral index map was blanked in places where the uncertainty equals or exceeds $0.3$.

The mean spectral index of the disk between C-band and L-band (top of Fig.~\ref{fig:N4013_SpI}) of $\upalpha_\mathrm{nt}$~=~$-$0.84~$\pm$~0.05 agrees well with the expected theoretical spectral slope of synchrotron emission. The central region shows a slightly flatter spectral index of $\upalpha_\mathrm{nt}$~=~$-$0.63~$\pm$~0.03. At the south-western end of the disk, a region with a flat spectral index of about $\upalpha_\mathrm{nt}$~=~$-$0.3~$\pm$~0.2 is visible. The large error 
is a result of the weak intensity levels towards the edge of the galaxy. Nevertheless, the flat spectral index indicates a large thermal fraction. This is corroborated by the fact that this region coincides with a star-forming region seen in H$\upalpha$ emission. This origin is additionally supported by the lower nonthermal fraction in this region seen in L-band. 

The mean spectral index of the disk between L-band and LOFAR (bottom of Fig.~\ref{fig:N4013_SpI}) of $\upalpha_\mathrm{nt}$~=~$-$0.69~$\pm$~0.05 is flatter in comparison to the higher frequency spectral index. The central source shows a flat spectral index of $\upalpha_\mathrm{nt}$~=~$-$0.48~$\pm$~0.03.

\subsection{Magnetic field strength via the equipartition assumption}

The map of the magnetic field strength was determined using the revised equipartition formula by \citet{beckkrause2005}.
The calculations were done in a pixel-by-pixel fashion by combining the nonthermal intensity map of C-band and the spectral index map between C-band and L-band (top of Fig.~\ref{fig:N4013_SpI}). As path length, we used 90\% of the extent in the radio continuum of 15\,kpc. Further, a proton to electron ratio of K$_0$ = 100 is assumed. The uncertainties in the assumptions lead to an error in field strength of about 20\% \citep[e.g.,][]{beck2016}.

\begin{figure}
	 	\includegraphics[width=1.06\hsize]{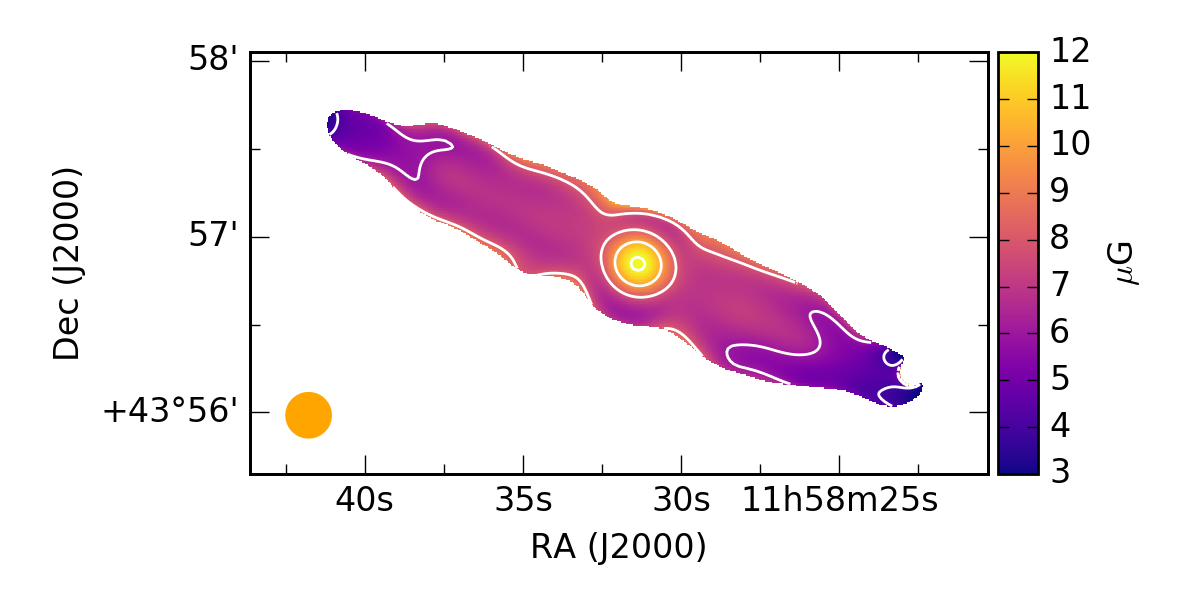}
	\caption{NGC~4013 magnetic field strength calculated pixel-by-pixel via the equipartition formula using the nonthermal spectral index map and the nonthermal intensity map. The beam size of 15"~$\times$~15" is shown in orange.}
	\label{fig:N4013_bfeld_genau}
\end{figure}

The resulting map of the magnetic field strength is presented in Figure~\ref{fig:N4013_bfeld_genau}. The magnetic field strength of the central point source is 12.1\,$\upmu$G. The mean field strength in the disk is weak with 6.6\,$\upmu$G, omitting the central region. In the inner disk, the magnetic field strength is 7.0\,$\upmu$G and the field strength decreases towards the edges of the disk. This map can explain our results from the analysis of the polarized emission. The polarized emission is mostly concentrated in the inner third of the disk along the major axis, where, according to the map shown here, the magnetic field is comparably strong.

\subsection{Cosmic-ray transport model}
\label{subsec:spinnaker}

\begin{figure}
	\centering
		\includegraphics[width=\hsize]{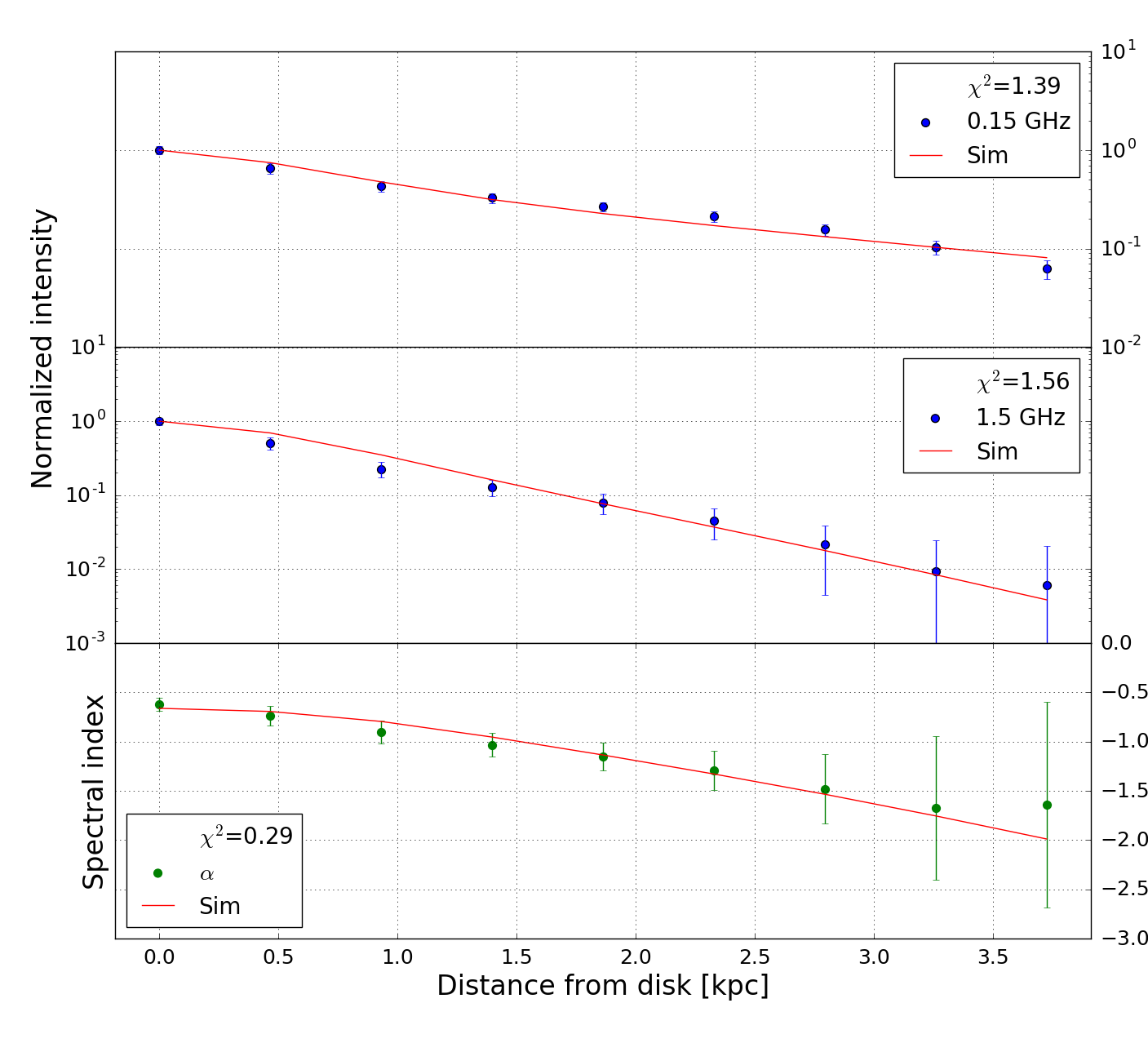}
	\caption{Best-fitting CRE advection model with an increasing speed for NGC~4013. The top panel shows the nonthermal intensities at 150\,MHz (LOFAR), the middle panel the 1.5\,GHz (L-band) data, and the bottom panel the nonthermal radio spectral index. In all panels, the solid red line represents the model simulation. The errors are represented by a weighted standard deviation. The best-fitting parameters are listed in Table~\ref{tab:N4013spinnaker}.}
	\label{fig:N4013-spinteractive-low}
\end{figure}

\begin{figure}
	\centering
		\includegraphics[width=\hsize]{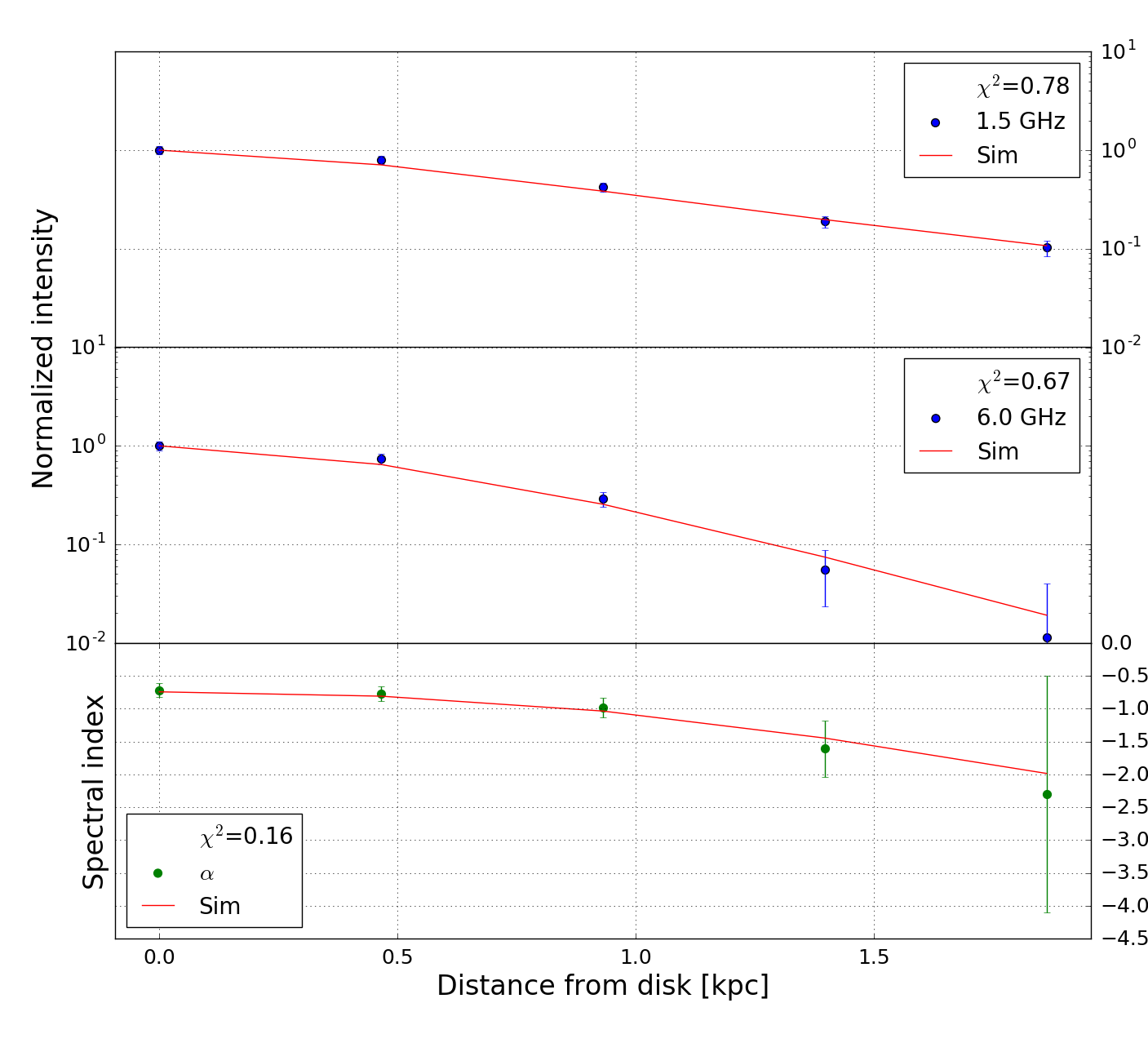}
	\caption{Best-fitting CRE advection model with an increasing speed for NGC~4013. The top panel shows the nonthermal intensities at 1.5\,GHz (L-band)}, the middle panel the 6\,GHz (C-band) data, and the bottom panel the nonthermal radio spectral index. In all panels, the solid red line represents the model simulation. The errors are represented by a weighted standard deviation. The best-fitting parameters are listed in Table~\ref{tab:N4013spinnaker}.
	\label{fig:N4013-spinteractive-high}
\end{figure}

Using  the  1D  CR  transport model SPectral  INdex  Numerical  Analysis of K(c)osmic-ray Electron Radio-emission \citep[\textsc{spinnaker},][]{heesenetal2016,heesenetal2018} the transport processes of CRs from the disk into the halo (perpendicular to  the  galaxy  disk)  are  derived. Analogous to the analysis of NGC~4666 \citep{steinetal2019}, we carried our an interactive analysis on the intensity profiles of NGC~4013 in order to find the best-fitting model. Because the galaxy is very thin, especially in C-band, only five data points could be measured in the vertical direction (box size of 6"~ $\times$~165") for the L- and C-band data (Fig.~\ref{fig:N4013-spinteractive-high}), and nine for the L-band and LOFAR data (Fig.~\ref{fig:N4013-spinteractive-low}). Since we are using the nonthermal radio continuum maps, we have an angular resolution of 15" ($1.1$~kpc).

The models assume either pure diffusion or advection for the CRE, which are injected with a power-law in the disk. The CRE number density is then $N(E)\propto E^{-\gamma}$, where $E$ is the CRE energy. The power-slope can be related to the nonthermal spectral index via $\alpha_{\rm nt} = (1-\gamma)/2$. 
Then, a steady state solution is reached between CRE injection and energy losses (synchrotron and inverse Compton losses). The magnetic field strength is described as a double exponential function with the superposition of a thin and thick disk:
\begin{equation}
    B(z) = B_1 \exp(-z/h_{\rm B1}) + (B_0-B_1) \exp(-z/h_{\rm B2}),
\end{equation}
where $B_0$ is the magnetic field strength in the disk, $B_1$ is the thin disk magnetic field strength, and $h_{\rm B1}$ and $h_{\rm B2}$ are the scale heights of the thin and thick disk, respectively.

To find the best-fitting model, we minimize $\chi_T^2$, which is the quadratic sum of the $\chi^2$-values for the intensity and spectral index profiles. We tested three different models, which are:
\begin{itemize}
    \item [](i) Cosmic ray diffusion.
    \item [] (ii) Cosmic-ray advection with constant speed.
    \item [] (iii) Cosmic-ray advection with an increasing speed.
\end{itemize}
The accelerating wind speed is parametrized as follows:
\begin{equation}
    V(z) = V_0 \left (1+\frac{z}{h_{\rm V}}\right ).
\end{equation}
For the diffusion model, we assumed no energy dependence for the diffusion coefficient. We fit the three different frequencies simultaneously pairwise, combining LOFAR and L-band data as well as L-band and C-band data. 

In this section, we present the best-fitting profiles for advection with increasing speed [model (iii)] in Figs.~\ref{fig:N4013-spinteractive-low}~and~\ref{fig:N4013-spinteractive-high}, with the other models presented in Appendix~\ref{app:cr_transport}. We find that diffusion [model (i)] can describe our L-band and C-band data equally well as advection [models (ii) and (iii)] with each $\upchi^2_T$ less than one
and thus no possible quantitative statement about which model is the best. While the profile of the radio spectral index between L-band and C-band has a parabolic shape that is typical for diffusion, this can also be explained with advection due to the convolution with the large effective beam ($FWHM=1.1~\rm kpc$). For advection, we expect the spectral index profile to be linear, but this assumes the vertical profile to be well resolved. For the profile in Figure~\ref{fig:N4013-spinteractive-high} the influence of the angular resolution becomes visible, so that, due the Gaussian convolution with an effective beam size of $\rm FWHM=1.1~kpc$, the resulting spectral index profiles have a shape similar to the one for diffusion. 

In contrast, the shape of the spectral index profile between LOFAR and L-band has a linear shape and thus cannot be well fit with diffusion. In Figure~\ref{fig:N4013-spinteractive-low} the best-fitting advection model assuming an increasing advection speed is shown, which is preferred over the constant speed since it ensures that energy equipartition is approximately fulfilled in the halo. Nevertheless, the model fits to the LOFAR data are better approximated with diffusion (first row in Fig.~\ref{fig:diff_high}) with a smaller $\upchi^2$ in comparison to advection. The best-fitting advection speed of $V_0=18-22$\,km\,s$^{-1}$ (increasing to only $\sim50$\,km\,s$^{-1}$ at 4\,kpc height) is very low, not comparable to wind speeds of typically hundreds of kilometers per second. Furthermore, it is lower compared to the outflow velocity estimated in Section~3.4, which suggests a basic difference between the global transport mechanisms averaged over the entire galaxy (modeled with \textsc{spinnaker}) and local mechanisms that can lead to higher velocities. The best-fitting advection model has $\chi_T^2=0.5$--$1.1$, which shows that the data can be adequately fit with this simple model. 

\footnotetext{\raggedright \textsc{spinnaker} and \textsc{spinteractive} are available at: www.github.com/vheesen/Spinnaker}

\begin{table}
 \captionof{table}{NGC~4013 \textsc{spinnaker} best-fitting parameters.}  
\label{tab:N4013spinnaker}
\centering
\begin{tabular}{lc}
\hline \hline
 Parameter &\\
\hline
B$_0$ ($\upmu$G)  & $6.6$  (fixed) \\
\hline

\multicolumn{2}{c}{Model (i) -- Diffusion} \\
$\gamma$ & $2.6$ \\
$B_1$ ($\upmu$G) &  $4.9$  \\
$h_{\rm B1}$ (kpc)        &  $0.1$  \\
$h_{\rm B2}$ (kpc)        &  $9.5$  \\
D (10$^{28}$\,cm$^2$\,s$^{-1}$)      &  $0.55$--$0.65$ \\
$\upchi_T^2$      &    $0.9$--$2.0$ \\
\hline
\multicolumn{2}{c}{Model (ii) -- Advection (constant speed)} \\
$\gamma$ & $2.4$  \\
$B_1$ ($\upmu$G) &  $4.7$  \\
$h_{\rm B1}$ (kpc)        &  $0.1$  \\
$h_{\rm B2}$ (kpc)        &  $5.5$  \\
$V_0$ (km\,s$^{-1}$) & 19--22 \\
$\upchi_T^2$      &    $0.5$--$1.2$  \\
\hline
\multicolumn{2}{c}{Model (iii) -- Advection (increasing speed)} \\
$\gamma$ & $2.2$  \\
$B_1$ ($\upmu$G) &  $2.6$  \\
$h_{\rm B1}$ (kpc)        &  $0.4$  \\
$h_{\rm B2}$ (kpc)        &  $15.0$  \\
$V_0$ (km\,s$^{-1}$)     &   18--22\\
$h_{\rm V}$ (kpc) & 3 \\
$\upchi_T^2$      &    $0.5$--$1.1$  \\
\hline
\end{tabular}
\end{table}

\section{Discussion}

\subsection{Cosmic ray transport}
The main cosmic ray transport of this galaxy is analysed via various indicators discussed in this section:

\noindent\textit{Halo scale heights}\\
The frequency dependence of the halo scale height ($\bar{z_0} \propto \upnu^x$) gives information about the type of CRE propagation, in the case of dominating synchrotron losses \citep{mulcahy2018,krauseetal2018,steinetal2019}. An exponent $x=-0.25$ is expected for diffusive transport and $x=-0.5$ for advective transport. A scenario, where the CREs emit most of their energy within the galaxy halo, is referred to as an \emph{electron calorimeter}. We are working with this assumption (even though it may not be fully valid for the LOFAR 150~MHz data) while keeping in mind that this galaxy has local outflows which are not characterizing the global mechanisms. Using $\bar{z_0}$ for the halo from Table~\ref{tab:N4013albflux_guassian} gives $x > -0.38 \pm 0.15$ between 6\,GHz and 1.5\,GHz. Considering the nonthermal Gaussian scale height in C-band it is $x = -0.20 \pm 0.10$. Thus the smaller C-band/L-band exponent can be attributed to the influence of thermal emission on the scale height at C-band. Together with an exponent of $x=-0.19 \pm 0.05$
between 1.5\,GHz and 150\,MHz this gives a clear preference for diffusive CRE propagation.\\

\noindent\textit{Fitting the total intensity distribution}\\
The scale height analysis performed here shows that Gaussian functions fit better to the vertical radio intensity profiles than exponential functions at L-band and the LOFAR frequency. Although the quality of both fits was similar for C-band, the Gaussian profile seems to represent the data slightly better. This suggests that the main transport mechanism of the CREs is diffusion, whereas exponential intensity profiles suggest advection as the main transport process \citep{heesenetal2016}. Dominating CRE diffusion implies that no wind or only a slow advection velocity is present in NGC~4013. This results in slow escape of CREs, so that the calorimetric assumption may be used. Therefore, the scale height ratio corroborates the finding of dominating diffusion.\\ 

\noindent\textit{Integrated flux densities of disk and halo}\\
The derived ratios of the integrated flux densities between disk and halo (Table~\ref{tab:N4013albflux_guassian}) decrease from C-band towards lower frequencies with values of 1.7 for C-band, 0.8 for L-band, and 0.3 for the LOFAR frequency. This result shows that the halo becomes more dominant towards lower frequencies. To compare the determined values with another galaxy, we have undertaken the same analysis for the two-component exponential fit of NGC~4666. \citet{steinetal2019} derived scale heights for the upper and the lower halo separately for one large strip. The mean amplitudes and scale heights for the two CHANG-ES bands are presented in Table~\ref{tab:N4666albflux_exp}. Interestingly, the ratios of the integrated flux densities for the thin disk and the halo of NGC~4666 are 0.35 in C-band and 0.33 in L-band, much smaller than in NGC~4013 and almost independent of frequency. The halo of NGC~4666 contributes about 74\% and 75\% of the total flux density in C-band and L-band, respectively, while the corresponding values for the halo of NGC~4013 are about 37\% and 56\%. The contribution of the halo of NGC~4013 becomes dominant at 150\,MHz with 77\% of the total flux density. NGC~4666 is transporting the CRs much faster into the halo by an advective wind emerging from the entire galaxy. The halo of NGC~4666 is relatively bright and the disk/halo flux density ratio does not vary with frequency. NGC~4013 is clearly different, with a relatively faint halo and a strong frequency dependence of the disk/halo flux density ratio. A possible reason is diffusive transport of CREs which leads to a more rapid aging due to stronger (energy-dependent) synchrotron losses, explaining why the contribution of the halo decreases with increasing frequency.

As a result we propose to use the multiplication of the mean amplitude ratios  $\bar{w}\,(\bar{w}_0/\bar{w}_1)$ (Disk/Halo) and mean scale height ratios $\bar{z}\,(\bar{z}_0/\bar{z}_1)$ (Disk/Halo) of different frequencies to get hints on the propagation of cosmic rays in the galaxy. In Table~\ref{tab:N4013uN4666amp} the main values are given. The values of the last row ($\bar{w} * \bar{z}$) show different frequency dependencies for the two galaxies NGC~4013 (diffusion-dominated) and NGC~4666 (wind-dominated).\\

\begin{table}
\captionof{table}{Mean amplitude ratios $\bar{w}\,(w_0/w_1)$ (Disk/Halo) and mean scale height ratios $\bar{z}\,(z_0/z_1)$ (Disk/Halo) of the thin disk/halo from Gaussian fits for NGC~4013 and from exponential fits for NGC~4666.}  
\label{tab:N4013uN4666amp}
\centering
\begin{tabular}{lccc}
\hline \hline
   &  C-band & L-band & LOFAR \\
\hline
NGC 4013 &&&\\
$\bar{w}\,(w_0/w_1)$ &  9.40 & 4.50 & 1.90 \\
$\bar{z}\,(z_0/z_1)$ &  0.18 & 0.18 & 0.15 \\
 $\bar{w} * \bar{z} $ &  1.7  & 0.81 & 0.29 \\
\hline
NGC 4666 &&&\\
$\bar{w}\,(w_0/w_1)$ & 1.38 & 0.97 &-  \\
$\bar{z}\,(z_0/z_1)$ & 0.26 & 0.34 & - \\
$\bar{w} * \bar{z} $ & 0.36 & 0.33 &- \\
\hline
\end{tabular}
\end{table}

\noindent\textit{Velocity of cosmic rays}\\
A local transport velocity was determined in Section~\ref{subsec:pol} via the extent of an emission blob and the synchrotron lifetime, resulting in a value of 160\,km\,s$^{-1}$. This velocity is lower than the rotational velocity of the galaxy (see Table~\ref{tab:bparameter}), whereas other galaxies have advection speeds of (2--3) $\times$ v$_{\mathrm rot}$ \citep{heesenetal2018}. Consequently, NGC~4013 hosts only a slow wind that does not reach the escape velocity within a few kpc from the disk. Furthermore, the calculated velocity refers to a local outflow and is therefore different from global advection velocities found via \textsc{spinnaker}.\\

\noindent\textit{Transition from diffusion to advection}\\
The CRE transport is a superposition of diffusion and advection. Our combined C- and L-band data are equally well modeled by diffusion and advection. In contrast, our best-fitting model for the combined L-band and LOFAR data is advection with an accelerating velocity, which fits slightly better than diffusion. Even then, diffusion may play a large role beside advection, as the resulting advection speed of about 20~$\rm km\,s^{-1}$ is fairly slow near the disk, but accelerates to about $50~\rm km\,s^{-1}$ at the edge of the observed halo at 4~kpc height.
More precisely, diffusion is expected to dominate over advection near the disk, where the CREs are relatively young (observed at L- and C-band), while at larger heights, where the CREs are older (as observed with LOFAR), advection is the dominant transport. The reason for this is that the height that cosmic rays travel scales as $z\propto \sqrt{D\,t}$ for diffusion\footnote{For 1D anisotropic diffusion, or $z\propto \sqrt{4D\,t}$ for 3D isotropic diffusion}, whereas for advection the travelling distance is $z\propto V\,t$. Hence, diffusion dominates over advection for young CRE, whereas for old CRE the opposite is true. Consequently, there is a certain time (CRE age), from which on advection dominates over diffusion.

For a constant wind speed, the transition from the diffusion-dominated regime near the disk to the advection-dominated regime in the halo occurs at a height of \citep{recchiaetal2016}:
\begin{equation}
    s_{\star} \approx (0.3\mathrm{-}0.6) \times \frac{D_{28}}{V_{100}}~\rm kpc,
\end{equation}
where $D_{28}$ is the diffusion coefficient in units of $10^{28}~\rm cm^2\,s^{-1}$ and $V_{100}$ is the advection speed in units of $100~\rm km\,s^{-1}$. With an advection speed of $V_{100}=0.2$ and a diffusion coefficient of $D_{28}=0.6$, we see that for our case $s_{\star}\approx 1$--2\,\rm kpc. Hence, for the low advection speeds we measure, diffusion would dominate the cosmic-ray transport up to a height of about $2$\,kpc. \\

\noindent\textit{X-ray emission}\\
Due to sensitivity of the data and the overall low X-ray brightness of NGC~4013, it was only possible to obtain a reliable spectral model fit to the emission from the galactic core. Nevertheless, since most of the X-ray emission seems to be associated with the inner part of the disk, the derived parameters of the hot gas should be representative for the entire area of the central disk, especially within uncertainties.

The derived temperature of the hot gas in the central region of NGC~4013 of about 3$\times$\,10$^6$\,K (Table~\ref{tab:XMMfluxes_n4013n4666}) is surprisingly low for the disk of a star-forming galaxy, as such temperatures are more typical for galactic halos \citep[e.g.,][]{strickland04,tuellmann06}. The low thermal energy density and hence low internal pressure may not be sufficient to drive an efficient outflow. This would confirm the findings from the analysis of the radio data.

\subsection{Correlation of magnetic field strength with star formation rate}

Comparing NGC~4013 to other galaxies, we found NGC~4594 (M104; the Sombrero galaxy) to be quite similar. They both have a comparably low star formation rate (SFR). They both have a dust lane in the optical and have undergone some interaction, with NGC~4013 having a huge warp and tidal stellar streams and NGC~4594 having a dominant nuclear bulge. The radio continuum and polarization intensities of both galaxies are comparable, without signs of an ongoing interaction. The radio continuum observations of NGC~4594 \citep{bajajaetal1988} show a similarly thin disk but with polarization only observable in the inner part of the galaxy rather than along the entire disk, as shown in our RM-synthesis results. In more recent observations by \citet{krauseetal2006}, a plane-parallel disk field, as well as vertical magnetic field components were found, comparable to our findings. In their study the magnetic field strength of NGC~4594 was determined to be~5.0\,$\upmu$G, which is quite small in comparison to other star-forming spiral galaxies \citep[][for edge-on galaxies]{beck2016,krauseetal2018}. We found here a similarly small magnetic field strength for NGC~4013 of 6.6\,$\upmu$G.

A possible correlation between the SFR and the magnetic field strength is revealed from the low magnetic field strength value in NGC~4013 and NGC~4594 (the Sombrero galaxy) and their low SFR of 0.5\,M$_{\odot}$~yr$^{-1}$ in NGC~4013 \citep{wiegertetal2015} and 0.13\,M$_{\odot}$~yr$^{-1}$ in NGC~4594 \citep{kharbetal2016}. Other galaxies following this trend are UGC~10288, with a SFR of 0.4\,M$_{\odot}$~yr$^{-1}$ and a mean magnetic field strength in the disk of~6.3\,$\upmu$G \citep{krauseetal2018}, and NGC~4565, with a SFR of 0.74\,M$_{\odot}$~yr$^{-1}$ \citep{wiegertetal2015} and a mean magnetic field strength in the disk of~6.0\,$\upmu$G (Schmidt et al. submitted). A correlation between the radio continuum and the SFR is expected, which follows from the well-defined FIR-radio-correlation \citep[e.g.,][]{schleicher2013}, where the emission is believed to originate in both wavelength regimes from the same population of young stars. To further study this correlation, a larger sample of galaxies should be analyzed and the different components of the magnetic field, namely the turbulent and ordered magnetic field component in galaxies, should be considered \cite[e.g.,][]{drzazgaetal2011}.

The efficiency of the $\upalpha$--effect of the mean-field $\upalpha$--$\upomega$ dynamo is coupled to the SFR of a galaxy \citep[e.g.,][]{becketal1996}. The SFR influences the supernova rate and thus the SN-induced turbulence of the ISM and the magnetic field. The dependence of the magnetic field strength on the SFR is also expected from theoretical simulations \citep[e.g.,][]{brendreetal2015}. A small-scale dynamo can efficiently amplify the turbulent magnetic field that serves as the seed for the large-scale dynamo \citep{arshakianetal2009}. If the SFR is small, the small-scale dynamo is working less effectively.

\subsection{Central source}
The total intensity distribution of NGC~4013 is dominated by the central point source, with 40\% of total flux density in C-band, 30\% in the L-band, and 13\% at LOFAR 150\,MHz  coming from that region. The radio continuum observations show a central intensity peak at this position. The radio spectral index flattens towards the lower frequencies to $\upalpha = -0.44$ between L-band and the LOFAR frequency. The high resolution L-band B-configuration data of CHANG-ES (including the data of NGC~4013 we use here), are used in \citet{irwinetal2019} to search for AGN features in 35 edge-on spiral galaxies, in the radio regime. NGC~4013 does not show any of the characteristic features associated with AGNs and was concluded not to be a clear AGN candidate.

The complementary XMM-Newton data show a hard X-ray spectrum and a relatively flat photon index but no clear AGN fingerprint. A dominant X-ray point source coinciding with the location of the radio peak is detected by \emph{Chandra}. These data are suggestive of an AGN.

The fact that we do not see clear signs of an AGN in NGC~4013 could be related to large amounts of dust in the galactic disk \citep{mosenkov18}, with the highest concentration in the galactic core. According to \citet{dudiketal2005} the central source is possibly a Compton-thick AGN. Due to the high internal absorption in such a case, the obtained X-ray luminosity for the central source of 7.1$^{+5.2}_{-3.0}$\,$\times$\,10$^{38}$\,{\rm erg}\,s$^{-1}$ can be considered as a lower limit.

\section{Summary and conclusions}

We present radio continuum CHANG-ES VLA and LOFAR data of the edge-on spiral galaxy NGC~4013 to analyze the total intensity characteristics, the central source and the large scale magnetic field as well as the cosmic ray transport. The total radio continuum intensity distribution of NGC~4013 is dominated by the central point source. The radio and supplemental X-ray data hint at the central source being a Compton-thick AGN as suggested by \citet{dudiketal2005}. The disk of NGC~4013 shows faint radio emission extending into the halo in C-band with a Gaussian halo scale height of 1.2\,kpc. The halo increases in L-band to a scale height of 2.0\,kpc and at 150\,MHz to 3.1\,kpc, so that at low frequencies a large radio halo is observed. The frequency dependence shows a clear preference for diffusive CRE transport. The appearance of a small radio halo in C- and L-band is very different from other spiral galaxies with higher star formation rates, e.g., in NGC~4666 \citep{steinetal2019}, where the radio halo at the CHANG-ES frequencies is more extended with a fast advective wind being apparent. It seems that in NGC~4013 fewer or less energetic CREs are reaching the halo, with probably a much weaker wind across the galaxy. This is confirmed by the analysis of the X-ray data, showing a low temperature of the hot gas in the central region of NGC~4013. The resulting lower thermal energy densities of the hot gas could prevent it from leaving the disk plane efficiently.

In this paper, the total flux densities have been derived separately for the thin disk and the halo of a galaxy for the first time. The radio halo of NGC~4013 is relatively faint and contributes only 37\% and 56\% of the total flux density in C-band and L-band, respectively, compared to 74\% and 75\% for the bright, wind-driven halo of NGC~4666. The decreasing trend of the ratio of the thin disk to halo with frequency, from C-band to LOFAR 150\,MHz, also suggests diffusion to be the main transport process in NGC~4013.

An advective outflow has also been found in the halo of our Milky Way \citep{everett2008}, which is consistent with the bright radio halo that contributes about 90\% of the total flux density at 408\,MHz \citep{beuermann1985}. We propose using either the multiplication of the mean amplitude ratios  $\bar{w}\,(w_0/w_1)$ (Disk/Halo) and mean scale height ratios $\bar{z}\,(z_0/z_1)$ (Disk/Halo) or the ratio of flux densities between disk and halo at different frequencies as another indicator of the type of CRE transport that is occurring.

Previous observations of \citet{condon1987} suggested a boxy halo around NGC~4013. This is not found here, but hints of a slightly dumbbell-shaped halo in L-band are seen, where the extent of radio emission perpendicular to the major axis is smaller towards the galaxy's center. This could be explained by synchrotron losses as the main energy loss of the CRE particles. The smaller extent of observed radio emission towards the center is due to the larger magnetic fields near the center of the galaxy, which leads to stronger energy losses.

Our 1D model suggests that diffusion or a slow and accelerated advective velocity of $\sim$ 20-50\,km s$^{-1}$ fit equally well as the main CR transport. While a slow advection velocity cannot be ruled out, the cosmic ray transport would be still dominated by diffusion up to 2\,kpc height in the halo. Diffusion was found for NGC~7462 by \citet{heesenetal2016} with the radio profile being better approximated by a Gaussian than by an exponential fit like in NGC~4013. Scale heights determined by fitting exponential functions are different from Gaussian ones, as a comparison in the three frequencies shows.

Polarized emission in the halo with vertical magnetic field orientations was observed up to about 6\,kpc above the plane, indicating an outflow in addition to diffusion. The velocity needed to transport CREs to such height is about 160\,km s$^{-1}$, less than observed in halos driven by a galactic wind. Considering this outflows as a local feature, they are consistent with the result from our transport model for the global transport properties.

In summary, the arguments for a synchrotron-loss dominated CRE halo of NGC~4013 with diffusive propagation are:
\begin{itemize}
\item the characteristic frequency dependence of the halo scale heights (Table~\ref{tab:N4013albflux_guassian}),
\item the large disk/halo flux density ratio in C-band, decreasing towards lower frequencies, showing that the halo becomes more dominant over the disk at lower frequencies (Table~\ref{tab:N4013albflux_guassian}),
\item the parabolic shape of the spectral index profile (Fig.~\ref{fig:N4013-spinteractive-high}),
\item the low temperature of the X-ray emitting gas (Table ~\ref{tab:XMMfluxes_n4013n4666}).
\end{itemize}

A regular, plane-parallel magnetic field was found to be apparent in the disk of NGC~4013. Additionally, in the low resolution polarized intensity image of C-band (robust~2), halo emission and vertical magnetic field orientations were found. The XMM-Newton image shows corresponding emission in the inner part of the galaxy as well as the halo, especially the south and the north-eastern parts show a similar distribution. We propose that the magnetic field has influenced the hot gas dynamics. This is supported by previous findings of other CHANG-ES galaxies like NGC~4631 (Mora-Partiarroyo et al., submitted.) or NGC~4666 \citep{steinetal2019}, where the magnetic field and X-ray morphologies are also similar.

The mean disk magnetic field strength derived with the equipartition formula of 6.6\,$\upmu$G is rather low. NGC~4013 has low intensity radio continuum and polarization emission and a rather small radio halo. This is probably caused by the low SFR leading to a less effectively working small-scale dynamo. Nevertheless, considering the strong impact of the past interaction(s) on NGC~4013, it is quite remarkable that a large-scale plane-parallel field in the entire galaxy disk as well as a vertical field in the halo is observed in polarization. Our RM data indicate that the large-scale dynamo is still able to generate an axisymmetric spiral field in the disk.

In conclusion, the interaction and the low SFR across the disk probably influence the appearance in the radio continuum of NGC~4013. In the inner third of the galaxy somewhat stronger star formation is apparent, with hints of a star-forming ring, where the polarized emission is strongest and the X-ray emission reaches into the halo. 
   
\begin{acknowledgements}
We thank the anonymous referee as well as Hans-Rainer Kl\"ockner for constructive and helpful suggestions to improve this work. This research is kindly supported and funded by the Hans-B\"ockler Foundation. This research was also supported and funded by the DFG Research Unit 1254 "Magnetisation of Interstellar and Intergalactic Media: The Prospects of Low-Frequency Radio Observations".
The LOFAR related research at Ruhr-University Bochum is supported by BMBF Verbundforschung under D-LOFAR IV – FKZ: 05A17PC1. The project has in part also benefit from the exchange programme between Jagiellonian University Krak\'ow and Ruhr-University Bochum. 
This research has made use of the NASA's Astrophysics Data System Bibliographic Services, and the NASA/IPAC Extragalactic Database (NED) which is operated by the Jet Propulsion Laboratory, California Institute of Technology, under contract with the National Aeronautics and Space Administration and has made use of "Aladin sky atlas" developed at CDS, Strasbourg Observatory, France. 
\end{acknowledgements}

\bibliography{Bibliography}

\begin{thebibliography}{70}
\expandafter\ifx\csname natexlab\endcsname\relax\def\natexlab#1{#1}\fi

\bibitem[{Arnaud(1996)}]{xspec}
Arnaud, K.~A. 1996, in ASP Conf. Series, Vol. 101, Astronomical Data Analysis
  Software and Systems V, ed. G.~Jacoby \& J.~Barnes, p17

\bibitem[{{Arshakian} {et~al.}(2009){Arshakian}, {Beck}, {Krause}, \&
  {Sokoloff}}]{arshakianetal2009}
{Arshakian}, T.~G., {Beck}, R., {Krause}, M., \& {Sokoloff}, D. 2009, \aap,
  494, 21

\bibitem[{{Bajaja} {et~al.}(1988){Bajaja}, {Dettmar}, {Hummel}, \&
  {Wielebinski}}]{bajajaetal1988}
{Bajaja}, E., {Dettmar}, R.-J., {Hummel}, E., \& {Wielebinski}, R. 1988, \aap,
  202, 35

\bibitem[{Baldi {et~al.}(2018)Baldi, Williams, McHardy, Beswick, Argo, Dullo,
  Knapen, Brinks, Muxlow, Aalto, Alberdi, Bendo, Corbel, Evans, Fenech, Green,
  KlÃ¶ckner, KÃ¶rding, Kharb, Maccarone, MartÃ­-Vidal, Mundell, Panessa,
  Peck, PÃ©rez-Torres, Saikia, Saikia, Shankar, Spencer, Stevens, Uttley, \&
  Westcott}]{baldietal2018}
Baldi, R.~D., Williams, D. R.~A., McHardy, I.~M., {et~al.} 2018, \mnras, 476,
  3478

\bibitem[{{Beck}(2016)}]{beck2016}
{Beck}, R. 2016, \aapr, 24, 4

\bibitem[{{Beck} {et~al.}(1996){Beck}, {Brandenburg}, {Moss}, {Shukurov}, \&
  {Sokoloff}}]{becketal1996}
{Beck}, R., {Brandenburg}, A., {Moss}, D., {Shukurov}, A., \& {Sokoloff}, D.
  1996, \araa, 34, 155

\bibitem[{{Beck} \& {Krause}(2005)}]{beckkrause2005}
{Beck}, R. \& {Krause}, M. 2005, Astron. Nachr., 326, 414

\bibitem[{{Bendre} {et~al.}(2015){Bendre}, {Gressel}, \&
  {Elstner}}]{brendreetal2015}
{Bendre}, A., {Gressel}, O., \& {Elstner}, D. 2015, Astron. Nachr., 336, 991

\bibitem[{{Beuermann} {et~al.}(1985){Beuermann}, {Kanbach}, \&
  {Berkhuijsen}}]{beuermann1985}
{Beuermann}, K., {Kanbach}, G., \& {Berkhuijsen}, E.~M. 1985, \aap, 153, 17

\bibitem[{{Bottema}(1996)}]{bottema1996}
{Bottema}, R. 1996, \aap, 306, 345

\bibitem[{{Brentjens} \& {de Bruyn}(2005)}]{brentjensdebruyn2005}
{Brentjens}, M.~A. \& {de Bruyn}, A.~G. 2005, \aap, 441, 1217

\bibitem[{{Briggs}(1995)}]{briggs1995}
{Briggs}, D.~S. 1995, in BAAS, Vol.~27, American Astronomical Society Meeting
  Abstracts, 1444

\bibitem[{{Calzetti} {et~al.}(2007){Calzetti}, {Kennicutt}, {Engelbracht},
  {Leitherer}, {Draine}, {Kewley}, {Moustakas}, {Sosey}, {Dale}, {Gordon},
  {Helou}, {Hollenbach}, {Armus}, {Bendo}, {Bot}, {Buckalew}, {Jarrett}, {Li},
  {Meyer}, {Murphy}, {Prescott}, {Regan}, {Rieke}, {Roussel}, {Sheth}, {Smith},
  {Thornley}, \& {Walter}}]{calzettietal2007}
{Calzetti}, D., {Kennicutt}, R.~C., {Engelbracht}, C.~W., {et~al.} 2007, \apj,
  666, 870

\bibitem[{Carter \& Read(2007)}]{carter}
Carter, J.~A. \& Read, A.~M. 2007, \aap, 464, 1155

\bibitem[{{Chamandy}(2016)}]{chamandy2016}
{Chamandy}, L. 2016, \mnras, 462, 4402

\bibitem[{{Chy{\.z}y} {et~al.}(2018){Chy{\.z}y}, {Jurusik}, {Piotrowska},
  {Nikiel-Wroczy{\'n}ski}, {Heesen}, {Vacca}, {Nowak}, {Paladino}, {Surma},
  {Sridhar}, {Heald}, {Beck}, {Conway}, {Sendlinger}, {Cury{\l}o}, {Mulcahy},
  {Broderick}, {Hardcastle}, {Callingham}, {G{\"u}rkan}, {Iacobelli},
  {R{\"o}ttgering}, {Adebahr}, {Shulevski}, {Dettmar}, {Breton}, {Clarke},
  {Farnes}, {Orr{\'u}}, {Pandey}, {Pandey-Pommier}, {Pizzo}, {Riseley},
  {Rowlinson}, {Scaife}, {Stewart}, {van der Horst}, \& {van
  Weeren}}]{chyzyetal2018}
{Chy{\.z}y}, K.~T., {Jurusik}, W., {Piotrowska}, J., {et~al.} 2018, \aap, 619,
  A36

\bibitem[{{Comer{\'o}n} {et~al.}(2011){Comer{\'o}n}, {Elmegreen}, {Knapen},
  {Sheth}, {Hinz}, {Regan}, {Gil de Paz}, {Mu{\~n}oz-Mateos},
  {Men{\'e}ndez-Delmestre}, {Seibert}, {Kim}, {Mizusawa}, {Laurikainen},
  {Salo}, {Laine}, {Athanassoula}, {Bosma}, {Buta}, {Gadotti}, {Ho},
  {Holwerda}, {Schinnerer}, \& {Zaritsky}}]{comeronetal2011}
{Comer{\'o}n}, S., {Elmegreen}, B.~G., {Knapen}, J.~H., {et~al.} 2011, \apjl,
  738, L17

\bibitem[{{Comer\'on} {et~al.}(2018){Comer\'on}, {Salo}, \&
  {Knapen}}]{comeronet2018}
{Comer\'on}, S., {Salo}, H., \& {Knapen}, J.~H. 2018, A\&A, 610, A5

\bibitem[{{Condon}(1987)}]{condon1987}
{Condon}, J.~J. 1987, \apjs, 65, 485

\bibitem[{{Condon} {et~al.}(1998){Condon}, {Cotton}, {Greisen}, {Yin},
  {Perley}, {Taylor}, \& {Broderick}}]{condonetal1998}
{Condon}, J.~J., {Cotton}, W.~D., {Greisen}, E.~W., {et~al.} 1998, \aj, 115,
  1693

\bibitem[{{Damas-Segovia} {et~al.}(2016){Damas-Segovia}, {Beck}, {Vollmer},
  {Wiegert}, {Krause}, {Irwin}, {We{\.z}gowiec}, {Li}, {Dettmar}, {English}, \&
  {Wang}}]{damasetal2016}
{Damas-Segovia}, A., {Beck}, R., {Vollmer}, B., {et~al.} 2016, \apj, 824, 30

\bibitem[{{Drzazga} {et~al.}(2011){Drzazga}, {Chy{\.z}y}, {Jurusik}, \&
  {Wi{\'o}rkiewicz}}]{drzazgaetal2011}
{Drzazga}, R.~T., {Chy{\.z}y}, K.~T., {Jurusik}, W., \& {Wi{\'o}rkiewicz}, K.
  2011, \aap, 533, A22

\bibitem[{{Dudik} {et~al.}(2005){Dudik}, {Satyapal}, {Gliozzi}, \&
  {Sambruna}}]{dudiketal2005}
{Dudik}, R.~P., {Satyapal}, S., {Gliozzi}, M., \& {Sambruna}, R.~M. 2005, \apj,
  620, 113

\bibitem[{{Dumke} {et~al.}(1995){Dumke}, {Krause}, {Wielebinski}, \&
  {Klein}}]{dumke1995}
{Dumke}, M., {Krause}, M., {Wielebinski}, R., \& {Klein}, U. 1995, \aap, 302,
  691

\bibitem[{{Everett} {et~al.}(2008){Everett}, {Zweibel}, {Benjamin}, {McCammon},
  {Rocks}, \& {Gallagher}}]{everett2008}
{Everett}, J.~E., {Zweibel}, E.~G., {Benjamin}, R.~A., {et~al.} 2008, \apj,
  674, 258

\bibitem[{Gabriel {et~al.}(2004)Gabriel, Denby, Fyfe, Hoar, Ibarra, Ojero,
  Osborne, Saxton, Lammers, \& Vacanti}]{sas}
Gabriel, C., Denby, M., Fyfe, D.~J., {et~al.} 2004, ASPC, 314, 759

\bibitem[{{Garc{\'{\i}}a-Burillo} {et~al.}(1999){Garc{\'{\i}}a-Burillo},
  {Combes}, \& {Neri}}]{garcia-burilloetal1999}
{Garc{\'{\i}}a-Burillo}, S., {Combes}, F., \& {Neri}, R. 1999, \aap, 343, 740

\bibitem[{{Han}(2006)}]{han06}
{Han}, J.~L. 2006, Chinese Journal of Astronomy and Astrophysics Supplement, 6,
  211

\bibitem[{{Han} {et~al.}(1997){Han}, {Manchester}, {Berkhuijsen}, \&
  {Beck}}]{han97}
{Han}, J.~L., {Manchester}, R.~N., {Berkhuijsen}, E.~M., \& {Beck}, R. 1997,
  \aap, 322, 98

\bibitem[{{Heesen} {et~al.}(2016){Heesen}, {Dettmar}, {Krause}, {Beck}, \&
  {Stein}}]{heesenetal2016}
{Heesen}, V., {Dettmar}, R.-J., {Krause}, M., {Beck}, R., \& {Stein}, Y. 2016,
  \mnras, 458, 332

\bibitem[{{Heesen} {et~al.}(2018){Heesen}, {Krause}, {Beck}, {Adebahr},
  {Bomans}, {Carretti}, {Dumke}, {Heald}, {Irwin}, {Koribalski}, {Mulcahy},
  {Westmeier}, \& {Dettmar}}]{heesenetal2018}
{Heesen}, V., {Krause}, M., {Beck}, R., {et~al.} 2018, \mnras, 476, 158

\bibitem[{{Henriksen}(2017)}]{henriksen2017}
{Henriksen}, R.~N. 2017, \mnras, 469, 4806

\bibitem[{{Henriksen} {et~al.}(2018){Henriksen}, {Woodfinden}, \&
  {Irwin}}]{henriksenetal2018}
{Henriksen}, R.~N., {Woodfinden}, A., \& {Irwin}, J.~A. 2018, \mnras, 476, 635

\bibitem[{{Ho} {et~al.}(1997){Ho}, {Filippenko}, \& {Sargent}}]{hoetal1997}
{Ho}, L.~C., {Filippenko}, A.~V., \& {Sargent}, W.~L.~W. 1997, \apjs, 112, 315

\bibitem[{{Ho} {et~al.}(1993){Ho}, {Shields}, \& {Filippenko}}]{hoetal1993}
{Ho}, L.~C., {Shields}, J.~C., \& {Filippenko}, A.~V. 1993, \apj, 410, 567

\bibitem[{{Hummel} {et~al.}(1991){Hummel}, {Beck}, \&
  {Dettmar}}]{hummeletal1991}
{Hummel}, E., {Beck}, R., \& {Dettmar}, R.-J. 1991, \aaps, 87, 309

\bibitem[{{Irwin} {et~al.}(2012){Irwin}, {Beck}, {Benjamin}, {Dettmar},
  {English}, {Heald}, {Henriksen}, {Johnson}, {Krause}, {Li}, {Miskolczi},
  {Mora}, {Murphy}, {Oosterloo}, {Porter}, {Rand}, {Saikia}, {Schmidt},
  {Strong}, {Walterbos}, {Wang}, \& {Wiegert}}]{irwinetal2012}
{Irwin}, J., {Beck}, R., {Benjamin}, R.~A., {et~al.} 2012, \aj, 144, 43

\bibitem[{{Irwin} {et~al.}(2013){Irwin}, {Krause}, {English}, {Beck}, {Murphy},
  {Wiegert}, {Heald}, {Walterbos}, {Rand}, \& {Porter}}]{irwinetal2013}
{Irwin}, J., {Krause}, M., {English}, J., {et~al.} 2013, \aj, 146, 164

\bibitem[{{Irwin} {et~al.}(2019){Irwin}, {Wiegert}, {Merritt}, {Wezgowiec},
  {Hunt}, {Woodfinden}, {Stein}, {Damas-Segovia}, {Li}, {Wang}, {Johnson},
  {Krause}, {Dettmar}, {Im}, {Schmidt}, {Miskolczi}, {Braun}, {Saikia},
  {English}, \& {Richardson}}]{irwinetal2019}
{Irwin}, J., {Wiegert}, T., {Merritt}, A., {et~al.} 2019, arXiv e-prints,
  arXiv:1905.05160

\bibitem[{{Irwin} {et~al.}(2018){Irwin}, {Henriksen}, {We{\.z}gowiec},
  {Damas-Segovia}, {Wang}, {Krause}, {Heald}, {Dettmar}, {Li}, {Wiegert},
  {Stein}, {Braun}, {Im}, {Schmidt}, {Macdonald}, {Miskolczi}, {Merritt},
  {Mora-Partiarroyo}, {Saikia}, {Sotomayor}, \& {Yang}}]{irwinetal2018}
{Irwin}, J.~A., {Henriksen}, R.~N., {We{\.z}gowiec}, M., {et~al.} 2018, \mnras,
  476, 5057

\bibitem[{Kaastra(1992)}]{kaastra}
Kaastra, J.~S. 1992, An X-Ray Spectral Code for Optically Thin Plasmas
  (Internal SRON-Leiden Report, updated version 2.0)

\bibitem[{Kalberla {et~al.}(2005)Kalberla, Burton, Hartmann, Arnal, Bajaja,
  Morras, \& P{\"o}ppel}]{lab}
Kalberla, P. M.~W., Burton, W.~B., Hartmann, D., {et~al.} 2005, \aap, 440, 775

\bibitem[{{Kharb} {et~al.}(2016){Kharb}, {Srivastava}, {Singh}, {Gallimore},
  {Ishwara-Chandra}, \& {Ananda}}]{kharbetal2016}
{Kharb}, P., {Srivastava}, S., {Singh}, V., {et~al.} 2016, \mnras, 459, 1310

\bibitem[{{Krause} {et~al.}(2018){Krause}, {Irwin}, {Wiegert}, {Miskolczi},
  {Damas-Segovia}, {Beck}, {Li}, {Heald}, {M{\"u}ller}, {Stein}, {Rand},
  {Heesen}, {Walterbos}, {Dettmar}, {Vargas}, {English}, \&
  {Murphy}}]{krauseetal2018}
{Krause}, M., {Irwin}, J., {Wiegert}, T., {et~al.} 2018, \aap, 611, A72

\bibitem[{{Krause} {et~al.}(2006){Krause}, {Wielebinski}, \&
  {Dumke}}]{krauseetal2006}
{Krause}, M., {Wielebinski}, R., \& {Dumke}, M. 2006, \aap, 448, 133

\bibitem[{{Li} \& {Wang}(2013)}]{liwang2013a}
{Li}, J.-T. \& {Wang}, Q.~D. 2013, \mnras, 428, 2085

\bibitem[{{Mart{\'{\i}}nez-Delgado} {et~al.}(2009){Mart{\'{\i}}nez-Delgado},
  {Pohlen}, {Gabany}, {Majewski}, {Pe{\~n}arrubia}, \&
  {Palma}}]{martinezetal2009}
{Mart{\'{\i}}nez-Delgado}, D., {Pohlen}, M., {Gabany}, R.~J., {et~al.} 2009,
  \apj, 692, 955

\bibitem[{{McMullin} {et~al.}(2007){McMullin}, {Waters}, {Schiebel}, {Young},
  \& {Golap}}]{mcmullinetal2007}
{McMullin}, J.~P., {Waters}, B., {Schiebel}, D., {Young}, W., \& {Golap}, K.
  2007, in Astronomical Society of the Pacific Conference Series, Vol. 376,
  Astronomical Data Analysis Software and Systems XVI, ed. R.~A. {Shaw},
  F.~{Hill}, \& D.~J. {Bell}, 127

\bibitem[{Mewe {et~al.}(1985)Mewe, Gronenschild, \& van~den Oord}]{mewe}
Mewe, R., Gronenschild, E. H. B.~M., \& van~den Oord, G. H.~J. 1985, \aaps, 62,
  197

\bibitem[{{Mosenkov} {et~al.}(2018){Mosenkov}, {Allaert}, {Baes}, {Bianchi},
  {Camps}, {Clark}, {Decleir}, {De Geyter}, {De Looze}, {Fritz}, {Gentile},
  {Holwerda}, {Hughes}, {Lewis}, {Smith}, {Verstappen}, {Verstocken}, \&
  {Viaene}}]{mosenkov18}
{Mosenkov}, A.~V., {Allaert}, F., {Baes}, M., {et~al.} 2018, \aap, 616, A120

\bibitem[{{Mulcahy} {et~al.}(2018){Mulcahy}, {Horneffer}, {Beck}, {Krause},
  {Schmidt}, {Basu}, {Chy{\.z}y}, {Dettmar}, {Haverkorn}, {Heald}, {Heesen},
  {Horellou}, {Iacobelli}, {Nikiel-Wroczy{\'n}ski}, {Paladino}, {Scaife},
  {Sridhar}, {Strom}, {Tabatabaei}, {Cantwell}, {Carey}, {Grainge}, {Hickish},
  {Perrot}, {Razavi-Ghods}, {Scott}, \& {Titterington}}]{mulcahy2018}
{Mulcahy}, D.~D., {Horneffer}, A., {Beck}, R., {et~al.} 2018, \aap, 615, A98

\bibitem[{{M{\"u}ller} {et~al.}(2017){M{\"u}ller}, {Krause}, {Beck}, \&
  {Schmidt}}]{mulleretal2017}
{M{\"u}ller}, P., {Krause}, M., {Beck}, R., \& {Schmidt}, P. 2017, \aap, 606,
  A41

\bibitem[{{Parker}(1992)}]{parker1992}
{Parker}, E.~N. 1992, \apj, 401, 137

\bibitem[{{Recchia} {et~al.}(2016){Recchia}, {Blasi}, \&
  {Morlino}}]{recchiaetal2016}
{Recchia}, S., {Blasi}, P., \& {Morlino}, G. 2016, \mnras, 462, 4227

\bibitem[{{Ruzmaikin} {et~al.}(1988){Ruzmaikin}, {Sokolov}, \&
  {Shukurov}}]{ruzmaikinetal1988}
{Ruzmaikin}, A.~A., {Sokolov}, D.~D., \& {Shukurov}, A.~M., eds. 1988,
  Astrophysics and Space Science Library, Vol. 133, {Magnetic fields of
  galaxies}

\bibitem[{{Schleicher} \& {Beck}(2013)}]{schleicher2013}
{Schleicher}, D.~R.~G. \& {Beck}, R. 2013, \aap, 556, A142

\bibitem[{{Shimwell} {et~al.}(2017){Shimwell}, {R{\"o}ttgering}, {Best},
  {Williams}, {Dijkema}, {de Gasperin}, {Hardcastle}, {Heald}, {Hoang},
  {Horneffer}, {Intema}, {Mahony}, {Mandal}, {Mechev}, {Morabito}, {Oonk},
  {Rafferty}, {Retana-Montenegro}, {Sabater}, {Tasse}, {van Weeren},
  {Br{\"u}ggen}, {Brunetti}, {Chy{\.z}y}, {Conway}, {Haverkorn}, {Jackson},
  {Jarvis}, {McKean}, {Miley}, {Morganti}, {White}, {Wise}, {van Bemmel},
  {Beck}, {Brienza}, {Bonafede}, {Calistro Rivera}, {Cassano}, {Clarke},
  {Cseh}, {Deller}, {Drabent}, {van Driel}, {Engels}, {Falcke}, {Ferrari},
  {Fr{\"o}hlich}, {Garrett}, {Harwood}, {Heesen}, {Hoeft}, {Horellou},
  {Israel}, {Kapi{\'n}ska}, {Kunert-Bajraszewska}, {McKay}, {Mohan},
  {Orr{\'u}}, {Pizzo}, {Prandoni}, {Schwarz}, {Shulevski}, {Sipior}, {Smith},
  {Sridhar}, {Steinmetz}, {Stroe}, {Varenius}, {van der Werf}, {Zensus}, \&
  {Zwart}}]{shimwelletal2017}
{Shimwell}, T.~W., {R{\"o}ttgering}, H.~J.~A., {Best}, P.~N., {et~al.} 2017,
  \aap, 598, A104

\bibitem[{{Shimwell} {et~al.}(2019){Shimwell}, {Tasse}, {Hardcastle}, {Mechev},
  {Williams}, {Best}, {R{\"o}ttgering}, {Callingham}, {Dijkema}, {de Gasperin},
  {Hoang}, {Hugo}, {Mirmont}, {Oonk}, {Prandoni}, {Rafferty}, {Sabater},
  {Smirnov}, {van Weeren}, {White}, {Atemkeng}, {Bester}, {Bonnassieux},
  {Br{\"u}ggen}, {Brunetti}, {Chy{\.z}y}, {Cochrane}, {Conway}, {Croston},
  {Danezi}, {Duncan}, {Haverkorn}, {Heald}, {Iacobelli}, {Intema}, {Jackson},
  {Jamrozy}, {Jarvis}, {Lakhoo}, {Mevius}, {Miley}, {Morabito}, {Morganti},
  {Nisbet}, {Orr{\'u}}, {Perkins}, {Pizzo}, {Schrijvers}, {Smith}, {Vermeulen},
  {Wise}, {Alegre}, {Bacon}, {van Bemmel}, {Beswick}, {Bonafede}, {Botteon},
  {Bourke}, {Brienza}, {Calistro Rivera}, {Cassano}, {Clarke}, {Conselice},
  {Dettmar}, {Drabent}, {Dumba}, {Emig}, {En{\ss}lin}, {Ferrari}, {Garrett},
  {G{\'e}nova-Santos}, {Goyal}, {G{\"u}rkan}, {Hale}, {Harwood}, {Heesen},
  {Hoeft}, {Horellou}, {Jackson}, {Kokotanekov}, {Kondapally},
  {Kunert-Bajraszewska}, {Mahatma}, {Mahony}, {Mandal}, {McKean}, {Merloni},
  {Mingo}, {Miskolczi}, {Mooney}, {Nikiel-Wroczy{\'n}ski}, {O'Sullivan},
  {Quinn}, {Reich}, {Roskowi{\'n}ski}, {Rowlinson}, {Savini}, {Saxena},
  {Schwarz}, {Shulevski}, {Sridhar}, {Stacey}, {Urquhart}, {van der Wiel},
  {Varenius}, {Webster}, \& {Wilber}}]{shimwelletal2019}
{Shimwell}, T.~W., {Tasse}, C., {Hardcastle}, M.~J., {et~al.} 2019, \aap, 622,
  A1

\bibitem[{{Stein} {et~al.}(2019){Stein}, {Dettmar}, {Irwin}, {Beck},
  {We{\.z}gowiec}, {Miskolczi}, {Krause}, {Heesen}, {Wiegert}, {Heald},
  {Walterbos}, {Li}, \& {Soida}}]{steinetal2019}
{Stein}, Y., {Dettmar}, R.-J., {Irwin}, J., {et~al.} 2019, \aap, 623, A33

\bibitem[{{Strickland} {et~al.}(2004){Strickland}, {Heckman}, {Colbert},
  {Hoopes}, \& {Weaver}}]{strickland04}
{Strickland}, D.~K., {Heckman}, T.~M., {Colbert}, E.~J.~M., {Hoopes}, C.~G., \&
  {Weaver}, K.~A. 2004, \apjs, 151, 193

\bibitem[{Str{\"u}der {et~al.}(2001)Str{\"u}der, Briel, Dennerl., Hartmann,
  Kendziorra, Meidinger, Pfeffermann, Reppin, \& Aschenbach}]{strueder}
Str{\"u}der, L., Briel, U., Dennerl., K., {et~al.} 2001, \aap, 365, 18

\bibitem[{{Tabatabaei} {et~al.}(2017){Tabatabaei}, {Schinnerer}, {Krause},
  {Dumas}, {Meidt}, {Damas-Segovia}, {Beck}, {Murphy}, {Mulcahy}, {Groves},
  {Bolatto}, {Dale}, {Galametz}, {Sandstrom}, {Boquien}, {Calzetti},
  {Kennicutt}, {Hunt}, {De Looze}, \& {Pellegrini}}]{tabatabaeietal2017}
{Tabatabaei}, F.~S., {Schinnerer}, E., {Krause}, M., {et~al.} 2017, \apj, 836,
  185

\bibitem[{{T{\"u}llmann} {et~al.}(2006){T{\"u}llmann}, {Pietsch}, {Rossa},
  {Breitschwerdt}, \& {Dettmar}}]{tuellmann06}
{T{\"u}llmann}, R., {Pietsch}, W., {Rossa}, J., {Breitschwerdt}, D., \&
  {Dettmar}, R.-J. 2006, \aap, 448, 43

\bibitem[{{Tully} {et~al.}(1996){Tully}, {Shaya}, \& {Pierce}}]{tullyetal1996}
{Tully}, R.~B., {Shaya}, E.~J., \& {Pierce}, M.~J. 1996, VizieR Online Data
  Catalog, 208, 479

\bibitem[{Turner {et~al.}(2001)Turner, Abbey, Arnaud, Balasini, Barbera, \&
  Belsole}]{turner}
Turner, M. J.~L., Abbey, A., Arnaud, M., {et~al.} 2001, \aap, 365, 27

\bibitem[{{Vargas} {et~al.}(2018){Vargas}, {Mora-Partiarroyo}, {Schmidt},
  {Rand}, {Stein}, {Walterbos}, {Wang}, {Basu}, {Patterson}, {Kepley}, {Beck},
  {Irwin}, {Heald}, {Li}, \& {Wiegert}}]{vargasetal2018}
{Vargas}, C.~J., {Mora-Partiarroyo}, S.~C., {Schmidt}, P., {et~al.} 2018, \apj,
  853, 128

\bibitem[{{Wang} {et~al.}(2015){Wang}, {Hammer}, {Puech}, {Yang}, \&
  {Flores}}]{wangetal2015}
{Wang}, J., {Hammer}, F., {Puech}, M., {Yang}, Y., \& {Flores}, H. 2015,
  \mnras, 452, 3551

\bibitem[{{Wiegert} {et~al.}(2015){Wiegert}, {Irwin}, {Miskolczi}, {Schmidt},
  {Mora}, {Damas-Segovia}, {Stein}, {English}, {Rand}, {Santistevan},
  {Walterbos}, {Krause}, {Beck}, {Dettmar}, {Kepley}, {Wezgowiec}, {Wang},
  {Heald}, {Li}, {MacGregor}, {Johnson}, {Strong}, {DeSouza}, \&
  {Porter}}]{wiegertetal2015}
{Wiegert}, T., {Irwin}, J., {Miskolczi}, A., {et~al.} 2015, \aj, 150, 81

\bibitem[{{Xu} \& {Han}(2014)}]{rm2014}
{Xu}, J. \& {Han}, J.-L. 2014, Research in Astronomy and Astrophysics, 14, 942

\bibitem[{{Zschaechner} \& {Rand}(2015)}]{zschaechnerrand2015}
{Zschaechner}, L.~K. \& {Rand}, R.~J. 2015, \apj, 808, 153

\end{thebibliography}
\bibliographystyle{aa}

\begin{appendix} 
\section{Additional tables and figures}
\subsection{Imaging values}

In Table~\ref{tab:rms4013} we list all achieved rms values for the different images used in this study.

\begin{table*}
    \centering
\begin{threeparttable}
\caption{rms values of NGC~4013}  
\label{tab:rms4013}
\begin{tabular}{lcccc} 
\hline\hline 
Observation       &Freq &  weighting  & resolution          & rms            	\\
\		          &(GHz)&           & (")             &($\upmu$Jy/beam)    \\
\hline 
LOFAR            & 0.15 & rob 0	  &   6~$\times$~6       &   94.3	     	\\
                 &      & rob 0	  &   10~$\times$~10     &   110          \\
                 &      & rob 0	  &   18~$\times$~18     &   204         \\
CHANG-ES L-band   & 1.58 & rob 0, 18~k$\lambda$ taper &   5.9~$\times$~6.0   &   16.7             	\\        
                &       &rob 0, 18~k$\lambda$ taper &    10~$\times$~10     &   21.9                   \\
                &       &rob 0, 18~k$\lambda$ taper &    18~$\times$~18     &   50.0                   \\
CHANG-ES C-band &  6.00 &  rob 0, 18~k$\lambda$ taper  &    5.2~$\times$~5.3 &	3.6	            \\
                 &      & rob 0, 18~k$\lambda$ taper &    10~$\times$~10    & 6.8                \\
                 &      &  rob 0, 18~k$\lambda$ taper &    18~$\times$~18      & 13.0                   \\
Polarization    & 6.00  & rob 2 &    5.2~$\times$~5.3   &	2.5	            \\
                &       & rob 2 &    15~$\times$~15     & 8.0                  \\	    
                &       & rob 2 &    18~$\times$~18     &  8.6                   \\
	  
\hline
\end{tabular}
\begin{tablenotes}
\footnotesize
\item \textbf{Notes.} Images with 10" resolution were used for the scale height analysis, the $\upsigma$ of the polarization was measured with the mean of Stokes Q and Stokes U. 
\end{tablenotes}
\end{threeparttable}
\end{table*}
\normalsize

\subsection{Exponential scale height fits of NGC 4013}
\begin{figure*}[ht]
 \centering
 \includegraphics[width=0.6\textwidth]{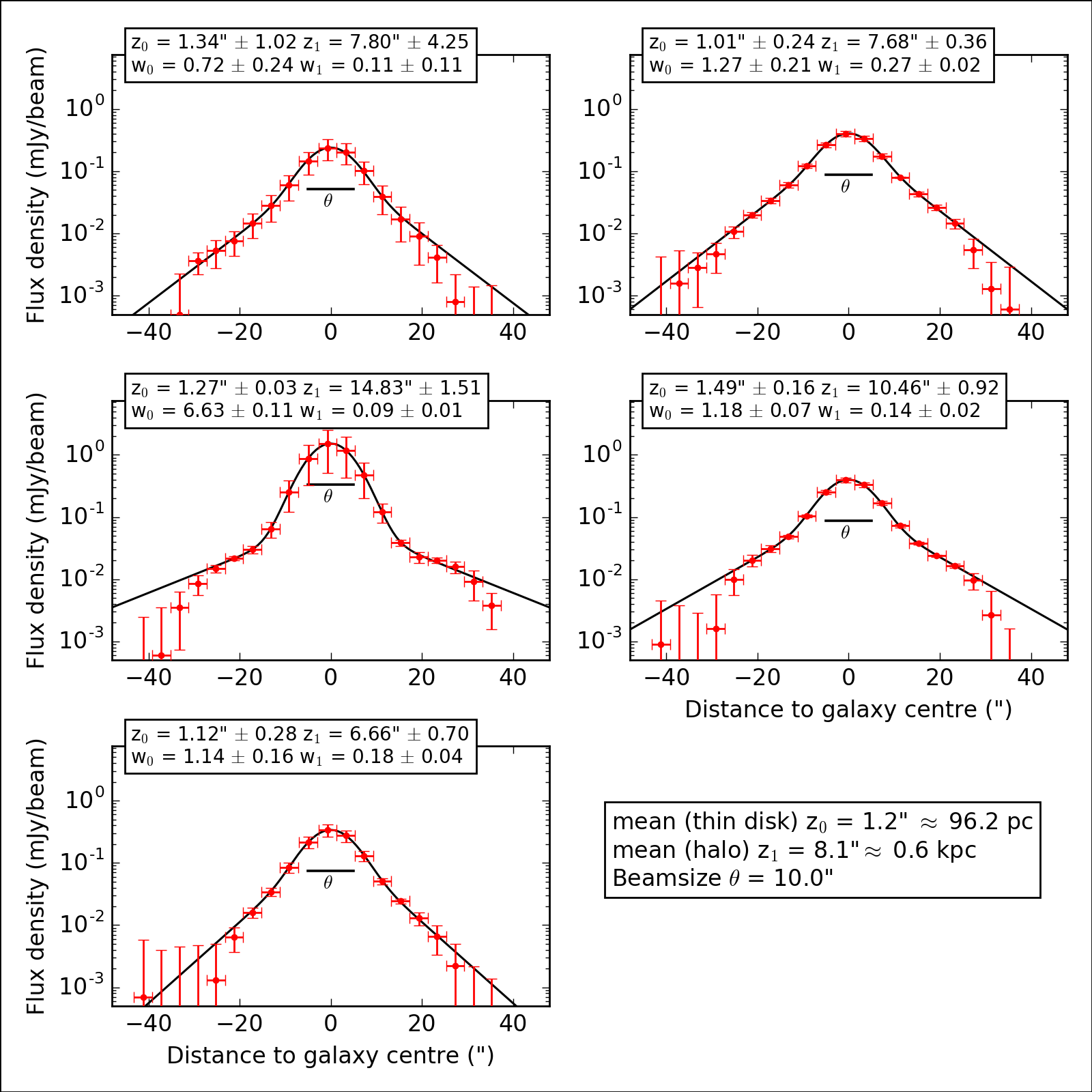}
 \caption{Strip fitting with NOD3 for five strips of NGC~4013 on the combined (C- and D-configuration) C-band data with a two-component exponential fit ($\upchi_{\text{total}}^2$~=~0.03) for comparison. Designations are the same as for Figure~\ref{fig:N4013_nod3_C}.}
 \label{fig:N4013_nod3_C_exp}
\end{figure*}

\begin{figure*}[ht]
 \centering
 \includegraphics[width=0.6\textwidth]{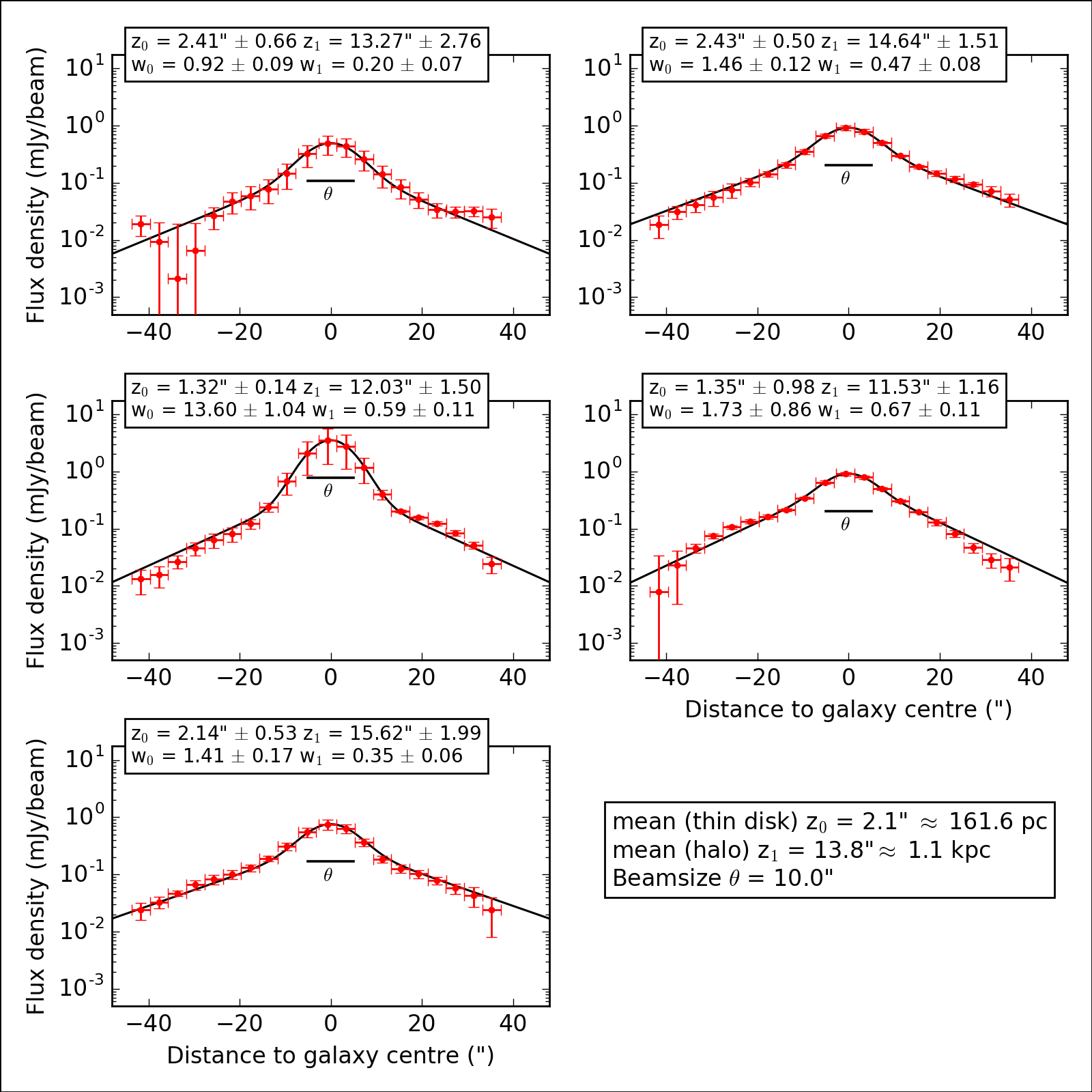}
 \caption{Strip fitting with NOD3 for five strips of NGC~4013 on the combined (B-, C- and D-configuration) L-band data with a two-component exponential fit ($\upchi_{\text{total}}^2$~=~1.11) for comparison. Designations are the same as for Figure~\ref{fig:N4013_nod3_C}.}
 \label{fig:N4013_nod3_L_exp}
\end{figure*}

\begin{figure*}[ht]
 \centering
 \includegraphics[width=0.6\textwidth]{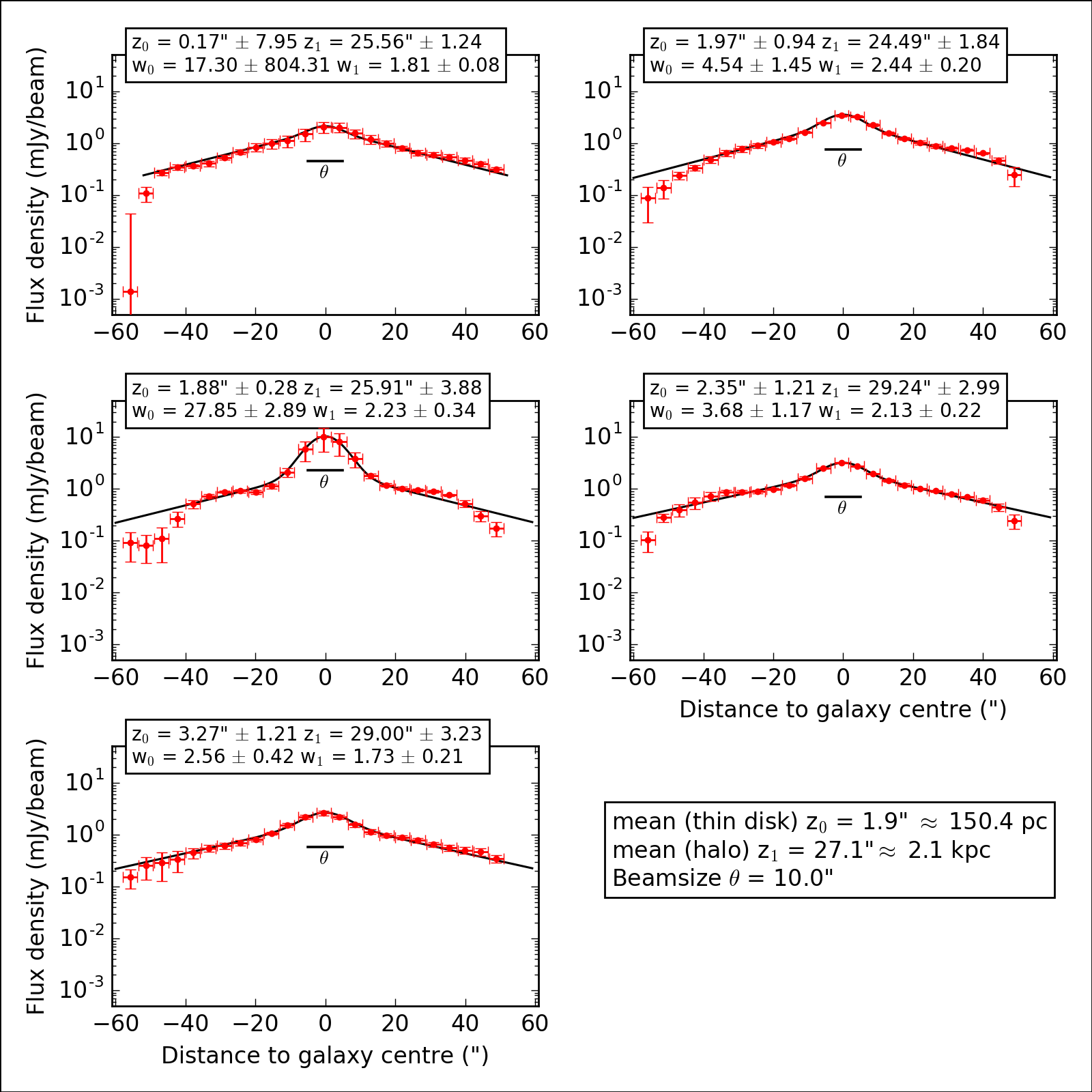} \caption{Strip fitting with NOD3 for five strips of NGC~4013 on the LOFAR data with a two-component exponential fit ($\upchi_{\text{total}}^2$~=~1.86) for comparison. Designations are the same as for Figure~\ref{fig:N4013_nod3_C}.}
 \label{fig:N4013_nod3_Lofar_exp}
\end{figure*}

In Figure~\ref{fig:N4013_nod3_C_exp} to Figure~\ref{fig:N4013_nod3_Lofar_exp} we show the exponential fits of the C-band, L-band and LOFAR data.

\subsection{Derived Gaussian flux densities of NGC 4666}

With the same approach as for NGC~4013, we derive exponential flux densities of the thin disk and halo for NGC~4666. 
With the exponential distribution of an intensity profile of a galaxy:
\begin{align}
w(z) = w_0 \ \text{exp}\,(-\,z/z_0)    
\end{align}
The integral between zero and infinity of this exponential is. 
\begin{align}
W (\text{mJy $\cdot$ arcsec /\,beam}) = w_0 \ \cdot \  z_0    
\end{align}
The full integral under the exponential curve is twice larger. $w_0$ is measured in mJy/beam and $z_0$ in arcsec. As the integrated flux density $W$ is in units per beam area, first we divide by the beam area in arcsec ($b$=200.4\,arcsec$^2$). To determine the total flux density $F$ of the thin disk and the halo, we multiply with the extent of the major axis, $e$ = 145" in C-band, $e$ = 155" in L-band, and use again the mean values of the amplitude and scale height: 

\begin{align}
F (\text{mJy}) = 2 \ \cdot  \bar{w_0} \ \cdot \ \bar{z_0} \cdot e\, /\, b    
\end{align}

The results are presented in Table~\ref{tab:N4666albflux_exp} where the entries to the right are taken from \citet{steinetal2019} for comparison.

\begin{table*}
\centering
\begin{threeparttable}
 \captionof{table}{Integrated flux densities of the thin disk and the halo of NGC~4666}  
\label{tab:N4666albflux_exp}
\label{tab:N4666scaleheight}
\begin{tabular}{lccccc|cc}
\hline \hline
\multicolumn{6}{c}{Exponential fits} & \multicolumn{2}{|c}{From total flux densities}\\  
     & $\bar{w}_0$  & $\bar{z}_0$ & $\bar{z}_0$& flux density& disk + halo & disk + halo & central source \\
     &(mJy/beam) & (") & (kpc) &  (mJy) &  (mJy) &  (mJy) &  (mJy) \\
\hline
C-band &        &     &   & && & \\
Disk  & 6.7~$\pm$~1.5 & 3.0 $\pm$ 1.2   & 0.4 $\pm$ 0.1 & 29 $\pm$ 13  & \multirow{2}{*}{111 $\pm$ 15}& \multirow{2}{*}{109 $\pm$ 1}& \multirow{2}{*}{1.8 $\pm$ 0.2} \\
Halo & 4.9~$\pm$~0.4 & 11.7 $\pm$ 0.5   & 1.6 $\pm$ 0.2 & 82 $\pm$ 8 & & &\\
Ratio &     1.38     & 0.26             &   &   0.35    & & & \\
\hline
L-band  &        &     & &  & & & \\
Disk  & 11.6~$\pm$~0.8 & 5.5 $\pm$ 0.9  & 0.74 $\pm$ 0.1  & 99 $\pm$ 18 & \multirow{2}{*}{397 $\pm$ 44} & \multirow{2}{*}{398 $\pm$ 2}& \multirow{2}{*}{4.4 $\pm$ 0.4} \\
Halo & 11.9~$\pm$~1.4 & 16.2 $\pm$ 0.9  & 2.1 $\pm$ 0.1    & 298 $\pm$ 40 & & & \\
Ratio  &  0.97        & 0.34            &                 & 0.33     &          &  &   \\
\hline
\end{tabular}
\begin{tablenotes}
\item \textbf{Notes.} $\bar{w}_0$ being the mean amplitude, $\bar{z}_0$ being the mean scale height. 
\end{tablenotes}
\end{threeparttable}
\end{table*}

\subsection{CR transport models}
\label{app:cr_transport}

\begin{figure}
	\centering
		\includegraphics[width=\hsize]{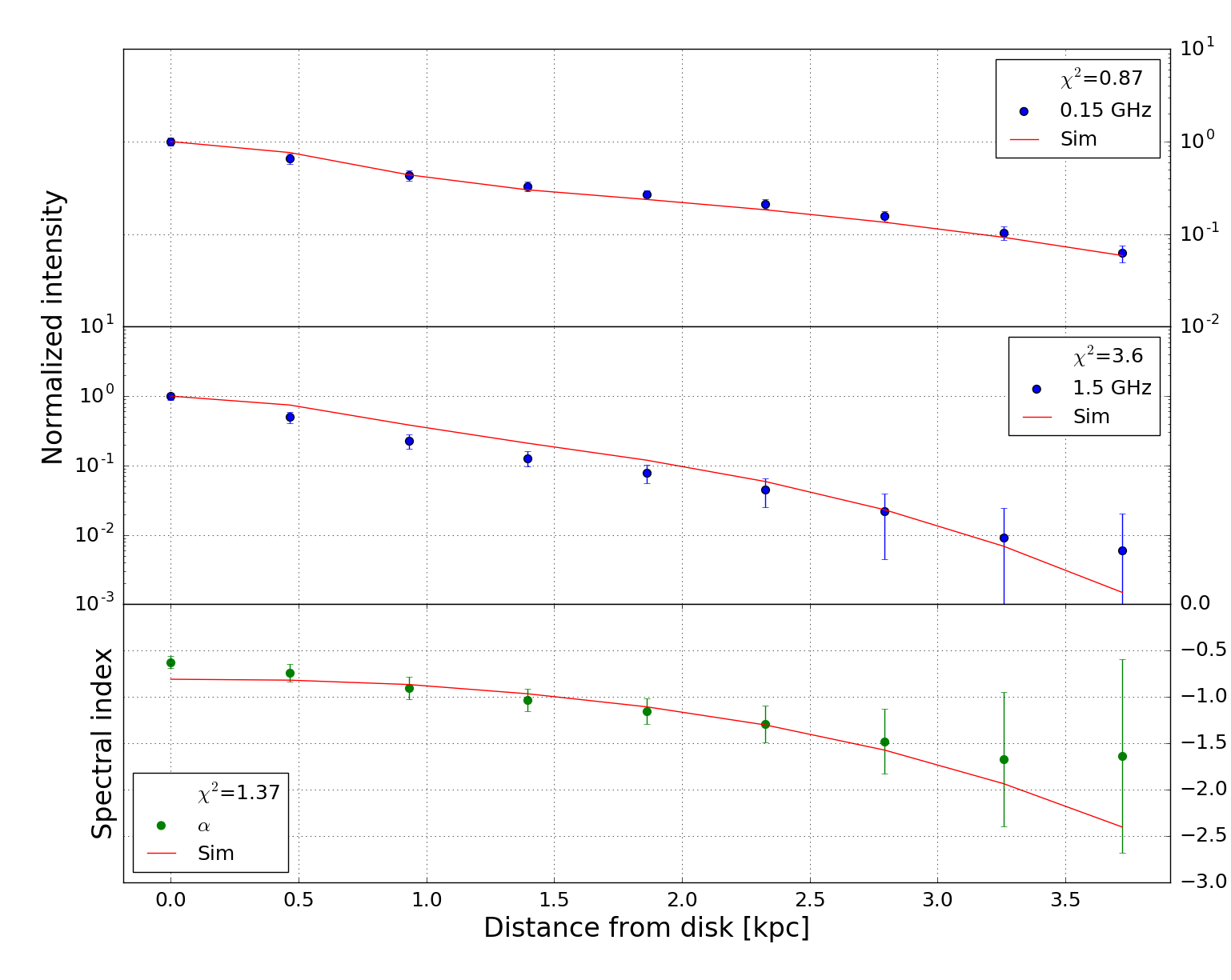}
	\caption{Best-fitting CRE diffusion model for NGC~4013. The top panel shows the 150\,MHz (LOFAR) data, the middle panel the 1.5\,GHz (L-band) data, and the bottom panel the nonthermal radio spectral index. In all panels, the solid red line represents the model simulation. The errors are represented by a weighted standard deviation. The best-fitting parameters are listed in Table~\ref{tab:N4013spinnaker}.}
	\label{fig:diff_low}
\end{figure}

\begin{figure}
	\centering
		\includegraphics[width=\hsize]{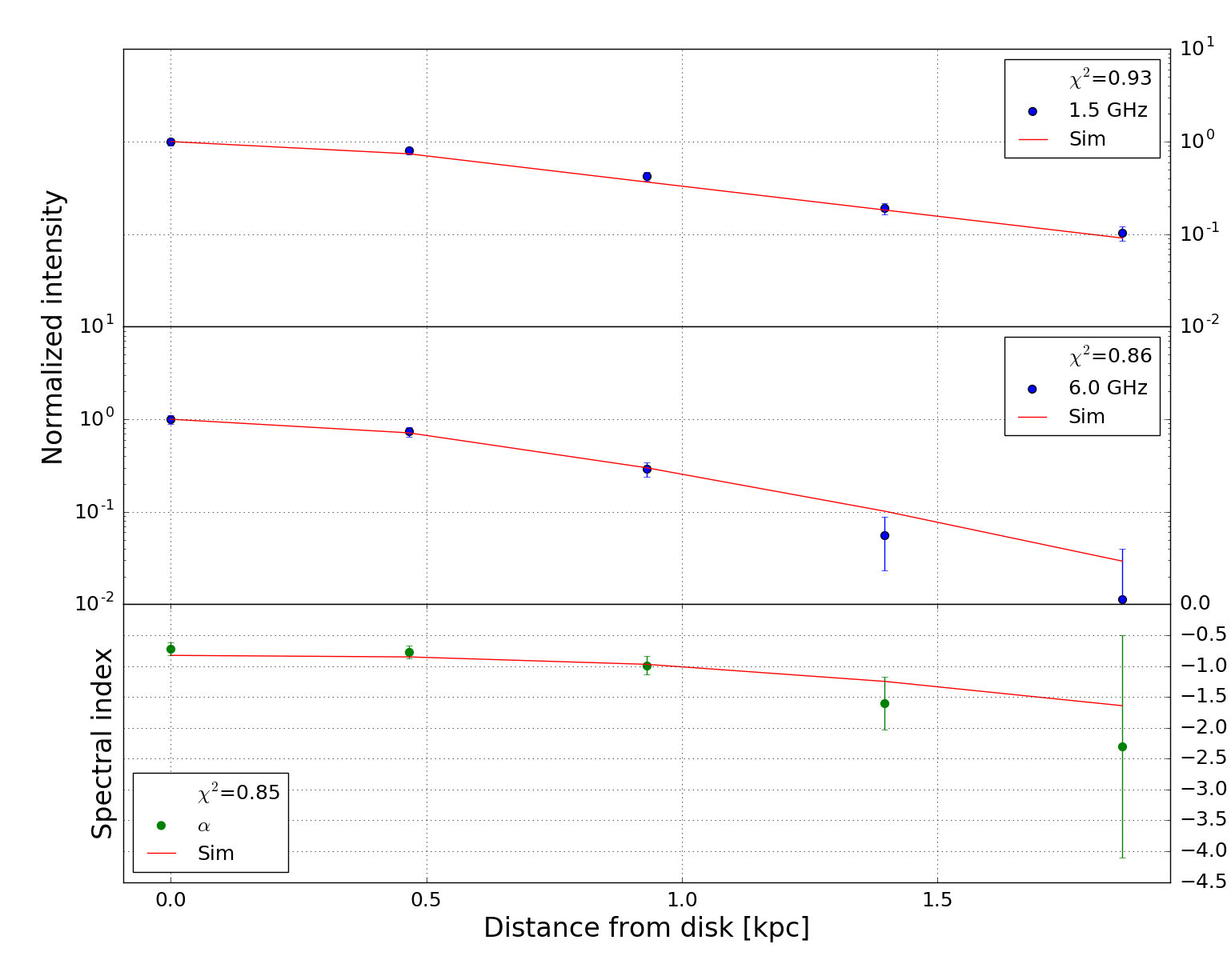}
	\caption{Best-fitting CRE diffusion model for NGC~4013. The top panel shows the 1.5\,GHz (L-band) data, the middle panel the 6\,GHz (C-band) data, and the bottom panel the nonthermal radio spectral index. In all panels, the solid red line represents the model simulation. The errors are represented by a weighted standard deviation. The best-fitting parameters are listed in Table~\ref{tab:N4013spinnaker}.}
	\label{fig:diff_high}
\end{figure}

\begin{figure}
	\centering
		\includegraphics[width=\hsize]{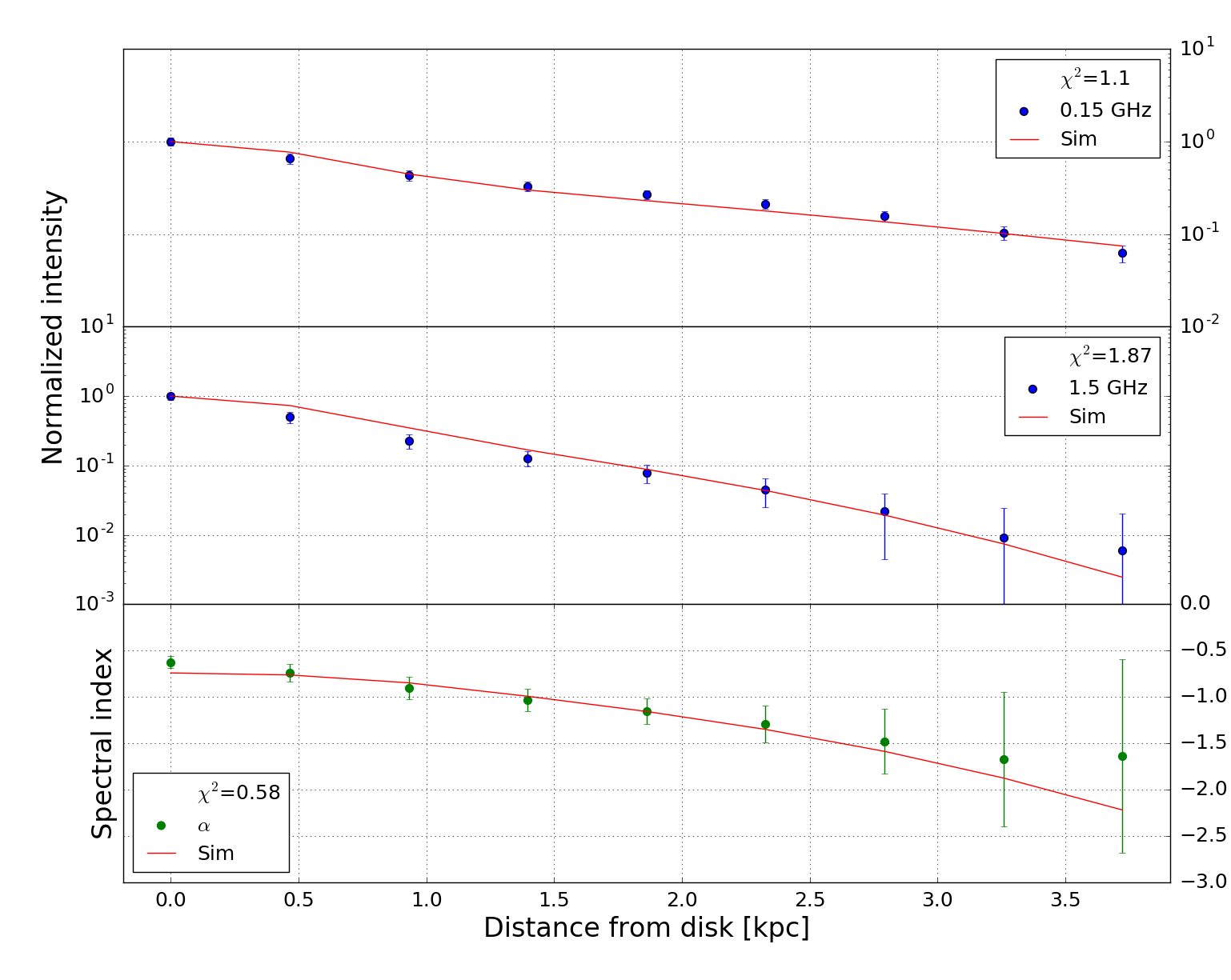}
	\caption{Best-fitting CRE advection model with a constant wind speed for NGC~4013. The top panel shows the 150\,MHz (LOFAR) data, the middle panel the 1.5\,GHz (L-band) data, and the bottom panel the nonthermal radio spectral index. In all panels, the solid red line represents the model simulation. The errors are represented by a weighted standard deviation. The best-fitting parameters are listed in Table~\ref{tab:N4013spinnaker}.}
	\label{fig:adv_const_low}
\end{figure}

\begin{figure}
	\centering
		\includegraphics[width=\hsize]{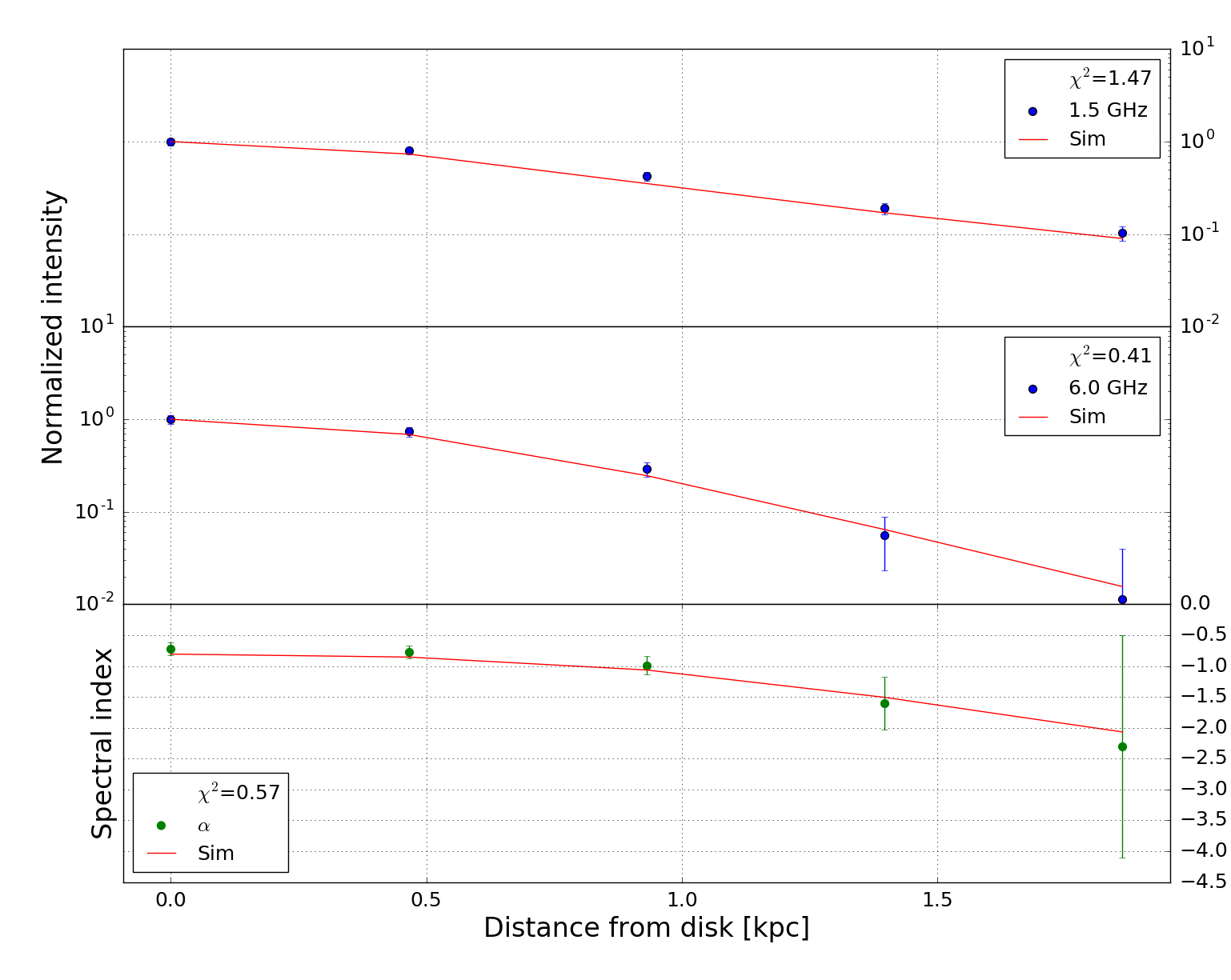}
	\caption{Best-fitting CRE advection model with a constant wind speed for NGC~4013. The top panel shows the 1.5\,GHz (L-band) data, the middle panel the 6\,GHz (C-band) data, and the bottom panel the nonthermal radio spectral index. In all panels, the solid red line represents the model simulation. The errors are represented by a weighted standard deviation. The best-fitting parameters are listed in Table~\ref{tab:N4013spinnaker}.}
	\label{fig:adv_const_high}
\end{figure}

In this section, we show the best-fitting diffusion and advection models with a constant wind speed. The diffusion models are shown in Figs.~\ref{fig:diff_low} and  \ref{fig:diff_high} and the advection models in Figs.~\ref{fig:adv_const_low} and \ref{fig:adv_const_high}.

\end{appendix} 

\end{document}